%% file: trotter-review_new.tex
\pdfoutput=1

\documentclass{iopjournal}

\usepackage{graphicx}
\usepackage[T1]{fontenc}
\usepackage[utf8]{inputenc}
\usepackage[UKenglish]{babel}
\usepackage{placeins}
\usepackage[shortlabels]{enumitem}
\usepackage[binary-units=true]{siunitx}
\usepackage{color}
\usepackage{amsmath}
\usepackage{amssymb}
\usepackage{amsthm}
\usepackage{amsfonts}
\usepackage{mathtools}
\usepackage{mathrsfs}
\usepackage{bbold}
\usepackage{dsfont}
\usepackage{braket}
\usepackage[ruled,lined]{algorithm2e}
\usepackage{nicefrac}
\usepackage{caption}
\usepackage[backend=bibtex8,style=phys,sorting=none,citestyle=numeric-comp,eprint=true,isbn=true]{biblatex}
\usepackage{cleveref}
\usepackage{tikz}
\usetikzlibrary{positioning}
\usetikzlibrary {arrows.meta}
\usepackage{pgfplots}
\usepackage{booktabs}

\newcolumntype{C}[1]{>{\centering\arraybackslash}m{#1}}
\newcolumntype{R}[1]{>{\raggedleft\arraybackslash}m{#1}}

\graphicspath{{figures/},{data/},{plots/}}
\newcommand{\inputtwofigs}[2]{\resizebox{0.98\textwidth}{!}{{\Large\input{#1}\input{#2}}}}
\newcommand{\inputepsfig}[2]{\resizebox{#2\textwidth}{!}{{\large\input{#1}}}}
\crefformat{footnote}{#2\footnotemark[#1]#3}

\DeclareMathOperator{\im}{i}
\DeclareMathOperator{\real}{Re}
\renewcommand{\Re}{\real}
\DeclareMathOperator{\imag}{Im}
\DeclareMathOperator{\res}{Res}
\DeclareMathOperator{\tr}{Tr}
\DeclareMathOperator{\diag}{diag}
\DeclareMathOperator{\sinc}{sinc}
\DeclareMathOperator{\asinh}{asinh}
\DeclareMathOperator{\acosh}{acosh}
\DeclareMathOperator*{\argmax}{arg\,max}
\DeclareMathOperator*{\argmin}{arg\,min}
\DeclareMathOperator{\ad}{ad}
\DeclarePairedDelimiter\abs{|}{|}
\DeclarePairedDelimiter\norm{\lVert}{\rVert}
\newcommand{\pdagger}{{\phantom{\dagger}}}
\newcommand{\matr}[1]{\begin{pmatrix}#1\end{pmatrix}}
\newcommand{\del}[2]{\ensuremath{\frac{\partial #1}{\partial#2}}}
\newcommand{\eto}[1]{\ensuremath{\mathrm{e}^{#1}}}
\newcommand{\trans}{\ensuremath{\mathsf{T}}}
\newcommand{\md}{\ensuremath{\mathrm{d}}}
\newcommand{\mD}{\ensuremath{\mathcal{D}}}
\newcommand{\id}{\ensuremath{\mathds{1}}}
\newcommand{\tint}{\ensuremath{\tau_\text{int}}}
\DeclareMathOperator{\ord}{\mathcal{O}}
\newcommand{\ordnung}[1]{\ensuremath{\ord\left(#1\right)}}
\newcommand{\erwartung}[1]{\ensuremath{\left\langle#1\right\rangle}}
\newcommand{\br}[1]{\ensuremath{\left(#1\right)}}
\newcommand{\brr}[1]{\ensuremath{\left[#1\right]}}
\newcommand{\brc}[1]{\ensuremath{\left\{#1\right\}}}

\newcommand{\licenseBY}{\href{https://creativecommons.org/licenses/by/4.0/}{CC BY 4.0}}
\newcommand{\licenseBYSA}{\href{https://creativecommons.org/licenses/by-sa/4.0/}{CC BY-SA 4.0}}
\newcommand{\licenseBYNCSA}{\href{https://creativecommons.org/licenses/by-nc-sa/4.0/}{CC BY-NC-SA 4.0}}
\newcommand{\licenseBYNCND}{\href{https://creativecommons.org/licenses/by-nc-nd/4.0/}{CC BY-NC-ND 4.0}}
\newcommand{\licenseCC}{\href{https://creativecommons.org/share-your-work/public-domain/cc0/}{CC Zero}}

\newcommand{\intaut}{integrated autocorrelation time}

\theoremstyle{definition}

\newtheorem{theorem}{Theorem}

\theoremstyle{remark}
\newtheorem*{remark}{Remark}

\newcommand{\todo}[1]{\textbf{\color{red}TODO: #1}}

\definecolor{darkblue}{rgb}{0.1, 0.1, 0.6}
\definecolor{plotblue}{RGB}{0, 114, 178}
\definecolor{plotorange}{RGB}{255, 127, 14}
\definecolor{plotgreen}{RGB}{44, 160, 44}
\definecolor{plotred}{RGB}{214, 39, 40}
\definecolor{plotpurple}{RGB}{148, 103, 189}
\definecolor{plotbrown}{RGB}{140, 86, 75}
\definecolor{plotpink}{RGB}{227, 119, 194}

\definecolor{0}{RGB}{51, 51, 51}
\definecolor{1}{RGB}{49, 130, 189}
\definecolor{2}{RGB}{255, 102, 0}
\definecolor{3}{RGB}{51, 160, 44}
\definecolor{4}{RGB}{227, 26, 28}
\definecolor{5}{RGB}{106, 61, 154}
\definecolor{6}{RGB}{140, 81, 10}
 
\makeatletter%
\makeatletter%
\begin{document}
	
	\articletype{Topical Review}
	
	\title{Suzuki-Trotter Decompositions\\ and other Methods for Quantum Time Evolution}
	
	\author{Johann Ostmeyer\orcid{0000-0001-7641-8030}}
	
	\affil{Helmholtz-Institut f\"{u}r Strahlen- und
		Kernphysik,
		University of Bonn, 53115 Bonn, Germany}
	
	\email{\href{mailto:ostmeyer@hiskp.uni-bonn.de}{ostmeyer@hiskp.uni-bonn.de}}
	
	\begin{abstract}
		(Suzuki-)Trotter decompositions, splitting methods, (Lie) product formulae...
		The most common numerical methods for the time evolution of quantum systems come with many names.
		And they are used practically everywhere with applications ranging from the solution of classical equations of motion and various Monte Carlo simulations to the real and imaginary time evolution on classical as well as quantum computers.
		Here we review the state of the art of said methods, focussing especially on the progress made over the last few years.
		We highlight recently discovered efficient time evolution algorithms and explain how best to use them in practice.
		A central part of this work is the estimation of error bounds that has improved greatly within the past decade.
		The relevance of time evolution methods for quantum computing is discussed with a focus on noisy hardware.
		Finally, a comprehensive overview of generalisations, related methods and alternatives to Trotterization is provided.
		This includes time-dependent Hamiltonian dynamics, processed methods, multi-product formulae, symplectic integrators, TDVP for tensor networks, quantum signal processing, Crouch-Grossman methods and more.
		The overall perspective in this work is that of a theoretical physicist.
		All mathematical proofs as well as some technical details are omitted for easier readability.
		Instead, this review serves as a hands-on guide and, of course, as a starting point for references that provide further details.
	\end{abstract}


	\allowdisplaybreaks[1]
	\unitlength = 1em
	
	\tableofcontents
	
	\newpage
	
	\section{Introduction}\label{sec:intro}

In quantum mechanics the dynamics of a state $\ket{\psi(t)}$ with time $t$ are governed by the \textit{Schrödinger equation}
\begin{align}
	\im \hbar \frac{\md}{\md t}\ket{\psi(t)} &= H\ket{\psi(t)}\,,
\end{align}
where $H$ is the \textit{Hamiltonian} operator and we set $\hbar = 1$ from now on.
Given an initial state $\ket{\psi(0)}$ and assuming $H$ to be time-independent, the Schrödinger equation is formally solved by
\begin{align}
	\ket{\psi(t)} = \eto{-\im H t} \ket{\psi(0)}\,.
\end{align}
The \textit{time evolution operator}
\begin{align}
	U(-\im t) &\equiv \eto{-\im H t}
\end{align}
can be understood as a matrix exponential since $H$ is a linear operator.

Conceptually, \textit{exact diagonalisation (ED)} is the simplest way to calculate $U(\tau)$ for some real or imaginary time $\tau$.
It suffices to diagonalise
\begin{align}
	H &= Q D Q^\dagger\,
\end{align}
where $D$ is a diagonal matrix containing the energy levels $E_n$ of $H$ as eigenvalues.
The energies are real and the basis transformation $Q$ is unitary because $H$ is hermitian.
Thus, the time evolution operator simplifies
\begin{align}
	U(\tau) &= Q \eto{D\tau} Q^\dagger
\end{align}
and all the exponentials remaining are scalar.

In practice, ED is only feasible analytically for very simple systems and numerically for relatively small systems.
For problems beyond ED one typically has to resort to approximations.
This review focuses on one of the most prominent numerical approximation methods.
In physics, this method is usually referred to as \textbf{Suzuki-Trotter decomposition} or simply Trotterization.
In a more mathematical context, the same concepts are called \textbf{splitting methods} or \textbf{(Lie) product formulae}.
No matter the name, the idea is to split the Hamiltonian into a sum of simpler operators
\begin{align}
	H &= \sum_{k=1}^{\Lambda} A_k
\end{align}
so that the matrix exponential $\eto{A_k \tau}$ of every individual operator is relatively easy to calculate.
Then the time evolution operator can be approximated via the $n$-th order scheme $S_n$
\begin{align}
	U(\tau) &= U(h)^{N_\tau} = S_n(h)^{N_\tau} + \ordnung{h^n}\label{eq:approx_time_evol_op}
\end{align}
by $N_\tau$ steps of length $h\equiv \frac{\tau}{N_\tau}$ up to a systematically controllable error $\ordnung{h^n}$.
The error can be expressed through the \textit{Baker-Campbell-Hausdorff (BCH) formula}
\begin{align}
	\eto{A}\eto{B} &= \eto{A+B + \frac12 [A,B] + \frac{1}{12} \br{[A,[A,B]] + [B,[B,A]]} + \cdots}\,,\label{eq:BCH}
\end{align}
where $[A,B] \equiv AB-BA$ denotes the commutator.
An immediate consequence is that the Trotter error vanishes if all the operators $A_k$ commute pairwise.
Since this case is trivially solvable, we will assume from now on that the operators do not commute, that is $[A_k,A_{k'}]\neq0$ for $k\neq k'$.

The order $n$ of the error is dictated by the specific Suzuki-Trotter decomposition $S_n$ composed of a product of terms $\eto{c_i A_k h}$ with a set of coefficients $c_i$.
Crucially, $n$ refers to the \textit{order of the global error} after the entire time $\tau$.
The local error after a single $h$-step, on the other hand, is of order $\ordnung{h^{n+1}}$.
It accumulates over $N_\tau$ steps to $\ordnung{N_\tau h^{n+1}} = \ordnung{\tau h^n}$.

The simplest product formula
\begin{align}
	S_1^\uparrow(h) &= \eto{A_1 h}\eto{A_2 h}\cdots \eto{A_\Lambda h}\label{eq:1st_order_Trotter}
\end{align}
is obtained by choosing all coefficients $c=1$ equal.
It reproduces the Lie product formula~\cite{lie1888theorie} (generalised to $\Lambda$ operators)
\begin{align}
	\eto{\sum_{k=1}^{\Lambda} A_k \tau} &= \lim\limits_{N_\tau\rightarrow\infty} \br{\eto{A_1 h}\cdots \eto{A_\Lambda h}}^{N_\tau}.
\end{align}
While this original 1st order Trotter scheme~\cite{trotter_original} is very intuitive, it is also highly inefficient and should \textit{never} be used in practice.
Instead, the strictly superior 2nd order \textit{Strang} splitting\footnote{It is also called \textit{Verlet} decomposition~\cite{Verlet:1967}, \textit{Størmer's} method or \textit{Leapfrog} algorithm and was rediscovered multiple times over the centuries~\cite{NumericalRecipes:2007}.}~\cite{Strang:1968} is obtained by simply reverting the order of the operators $A_\Lambda,\dots,A_1$ in every other step
\begin{align}
	\begin{split}
		S_2(h) &\equiv S_1^\uparrow \br{\frac h2 }S_1^\downarrow \br{\frac h2}\\
		&= \eto{\frac12A_1 h}\eto{\frac12A_2 h}\cdots \eto{\frac12A_\Lambda h}\eto{\frac12A_\Lambda h}\cdots\eto{\frac12A_2 h}\eto{\frac12A_1 h}\,.
	\end{split}
	\label{eq:leapfrog}
\end{align}

This concept readily generalises: an odd-order $n$ scheme can always be elevated to the even order $n+1$ by applying it alternately in the original and in the reverse order.
Throughout most literature including this review, therefore, the focus lies exclusively on \textit{symmetric} Suzuki-Trotter decomposition.

\begin{figure}[ht]
	\begin{center}
		\includegraphics[width=0.99\textwidth]{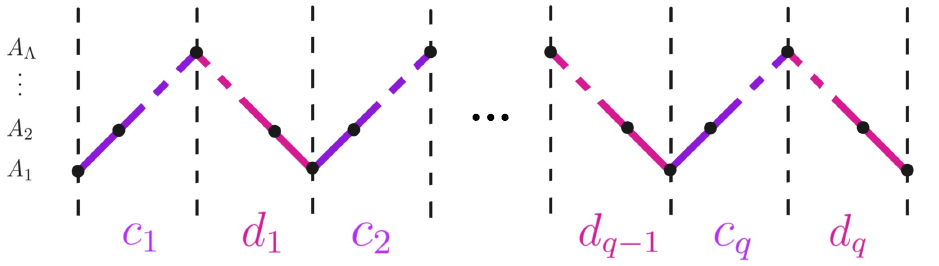}
	\end{center}
	\caption{Visualisation of a Suzuki-Trotter scheme~\cite{Ostmeyer:2022} with an arbitrary number $\Lambda$ of operators $A_k$ as in eq.~\eqref{eq:general_scheme}.
		There are $q$ \textit{cycles}, each consisting of two ramps (forward purple, backward pink).
		Symmetric schemes obey the identity $c_i=d_{q+1-i}$ for all $i\in\{1,\dots,q\}$.
		Reprinted figure from Ref.~\cite{Malezic:2026bds}.
	}\label{fig:ramp}
\end{figure}

Higher order Trotterizations are composed from lower order building blocks.
We have already seen this for $S_2$ which is composed out of two applications of $S_1$ (with opposite operator orders) each scaled by the coefficient $c=\frac12$.
The general form of the product formula~\eqref{eq:approx_time_evol_op} thus reads
\begin{equation}
	S_{n} (h) = \left( \prod_{k = 1}^{\Lambda} \eto{c_{1} A_{k} h} \right) \left( \prod_{k = \Lambda}^{1} \eto{d_{1} A_{k} h} \right) \cdots \left( \prod_{k = 1}^{\Lambda} \eto{c_{q} A_{k} h} \right) \left( \prod_{k = \Lambda}^{1} \eto{d_{q} A_{k} h} \right), \label{eq:general_scheme}
\end{equation}
where the $c_i,d_i$ coefficients correspond to forward and backward order of the operators, respectively.
In total, there are $q$ cycles through $A_1,\dots,A_\Lambda,\dots,A_1$ and as many independent coefficients for symmetric schemes, visualised in \cref{fig:ramp}.
It is essential that these products are ordered, e.g.\ $S_1^\downarrow (d_i h) = \prod_{k = \Lambda}^{1} \eto{d_{i} h A_{k}} = \eto{d_i A_\Lambda h}\cdots \eto{d_i A_1 h}$, since the operators $A_k$ do not commute and neither do their exponentials.
Depending on the convention, the products might be resolved right-to-left or left-to-right.
Both ways are used in the literature and this choice is irrelevant as long as it is consistent.
Moreover, for symmetric Suzuki-Trotter schemes the coefficients are, of course, the same read either way.

The construction of splitting methods as well as their selection, application, error estimation and the development of alternatives are challenging tasks and active areas of research.
In this review we collect the progress of the last several years made within the field of quantum time evolution.
Practical aspects relevant for an `end user' are covered as well as technical details of interest for a `developer'.

\subsection{Literature recommendations}\label{sec:literature}

Parts of this review might require non-trivial prior knowledge while others might motivate the reader to go on a quest for deeper understanding and further details.
In anticipation of both concerns, let us highlight several particularly relevant and useful reviews, books and lecture notes.

\subsubsection{Basics and prerequisites}

It goes without saying that a solid knowledge of linear algebra is essential not just for the time evolution of quantum systems but for practically any computation in physics.
Nevertheless, the efficient implementation of linear algebra operations on a (classical) computer can be surprisingly challenging.
The Numerical Recipes~\cite{NumericalRecipes:2007} cover all these basics and more in great detail.
Chapter~17 therein deals with the numerical integration of ordinary differential equations (ODEs) and is of particular interest in the context of this work.

\Cref{sec:quantum_comp} of this review deals with quantum computing and, unfortunately, we do not have the space to introduce this topic from scratch.
Instead, some textbook like that by Nielsen and Chuang~\cite{Nielsen_Chuang_2010} will have to do the heavy lifting.
Lin and Wiebe's lecture notes~\cite{Lin:2022vrd}, on the other hand, provide an invaluable collection of up-to-date quantum algorithms.

\subsubsection{Beyond this review}

The focus of this topical review lies on the progress made over the last few years.
This also means that well established results cannot be covered in detail.
It just so happens that Blanes, Casas and Murua have recently published a review~\cite{Blanes_Casas_Murua_2024} that perfectly fills this gap.
Ref.~\cite{Blanes_Casas_Murua_2024}  should be considered a first starting point by anyone interested in a broad overview and the mathematical background of splitting methods.
While the book by Hairer, Lubich and Wanner on geometric numerical integration~\cite{Hairer:2006gni} is less recent, it is certainly not outdated yet.
There is hardly any other reference as comprehensive on topics broadly related to symplectic integrators.

We hope that this review will prove useful in many branches of physics, but we also acknowledge that some topics will require further literature research for practical applicability.
In particular, tensor networks are only glanced over briefly in \cref{sec:tdvp}.
Fortunately, once more, there is a review at hand complementing this work: Paeckel et al.\ have collected time-evolution methods specifically for matrix-product states~\cite{Paeckel:2019yjf}.

\subsection{Outline of this review}

For starters, we will discuss the practical application of time evolution methods in \cref{sec:practice}.
This includes guidelines as to the optimal choice of method in a given setting.
In particular, alternatives to Suzuki-Trotter decompositions are highlighted before elaborating on specific Trotter schemes.
\Cref{sec:practice} ends with some hands-on tips and tricks concerning the efficient implementation of time evolution algorithms.

A brief summary of various techniques to construct higher order Trotter methods in \cref{sec:construction} is followed by a review of error estimators in \cref{sec:errors}.
Reliable error estimation can be crucial for the prediction of computational costs and, correspondingly, the required resources.
Upper and lower error bounds are discussed as well as generally applicable and model-specific estimators.

Subsequently, we provide an overview of aforementioned generalisations (sec.~\ref{sec:generalisations}), alternatives and further methods related to Trotterization (sec.~\ref{sec:alternatives}).
From a physical perspective, the main generalisation is the reintroduction of time dependence into the Hamiltonian $H(t)$.
Beyond that, we focus on approaches that might be more efficient than classical Trotterization in one setting or another.
Their applicability and advantages are covered as well as potential drawbacks.

Some of the alternatives like quantum signal processing (sec.~\ref{sec:qsp}) are optimised for the use on quantum computers.
Given the surging interest in quantum computing for time evolution, \cref{sec:quantum_comp} is dedicated to this topic in particular.
Quantum devices promise to provide a natural platform for efficient time evolution, but it is not yet mature enough for useful contributions to physics.
In \cref{sec:quantum_comp} we illuminate the long-term potential of quantum computers for time evolution without neglecting to emphasise the challenges currently posed by hardware noise and recent approaches to overcome these challenges.

We conclude in \cref{sec:conclusion}, ending with a number of open questions in the field. 	
	\section{Practical guidelines}\label{sec:practice}

Suzuki-Trotter decompositions are a tool for time evolution.
It is important to realise that they are not the goal in and of themselves.
Before proceeding with the implementation of a Trotterized time evolution, one should therefore consider whether there are superior alternatives.
Central questions concerning the application at hand are:
\begin{itemize}
	\item Can we get away with approximate unitarity?\\
	If yes, a polynomial \textit{Taylor} or \textit{Chebyshev} expansion (sec.~\ref{sec:polynomials}) of the time evolution operator %
	is typically computationally more efficient than any Trotter method.
	Alternatively, \textit{multi-product formulae} (MPF, sec.~\ref{sec:multi-product}) allow to obtain a higher order from linear combinations of lower order Trotterizations.
	\item Is a more specialised tool available?\\
	E.g.\ for \textit{tensor networks} the time-dependent variational principle (TDVP, sec.~\ref{sec:tdvp}), for \textit{quantum computing} optimised circuits based on quantum signal processing (QSP, sec.~\ref{sec:qsp}), or for \textit{classical equations of motion} (EOM) symplectic integrators (sec.~\ref{sec:symplectic}) might be applicable.
\end{itemize}
Having answered both questions with `no', it is likely that a Suzuki-Trotter decomposition is the method of choice.
In the following we will discuss how to choose and tune the most appropriate scheme for the desired time evolution.
For a visualisation of this decision tree, the flow chart at the end of Ref.~\cite{Ostmeyer:2022} is recommended.
Note also that the methods explained below are directly applicable for \textit{time-dependent Hamiltonians} following the prescription of \cref{th:time-dependent} in \cref{sec:time-dependent}.

\subsection{Method selection}\label{sec:methods}

During the development phase of a new project, it is usually a safe bet to start with the simple second order Strang splitting $S_2$ from equation~\eqref{eq:leapfrog}.
The method is sufficiently good for testing and very easy to implement.
Once correctness is ascertained and the computational cost of the time evolution becomes critical, higher order methods should be considered.

\begin{figure}[ht]
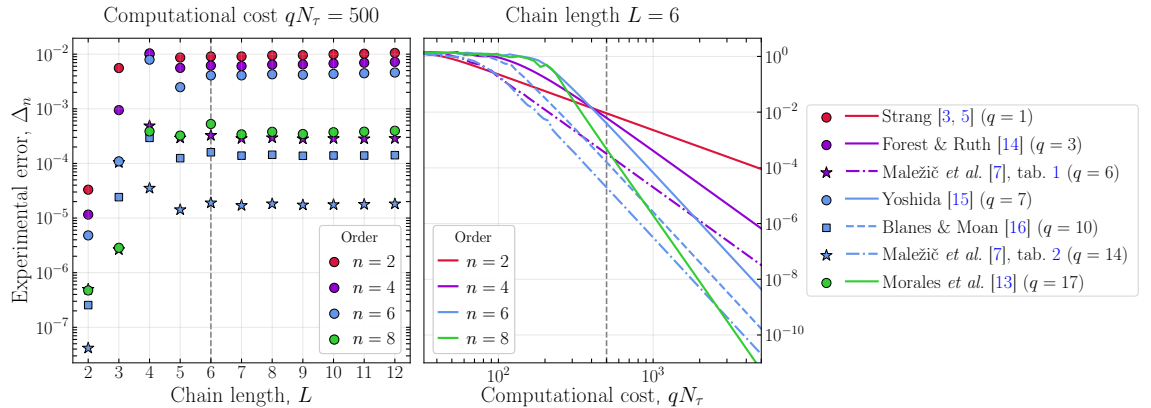

	\begin{center}
		\resizebox{0.99\textwidth}{!}{%
\begingroup%
\makeatletter%
%
\makeatother%
\endgroup%
 		}
	\end{center}
	\caption{Errors of Trotterized time evolution in the quantum Heisenberg model approximated by the relative error $\Delta_{n}$ in Frobenius norm as a function of the system size $L$ (left) and the computational cost $qN_\tau$ (right).
		All simulations used the fixed total time $t = 10$.
		The gray lines indicate where the two plots were simulated at, i.e.\ $q N_{\tau} = 500$ and $L = 6$, respectively.
		Adapted figure from Ref.~\cite{Malezic:2026xyi}.}\label{fig:Trotter_error}
\end{figure}

As a rule, in realistic time evolution problems the final time $\tau$ is dictated by the problem.
To reach this time after $N_\tau$ steps comes at a computational cost (number of floating point operations or gates) proportional to $qN_\tau$ where $q$ is the number of cycles within a single step.
Now a method needs to be found that allows to reach time $\tau$ with a minimal error for fixed computational resources or with minimal computational effort for a fixed target error.
Both approaches are easiest understood by applying them to the toy model simulations depicted in figure~\ref{fig:Trotter_error}.
The right panel shows the error after a fixed simulation time as a function of the computational cost.
Now, presume limited computational resources $qN_\tau\le 500$.
In that case one should choose the lowest-lying Trotter scheme with the smallest error along the $qN_\tau= 500$ line (here the order $n=6$ scheme in tab.~\ref{tab:recommened14} by Maležič et al.~\cite{Malezic:2026bds}).
Alternatively, one might require a relative error no larger than $\Delta_{n}\le \num{e-10}$.
Thus, the preferred method would be the left-most one along the $\Delta_{n}= \num{e-10}$ line (here the order $n=8$ scheme by Morales et al.~\cite{morales2022greatly}).

For the Suzuki-Trotter decompositions within a given order $n$ there is a clear efficiency ranking (smallest error at fixed cost and time, sec.~\ref{sec:omelyan}).
At the time of writing the best 4th and 6th order schemes are by Maležič et al.~\cite{Malezic:2026bds} with the coefficients provided in \cref{tab:recommened6,tab:recommened14}.
Highly efficient splitting methods of orders $n=8,10$ were found by Morales et al.~\cite{morales2022greatly}.
All of these schemes are implemented in the \texttt{qiskit-omelyan} package~\cite{qiskit-omelyan} as part of the Ecosystem for \texttt{Qiskit}~\cite{qiskit2024}.
Moreover, all the coefficients derived for Ref.~\cite{Malezic:2026bds} can be found together with the accompanying code~\cite{markomalezic_2026_18347430}.

\begin{table}[t]
	\begin{minipage}[t]{0.32\textwidth}
		\centering
		\captionof{table}{Coefficients for eq.~\eqref{eq:general_scheme} of the most efficient order $n = 4$ Suzuki-Trotter scheme to date~\cite{Malezic:2026bds}.%
		}\label{tab:recommened6}
		\begin{tabular}{S[table-format=1.0]S[table-format=-1.18]}
			\toprule
			$i$ & {$c_{i} = d_{q+1-i}$}   \\
			\midrule
			1  & 0.074082572180463262  \\
			2  & 0.232923088374338803  \\
			3  & 0.296820560634668408  \\
			4  & 0.122086989386933251  \\
			5  & -0.350153632343424469 \\
			6  & 0.124240421767020743  \\
			\bottomrule
		\end{tabular}
	\end{minipage}
	\hfill
	\begin{minipage}[t]{0.65\textwidth}
		\centering
		\captionof{table}{Coefficients for eq.~\eqref{eq:general_scheme} of the most efficient order $n = 6$ Suzuki-Trotter scheme to date~\cite{Malezic:2026bds}.%
		}\label{tab:recommened14}
		\begin{tabular}{S[table-format=1.0]S[table-format=1.18]|S[table-format=2.0]S[table-format=-1.18]}
			\toprule
			$i$ & {$c_{i} = d_{q+1-i}$} & $i$ & {$c_{i} = d_{q+1-i}$} \\
			\midrule
			1  & 0.037251326545569924 & 8  & -0.200016005078878524 \\
			2  & 0.120600278793781562 & 9  & 0.074145714537530386  \\
			3  & 0.266062994460763541 & 10 & 0.087345801243357893  \\
			4  & 0.163668553338143183 & 11 & 0.044234977360777830  \\
			5  & 0.071316838327437583 & 12 & -0.230821838291030424 \\
			6  & 0.058117508592333414 & 13 & -0.237197828922049295 \\
			7  & 0.188707697234255120 & 14 & 0.056583981858007803  \\
			\bottomrule
		\end{tabular}
	\end{minipage}
\end{table}

In contrast to the ranking above, a direct comparison of efficiencies between orders is not meaningful.
Asymptotically for small time steps $h\rightarrow0$, higher order methods will always beat lower order ones.
However, in practice we are rarely interested in the asymptotic regime and for large steps higher order methods often have a significant overhead.
This overhead is also highly dependent on the Hamiltonian $H$ and the choice of operators $A_k$ for the decomposition $H=\sum_k A_k$.
In addition, lower order methods allow more intermediate measurements because each step is shorter at the same overall cost.

That said, in many cases the decision is not as hard as it might seem at first.
For instance, the simulations shown in figure~\ref{fig:Trotter_error} feature a clear overall winner, namely the order $n=6$ scheme in tab.~\ref{tab:recommened14} by Maležič et al.~\cite{Malezic:2026bds}.
Up to a very slim high-error region and the very high-cost regime, this Trotterization always features the lowest error at a given computational cost.
Therefore, at least in simulations of systems similar to the quantum Heisenberg model used in~\cite{Malezic:2026bds} one can always use this decomposition without second thought.

\subsection{Tips and tricks}\label{sec:tips_and_tricks}

\subsubsection{Exploit sparsity of operators}\label{sec:sparsity}

In quantum computing local operators $\eto{A_k h}$ that act only on one or two spin-type degrees of freedom are naturally identified with local gates.
Similarly, in classical simulations it is essential to use sparse matrix implementations of these operators whenever possible.
Generally speaking, sparse matrix-vector multiplications reduce the runtime in $N$ dimensions from $\ordnung{N^2}$ to $\ordnung{N}$.
Often, it is not even required to store the operator $\eto{A_k h}$ explicitly (in matrix form).
Instead, a functional prescription can be the most efficient realisation.

Let us consider a simple example.
A quantum spin chain of length $L$ has the dimension $N=2^L$.
Working in the usual computational basis
\begin{align}
	\mathcal{B} &= \brc{\ket{0\cdots00},\, \ket{0\cdots01},\, \ket{0\cdots10},\dots,\ket{1\cdots11}}\,,
\end{align}
we can explicitly write down the action of a local operator on every state.
For instance, the Pauli matrix $\sigma^x$ acts on the $k$-th spin as
\begin{align}
	\eto{\im \sigma_k^x h} &= \cos(h) \mathds{1} + \im \sin(h) \sigma_k^x\\
	\Rightarrow \eto{\im \sigma_k^x h} & \ket{i_1,\dots,i_{k-1},i_k,i_{k+1},\dots,i_L}\\ &= \begin{cases}
		\cos(h) \ket{i_1,\dots,i_{k-1},0,i_{k+1},\dots,i_L} + \im \sin(h) \ket{i_1,\dots,i_{k-1},1,i_{k+1},\dots,i_L} & \text{if } i_k=0\,,\\
		\cos(h) \ket{i_1,\dots,i_{k-1},1,i_{k+1},\dots,i_L} + \im \sin(h) \ket{i_1,\dots,i_{k-1},0,i_{k+1},\dots,i_L} & \text{if } i_k=1\,.\\
	\end{cases}\nonumber
\end{align}
This prescription can be realised classically with a simple loop over all $N$ basis states without ever constructing the $N\times N$ matrix $\eto{\im \sigma_k^x h}$ explicitly.
Since $\sigma_k^x$ acts only on a single spin, the non-trivial operations can be reduced to a 2-dimensional subspace.

More generally, an operator involving $l$ spins requires $2^l$ non-trivial operations that have to be repeated $N$ times each, resulting in a runtime of order $\ordnung{2^l N}$.

\subsubsection{Tune on small systems}

Let us revisit figure~\ref{fig:Trotter_error}.
We have already learned from the right panel how to choose a Suzuki-Trotter decomposition or, more generally, any time evolution algorithm once representative benchmark simulations are available.
In practice, the target system is often too large and simulations thereof too expensive for full benchmarks.
Fortunately, according to Maležič et al.~\cite{Malezic:2026bds} benchmarks of the target system can be replaced by benchmarks on smaller systems using the same model.
This is demonstrated in the left panel of figure~\ref{fig:Trotter_error} where the relative error $\Delta_n$ is plotted for multiple lengths $L$ of the same model.
The total dimension of the system is $2^L$ in this example.
We observe that the error quickly approaches the thermodynamic limit.
In particular, the hierarchy of Trotter efficiencies remains constant for all $L\ge5$.
Thus, a viable procedure for finding the best model-dependent Trotterization is to first benchmark the model on a small and cheap system.
Whichever time evolution method performs best on the small systems, will likely be optimal on the target system as well.
This can be verified via simulations of a slightly larger (here e.g.\ $L=7$) system until convergence is reached.

\subsubsection{Choose the best operator decomposition $H=\sum_k A_k$}\label{sec:best_operators}

As a rule of thumb, one should try to find a decomposition of the Hamiltonian $H$ into as few operators $A_k$ as possible, i.e.\ minimise the number $\Lambda$ of operators under the constraint that all of them have to be cheap to exponentiate.
From a theoretical perspective, fewer operators come with tighter error bounds~\cite{Campbell:2020wqh,schubert2023trotter}, see \cref{th:Schubert_error_bounds}.
In addition, the first-same-as-last property of each cycle and of symmetric Trotterizations as a whole allows to save computational costs (gate count) by combining $\eto{d_i A_1 h}\eto{c_{i+1} A_1 h} = \eto{(d_i+c_{i+1})A_1 h}$.
The fewer intermediate operators cannot be merged, the larger this effect.
Moreover, efficient Suzuki-Trotter schemes like those in \cref{tab:recommened6,tab:recommened14} are (almost) always optimised for $\Lambda=2$ operators and their efficiency becomes less reliable for larger $\Lambda$.

There is no formal guarantee that this will lead to an optimal result, since the error in practice might be small even if theoretical error bounds are very loose.
Thus, it might be worth exploring some alternatives.
Different ways to decompose the Hamiltonian can be constructed systematically based on the commutation graph between the different operators~\cite{Simon:2026dtc}.
As has been demonstrated empirically in~\cite{Ostmeyer:2022}, there can be synergies between specific operator choices and Trotter schemes leading to a reduced error.
Multiple attempts have been made recently to optimise the operator decompositions in special cases~\cite{Tate:2026inm,Yan:2026jky,Negishi:2026ujz}.
Deviating from the minimal-$\Lambda$ rule should, however, remain the exception.

\subsubsection{Generalise efficient 2-operator Trotterizations}

As we will discuss in more detail in \cref{sec:construction}, it is much easier to construct a Suzuki-Trotter decomposition for exactly 2 operators,
\begin{align}
	\eto{(A+B)h + \ordnung{h^{n+1}}} &= \eto{Aa_1h}\eto{Bb_1h}\eto{Aa_2h}\cdots\eto{Bb_qh}\eto{Aa_{q+1}h}\label{eq:2-stage-decomposition}
\end{align}
than to build a product formula as in equation~\eqref{eq:general_scheme} for any number $\Lambda$ of operators $A_k$ directly.
Moreover, the efficiency of 2-operator schemes is well understood, allowing to optimise splitting methods over the decades~\cite{Omelyan_2002,BLANES2002313,Auzinger:2017,morales2022greatly,Ostmeyer:2022,Malezic:2026bds}.

A priori, it is not obvious that these methods are applicable to general Hamiltonians at all.
Fortunately, \cref{th:2-opt-to-many} provides a simple prescription that allows to construct the generally applicable coefficients $c_i,d_i$ for the decomposition~\eqref{eq:general_scheme} from the $a_i,b_i$ coefficients via a telescope sum.
This prescription has first been derived by McLachlan~\cite{McLachlan:1995otni} and an intuitive visual proof is provided in Ref.~\cite{Ostmeyer:2022}.

\begin{theorem}[Generalising coefficients for 2 operators~\cite{McLachlan:1995otni,Ostmeyer:2022}]\label{th:2-opt-to-many}
	Let the coefficients $a_i$ and $b_i$ as in equation~\eqref{eq:2-stage-decomposition} define a 2-operator decomposition of order $n$ and set
	\begin{align}
		c_1 &= a_1\,, & d_1 &= b_1 - c_1\,,\label{eq:c1_d1}\\
		c_2 &= a_2 - d_1\,,& d_2 &= b_2 - c_2\,,\\
		&\;\;\vdots & &\;\;\vdots\nonumber\\
		c_q &= a_q - d_{q-1}\,,& d_q &= b_q - c_q\,.\label{eq:cq_dq}
	\end{align}
	Then the decomposition $S_n(h)$ of $\Lambda\ge 2$ non-commuting operators $A_k$ as in equation~\eqref{eq:general_scheme} defined by these coefficients $c_i,d_i$ is exact to order $n$ as well.
\end{theorem}

We remark that specialised Suzuki-Trotter decompositions have been constructed for the case of exactly $\Lambda=3$ operators~\cite{Auzinger:2017}. 

\subsubsection{Exploit potential operators hierarchies $A\ll B$}

It is rather common in physics and related subjects that different operators come at vastly different scales, say $\norm{A} \ll \norm{B}$ in some norm.
In this case the smaller operator should be treated as a perturbation of the larger one.
Several methods have been derived to exploit such a separation of scales.
They are discussed in \cref{sec:hierarchy} in detail.
The principal idea is to elevate the error in $B$ to a higher order than $A$, or even to perform the time evolution in $B$ exactly and interpret perturbations by $A$ in the `interaction picture'.

\subsubsection{Think outside the box}

Any reader prepared to invest some additional tuning effort to reduce the error or computational cost by a few more percent is referred to \cref{sec:generalisations}.
In particular, complex coefficients (sec.~\ref{sec:complex}) allow for supreme efficiency at the cost of exact unitarity.
Alternatively, processed methods (sec.~\ref{sec:processed}) can be very useful for long-time evolution when results at intermediate time steps are not required.
Multi-product formulae (sec.~\ref{sec:multi-product}) construct higher order approximations of the time evolution operator from linear combinations of appropriately chosen standard Trotter product formulae.
Especially on noisy quantum hardware, randomised methods (sec.~\ref{sec:random}) might also be a viable option since they allow to use shallower circuit, trading hardware noise for algorithmic noise.
 	
	\section{Construction of higher order methods}\label{sec:construction}

Let us now dive into the details of product formula construction and optimisation.
For this we assume a Hamiltonian $H=A+B$ consisting of exactly two operators and refer to \cref{th:2-opt-to-many} for the generalisation to $\Lambda>2$ operators.
Repeatedly applying the BCH formula~\eqref{eq:BCH}, allows us to write
\begin{align}
	\eto{(A+B)h + \ord_1h + \ord_3 h^3 + \ord_5 h^5 + \cdots} &= \eto{Aa_1h}\eto{Bb_1h}\cdots\eto{Bb_qh}\eto{Aa_{q+1}h}\,,
\end{align}
where the error terms $\ord_{1,3,\dots}$ are defined by the coefficients $a_i,b_i$.
For instance
\begin{align}
	\ord_1 &= \br{\sum_{i=1}^{q+1} a_i-1} A + \br{\sum_{i=1}^q b_i-1}B\label{eq:explicit_1st_order_err}
\end{align}
denotes the first order \textit{local error}.
It needs to vanish so that the global error approaches zero in the limit of small steps $h\rightarrow0$.

Since we require the Suzuki-Trotter decomposition defined by the coefficients $a_i,b_i$ to be symmetric, all the local even-order errors $\ord_2=\ord_4=\cdots=0$ vanish immediately.
A scheme $S_n$ has the global error order $n$ if all the local errors $\ord_l=0$ with $l<n$ are zero and $\ord_{n+1}\neq 0$.
This leading order error $\ord_{n+1}$ then dominates in the so-called \textit{scaling region}, that is once the step size $h$ is small enough.

The error terms can always be written as a linear combination of nested commutators
\begin{align}
	\ord_1 &= (\nu-1) A + (\sigma-1)B\,,\label{eq:ord_1_errors}\\
	\ord_3 &=  \alpha [A, [A, B]] + \beta [B, [B, A]]\,,\label{eq:ord_3_errors}\\
	\begin{split}
		\ord_5 &= \gamma_1 [A, [A, [A, [A, B]]]] + \gamma_2 [A, [A, [B, [A, B]]]] + \gamma_3 [B, [A, [A, [A, B]]]]\\
		& + \gamma_6 [B, [B, [B, [B, A]]]] + \gamma_5 [B, [B, [A, [B, A]]]] + \gamma_4 [A, [B, [B, [B, A]]]]\,,
	\end{split}\label{eq:ord_5_errors}
\end{align}
with $\nu,\sigma$ defined by equation~\eqref{eq:explicit_1st_order_err} and the higher order error coefficients $\alpha,\beta,\gamma_k$ constructed in a similar manner.
The explicit derivation of the coefficients becomes very involved for higher orders and we refer to Ref.~\cite{Malezic:2026bds} for technical details.
There is no unique way from $\ord_3$ on to express the nested commutators and the coefficients $\alpha,\beta,\gamma_k$ depend on the specific basis choice.
Ref.~\cite{Arnal:2020xpt} lists the most common bases for the BCH splitting like the Hall~\cite{Hall1934} and the Lyndon basis~\cite{Lyndon1958}.
Casa et al.~\cite{Casas:2009,Arnal:2020xpt} have further introduced a very compact \textit{right-nested} basis which has been employed successfully for higher order Trotter scheme construction by Maležič et al.~\cite{Malezic:2026bds}.
In equation~\eqref{eq:ord_5_errors} we use this basis, featuring an aesthetically pleasing symmetry under $A\leftrightarrow B$ exchange and $\gamma_k$ reversal.

Counting the number of coefficients $\nu,\sigma,\alpha,\beta,\gamma,\dots$ yields the number of constraints $\# C$ that all need to be satisfied (all coefficients equal to zero) to obtain a Trotterization of given order $n$.
This constraint count has been summarised in \cref{tab:orders} up to order $n=10$.

\begin{table}[ht]
	\centering
	\caption{Number of constraints $\# C$ and free parameters $\# P$ for different orders $n$ and numbers of cycles $q$.
		Parentheses [] denote ranges of integer values and entries in []$_2$-ranges are duplicated (e.g.\ $[0, 1]_2 = \{0, 0, 1, 1\}$).
		Yoshida-type methods (sec.~\ref{sec:yoshida}) have the structure specified in eq.~\eqref{eq:construction_yoshida}.
		Table adapted from~\cite{Malezic:2026bds} based on~\cite{BLANES2002313} and expanded with formulae from~\cite{McLachlan:1995otni}.}
	\begin{tabular}{lccccccc}
		\toprule
		\multicolumn{1}{l}{Order $n$}        & \multicolumn{1}{c|}{2}      & \multicolumn{1}{c|}{4}      &                                  \multicolumn{2}{c|}{6}                                     & \multicolumn{2}{c|}{8}        &     10    \\
		\multicolumn{1}{l}{$\# C$} & \multicolumn{1}{c|}{2}      & \multicolumn{1}{c|}{4}      &                                  \multicolumn{2}{c|}{10}                                    & \multicolumn{2}{c|}{28}       &     84    \\
		\midrule
		\multicolumn{1}{l}{Cycles $q$}       & \multicolumn{1}{c|}{[1, 2]} & \multicolumn{1}{l|}{[3, 6]} & \multicolumn{1}{c|}{[7, 8]} & \multicolumn{1}{l|}{[9, 14]}  & \multicolumn{1}{c|}{[15, 26]} & \multicolumn{1}{c|}{[27, 30]} & [31, 66] \\
		\midrule
		\multicolumn{1}{l}{$\# P$ (Yoshida)}  & \multicolumn{1}{l|}{[0]$_2$} & \multicolumn{1}{l|}{[0, 1]$_2$} & \multicolumn{1}{l|}{[0]$_2$}      & \multicolumn{1}{l|}{[1, 3]$_2$}   & \multicolumn{1}{l|}{[0, 5]$_2$}        & \multicolumn{1}{l|}{[6, 7]$_2$}   &     [0, 17]$_2$     \\
		\multicolumn{1}{l}{$\# P$ (general symmetric)}  & \multicolumn{1}{c|}{[0, 1]} & \multicolumn{1}{l|}{[0, 3]} & \multicolumn{1}{c|}{-}      & \multicolumn{1}{l|}{[0, 5]}   & \multicolumn{1}{c|}{-}        & \multicolumn{1}{l|}{[0, 3]}   &     -     \\
		\bottomrule
	\end{tabular}\label{tab:orders}
\end{table}

A symmetric decomposition of $q$ cycles comes with $q+1$ independent parameters $a_i,b_i$.
Thus, the desired order conditions can always be fulfilled once $q\ge \# C-1$.
Intriguingly, from order $n\ge6$ on there exist solutions with fewer cycles than one would have expected~\cite{McLachlan:1995otni}.
These solutions are of Yoshida-type (sec.~\ref{sec:yoshida}) with the additional structure given in equation~\eqref{eq:construction_yoshida}.
The regions in which schemes of a given order and type can be realised are also collected in \cref{tab:orders} together with the number of free parameters.
Only Yoshida-type solutions exist from order $n\ge 10$ on.

\subsection{Omelyan's optimisation ansatz}\label{sec:omelyan}

The free parameters for $q\ge \# C$ can be optimised to derive maximally efficient Suzuki-Trotter schemes.
This approach has been pioneered by Omelyan et al.~\cite{Omelyan_2002,OMELYAN2003272} who have defined the efficiencies
\begin{align}
	\mathrm{Eff}_n &= \frac{1}{q^n \mathrm{Err}_n}
\end{align}
by means of the error
\begin{align}
	\mathrm{Err}_2 = \sqrt{|\alpha|^2+|\beta|^2}\,,\quad
	\mathrm{Err}_4 = \sqrt{\sum_{j=1}^6|\gamma_j|^2}
\end{align}
for orders $n=2,4$ and analogously for higher orders using the leading order error coefficients.
The factor $q^n$ accounts for the computational cost of a single step with the error $\mathrm{Err}_n$.

The (numerical) optimisation of these efficiencies quickly becomes very challenging which is why to date Omelyan's ansatz has not been applied for high orders $n>6$.
Whenever feasible, however, this method has produced the most efficient Trotter schemes of their time~\cite{Omelyan_2002,BLANES2002313,Ostmeyer:2022}.
For symplectic integrators (sec.~\ref{sec:symplectic}) the original work by Omelyan et al.~\cite{OMELYAN2003272} is still unsurpassed.
The current efficiency records for general Suzuki-Trotter decompositions of orders $n=4,6$ are held by Maležič et al.~\cite{Malezic:2026bds}.

\begin{figure}[ht]
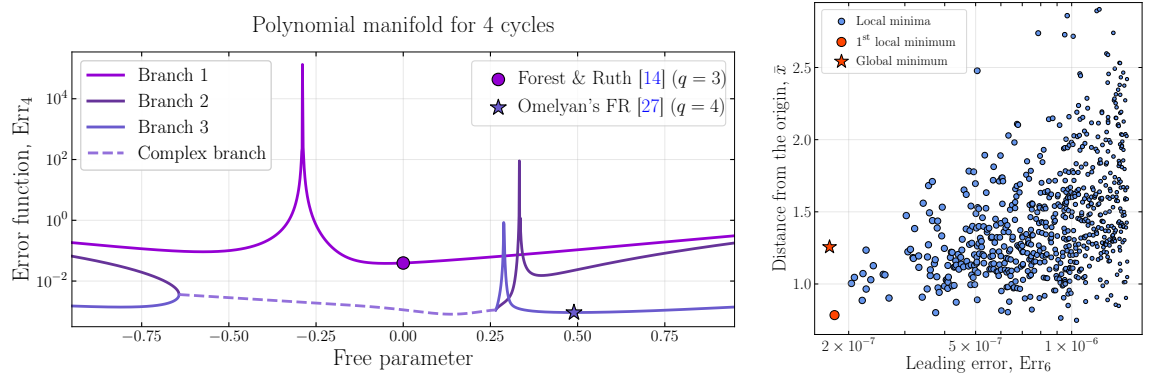

	\begin{center}
		\resizebox{0.65\textwidth}{!}{%
\begingroup%
\makeatletter%
%
\makeatother%
\endgroup%
 		}
	\end{center}
	\caption{Efficiency optimisation results.
		Left: Error manifolds of $4^{\textrm{th}}$ order schemes at $q = 4$ cycles,
		more manifold visualizations at $q = 5$ and $q = 6$ cycles are available~\cite{markomalezic_2026_18347430}.
		Right:
		A collection of order $n = 6$ schemes at $q = 14$ cycles plotted for their distance from the origin $\bar{x}$ against the leading-order error $\textrm{Err}_{6}$.
		The global and $1^{\textrm{st}}$ local minima are highlighted.
		The $1^{\textrm{st}}$ local minimum performs very well in practice and is provided in tab.~\ref{tab:recommened14}.
		Adapted figure from Ref.~\cite{Malezic:2026bds}.}\label{fig:origin_err}
\end{figure}

They have been obtained by a large-scale numerical search using the Levenberg-Marquardt algorithm~\cite{Levenberg:1944,Marquardt:1963} starting from multiple initial conditions.
As can be seen from \cref{fig:origin_err}, there are increasingly many local minima in the error landscape for higher numbers of cycles $q$.
With $q=4$ cycles (left) the error $\mathrm{Err}_4$ has 3 distinct real branches leading to 6 minima in total, the lowest of which is marked by a star.
There is an additional complex-valued minimum (sec.~\ref{sec:complex}) with even higher efficiency (listed in Sec.~3.2.6 of~\cite{Ostmeyer:2022}).
The picture becomes much less transparent for the error $\mathrm{Err}_6$ at $q=14$ cycles with hundreds of local minima.
Some of them are shown in the right panel of \cref{fig:origin_err} against the distance from the `origin'
\begin{equation}
	\bar{x}^{2} = 2 \sum_{i=1}^{q} \left( c_{i} - \frac{1}{2 q} \right)^{2}\,.\label{eq:origin}
\end{equation}

It has been observed by Maležič et al.~\cite{Malezic:2026bds} that the two local minima with the lowest theoretical error have very different magnitudes of the individual coefficients $c_i$, reflected in their $\bar x$ values.
As we will discuss in more detail in \cref{sec:long-time_errors}, the stability and long-time error accumulation is greatly enhanced for more evenly spread coefficients, that is smaller $c_i$.
Intuitively, this makes sense because fewer negative and overly large steps are taken, leading to a smoother interpolation of the dynamics.
This is why Maležič et al.\ ultimately recommend the 1st local minimum (tab.~\ref{tab:recommened14}) rather than the global one.

\subsection{Yoshida's method}\label{sec:yoshida}

As is vividly demonstrated by the increasing number of local minima from $q=4$ to $q=14$, there are practical limitations to the applicability of Omelyan's ansatz.
For orders as high as $n\ge8$ it makes sense to drop some degrees of freedom.
This is where the method pioneered by Yoshida~\cite{YOSHIDA1990262} comes in.
Rather than building the full Suzuki-Trotter decomposition from scratch with 2 operators (effectively from $S_1$ building blocks), Yoshida proposed to use 2nd order Strang splitting $S_2$ building blocks
\begin{align}
	S_{n}(h) &= S_2(w_mh)\cdots S_2(w_2h)S_2(w_1h)\, S_2\left(w_0h\right)\,S_2(w_1h)S_2(w_2h)\cdots S_2(w_mh)\,,\label{eq:construction_yoshida}\\
	w_0 &= 1-2\sum_{i=1}^m w_m\,.
\end{align}
The original motivation was to find the weights $w_i$ allowing for a minimal number of cycles $q=2m+1$ for a given order $n$ in the hope to simultaneously minimise the computational cost.
By now, it is well established that Trotter schemes with the minimal number of cycles are rarely efficient and it is worth increasing $m$ beyond the required minimum~\cite{Omelyan_2002,BLANES2002313}.
Nevertheless, the ansatz~\eqref{eq:construction_yoshida} is still well applicable and requires to solve far fewer order conditions than Omelyan's method.
This allowed Morales et al.~\cite{morales2022greatly} to find the most efficient product formulae of orders $n=8,10$ to date.

\subsection{Suzuki's recursive formula}\label{sec:suzuki}

A review of Suzuki-Trotter decompositions would not be complete without mentioning Suzuki's contributions formalising the theory~\cite{Suzuki:1976be} and developing a recursive construction of higher order splitting methods~\cite{suzuki_original,Hatano_2005}.
To this day the recursive construction
\begin{align}
	S_{n+2}(h) &= S_n(s_nh)^p\, S_n\left((1-2ps_n)h\right)\, S_n(s_nh)^p\,,\label{eq:construct_higher_order}\\
	s_n &= \frac{1}{2p-(2p)^\frac{1}{n+1}}
\end{align}
is very widely spread in time evolution applications.
This is understandable, not only because the construction is historically well established, but also because it is simple and very generally applicable.
In fact, any symmetric order-$n$ Trotterization $S_n$ and any integer $p\ge1$ lead to a valid scheme of order $n+2$.
However, only $p=2$ produces reasonably efficient product formulae and even they require $q=5^{n/2-1}$ cycles for order $n$ decompositions.
The sheer length of these splitting methods makes them highly impractical for $n+2 > 4$.

We recommend not to use this recursive construction any more.
Instead, the optimised Suzuki-Trotter decompositions discussed in the previous \cref{sec:omelyan,sec:yoshida} provide greatly superior alternatives. 	
	\section{Error estimation}\label{sec:errors}

Besides the derivation of ever more efficient time evolution methods, the last few years have seen a great collective effort in quantifying the errors of these methods.
The main challenge here is to replace the simple but very loose upper bounds based on the error terms in \cref{eq:ord_1_errors,eq:ord_3_errors,eq:ord_5_errors} by maximally tight estimators.
As a rule, some previously fixed target accuracy can be obtained using significantly larger Trotter time steps than naively expected~\cite{PhysRevX.11.011020,Heyl_2019}, especially for local observables.
Tighter error estimation allows to exploit this property and save compute time.

In \cref{sec:general_erros} we first review upper error bounds that are generally applicable for smooth bounded operators.
These bounds are broadly split into two classes: errors after a single Trotter step (sec.~\ref{sec:one-step_errors}) and error accumulation over long times (sec.~\ref{sec:long-time_errors}).
Lower error bounds are added in \cref{sec:lower_error_bounds}.
Next, in \cref{sec:non-dff_errors} we drop the ``smooth'' and ``bounded'' conditions and find that some state-depende bounds can still be derived.
General error bounds are typically not very tight since they do not take any specific features of the underlying model or measured observables into account.
For this reason a closer focus is laid on special observables, in particular the low-energy sector, in \cref{sec:low-energy_errors} and on some classes of models in \cref{sec:model-specific_errors}.
As opposed to these theoretical bounds, empirical error estimation methods are reviewed in \cref{sec:empirical_error}.

\subsection{General upper error bounds}\label{sec:general_erros}

As discussed in \cref{sec:construction}, the leading order Trotter error is given by a linear combination of higher order nested commutators.
The full evaluation of the resulting error is not feasible in most cases which is why the sum of operators is typically bounded using the triangle inequality.
A notable exception has been presented in~\cite{Blanes_2019} where embedded error estimation using two Trotter decompositions of different orders is derived.

Most of the errors estimates in the literature, both theoretical and empirical, are calculated for a specific operator norm.
For instance, Ref.~\cite{Malezic:2026bds} uses the Frobenius norm, Ref.~\cite{schubert2023trotter} relies on the spectral norm, and in~Ref.~\cite{morales2022greatly} the error of the eigenvalues is advocated.
It is an open problem how much can be concluded about specific observable errors from the various norms.
In practice, it has been observed that a Suzuki-Trotter decomposition with a small error in one norm will typically also perform well according to other metrics~\cite{morales2022greatly}.
This does not come as a surprise since on finite-dimensional spaces all norms are equivalent.
Therefore, the specific norm choice is typically less important than one might initially expect.

\subsubsection{Single-step error}\label{sec:one-step_errors}

The triangle inequality in its simplest form reproduces the error estimates minimised in Omelyan's method (sec.~\ref{sec:omelyan}).
The bounds obtained this way are, however, very loose and they are not applicable for more than two operators.
More general rigorous error bounds based on the same concept have been provided in~\cite{Childs:2018PNAS} and significantly improved in~\cite{Childs:2019PhRv}.
The series by (some of) the same authors culminates in~\cite{PhysRevX.11.011020}.

The results of~\cite{PhysRevX.11.011020} have yet again been significantly generalised by Schubert and Mendl~\cite{schubert2023trotter}.
They have derived rigorous error bounds for arbitrary Trotterizations.
Previously, similar error bounds had been found for the special case of the Hubbard model~\cite{Campbell:2020wqh}.
A large collection of error estimates has been derived in Ref.~\cite{Casas:2026krd} for general Trotter schemes as well as specifically for symmetric ones (sec.~\ref{sec:symmetric_errors}).
Generalised commutator-based error bounds for time-dependent Hamiltonians and Trotterized Lindbladian simulations have been derived in Refs.~\cite{An2021timedependent} and~\cite{Wang:2026qyq,Pillay:2026ojs}, respectively.

The bounds by Schubert and Mendl are presented in \cref{th:Schubert_error_bounds}.
They use the full product of $K$ operators.
That is, for regular Trotterizations with $q$ cycles and $\Lambda$ distinct operators this implies $K=2q(\Lambda-1)+1$ (using the first-same-as-last property).
Note also that the coefficients $c_i,d_i$ form equation~\eqref{eq:general_scheme} (or $a_i,b_i$ from eq.~\eqref{eq:2-stage-decomposition}) have already been absorbed into the operators $A_i$.
The bounds require to choose some splitting $s\in\{1,\dots,K\}$.
They are valid for every choice of $s$, but typically they are tightest for $s\approx \frac K2$.

\begin{theorem}[Generally applicable error bounds~\cite{schubert2023trotter,Schubert:2023code}]\label{th:Schubert_error_bounds}
	Let $S_n(h) = \eto{A_K h}\cdots \eto{A_1 h}$ be a product formula of order $n$.
	Then for every $s\in\{1,\dots,K\}$ the error of the product formula
	\begin{equation}
		\label{eq:product_formula_error_bound}
		\begin{split}
			\norm*{S_n(h) - \eto{H h}}_2 
			&\le \frac{\abs{h}^{n+1}}{(n+1)!} \Bigg( \sum_{j=2}^s \sum_{\substack{p_j + \dots + p_s = n \\ p_j \neq 0}} \binom{n}{p_j, \dots, p_s} \norm*{\ad_{A_s}^{p_s} \cdots \ad_{A_j}^{p_j} B_j}_2 \\
			&\hphantom{\le}\:\: + \sum_{j=s+1}^K \sum_{\substack{p_{s+1} + \dots + p_j = n \\ p_j \neq 0}} \binom{n}{p_{s+1}, \dots, p_j} \norm*{\ad_{A_{s+1}}^{p_{s+1}} \cdots \ad_{A_j}^{p_j} B_j}_2 \Bigg)
		\end{split}
	\end{equation}
	is bounded in spectral norm $\norm*{\cdot}_2$ with the $p$-fold adjoint action
	\begin{align}
		\ad_{C}^{p} D &\coloneqq \underbrace{[C,\cdots[C}_{p\text{-times}},D]\cdots]\quad \text{and}\label{eq:adjoint}\\
		B_j &\coloneqq \sum_{\ell=1}^{j-1} A_{\ell}, \quad j = 2, \dots, K\,.
	\end{align}
\end{theorem}

\begin{remark}
	The formula appears daunting at first, but there is no need to derive all the terms from scratch.
	The authors of~\cite{schubert2023trotter} also provide a python package~\cite{Schubert:2023code} automating the symbolic evaluation of the nested commutators.
	Moreover, a practical approach using projector Monte Carlo for the estimation of commutator norms has been presented in~\cite{Blunt:2025zul}.
\end{remark}

\begin{figure}
	\centering
	\begin{minipage}{0.53\textwidth}
		\centering
		\includegraphics[width=\linewidth,trim={0 0.7cm 0 1.2cm},clip]{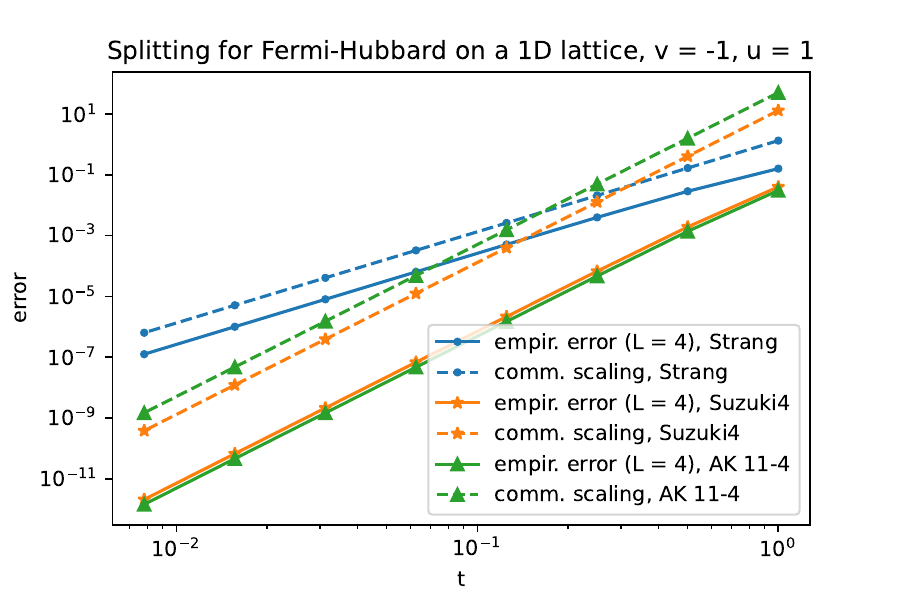}

		{\footnotesize $h$}
	\end{minipage}
	\hfill
	\begin{minipage}{0.46\textwidth}
		\centering
		\resizebox{1\textwidth}{!}{{\large\input{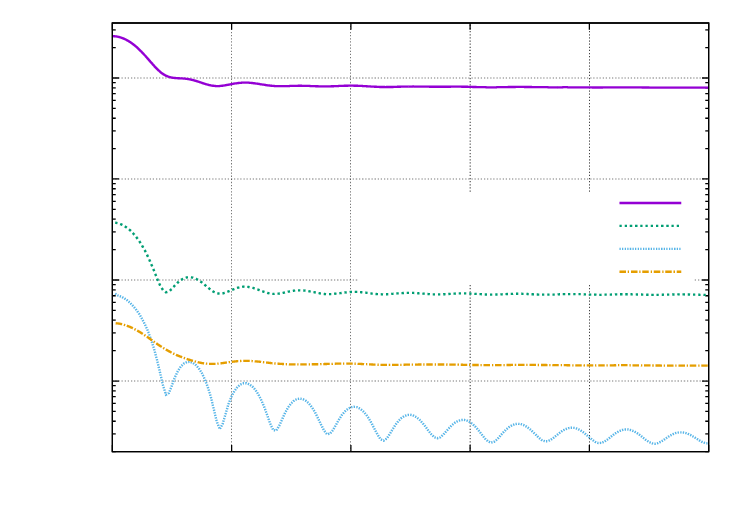}}}
	\end{minipage}
	\caption{Trotter error after a single time step $\norm*{S_n(-\im h) - \eto{-\im H h}}_2$ with the error bounds from \cref{th:Schubert_error_bounds} for the Fermi-Hubbard model (left) and relative long-time error $\frac1t \norm*{S_n(-\im t) - \eto{-\im H t}}_F$ for the Heisenberg model (right).
		The Strang splitting~\eqref{eq:leapfrog} is of order $n=2$, all the other Trotter schemes are of order $n=4$.
		Reprinted figures from Ref.~\cite{schubert2023trotter} (left) and Ref.~\cite{Ostmeyer:2022} (right) with adapted labels.
	}
	\label{fig:Schubert_comm_1d_error}
\end{figure}

\Cref{th:Schubert_error_bounds} does not provide tight bounds.
The discrepancy between empirical and estimated errors can be observed in the left panel of figure~\ref{fig:Schubert_comm_1d_error} as the distance between the solid and the dashed lines.
Therein three different Trotterizations have been used for a simple example Hamiltonian.
In the very same work~\cite{schubert2023trotter} Schubert and Mendl also provide strictly tighter bounds for the 2nd order Strang splitting (Proposition~1 in the published version and prop.~2 on arXiv).
Even these bounds still overestimate the true error as can be seen from the blue lines in figure~\ref{fig:Schubert_comm_1d_error}.

That said, at the time of writing, \cref{th:Schubert_error_bounds} provides the tightest known generally applicable bounds.

\subsubsection{Long-time error accumulation}\label{sec:long-time_errors}

An upper bound for the total Trotter error after some final time $\tau=N_\tau h$ can be obtained directly from the error of a single time step
\begin{align}
	\text{error}(\tau) &\le N_\tau\, \text{error}(h)
\end{align}
by triangle inequality.
For this reason the long-term $n$-th order Trotter error scales as $\ordnung{N_\tau h^{n+1}}=\ordnung{\tau^{n+1}/N_\tau^n}=\ordnung{\tau h^n}$.

The error accumulation in practice is often much milder than one might expect from the triangle inequality.
For instance, at long times only the diagonal parts of the error terms \eqref{eq:ord_1_errors} to \eqref{eq:ord_5_errors} etc.\ in the Hamiltonian eigenbasis contribute to the total error~\cite{Maxwell:2026ewi} (see Theorem~3 therein and note that their $\delta^d=t\Delta t^d$ corresponds to $\tau h^n$ in our notation).

It has been shown independently in Refs.~\cite{chen2024trottererrortimescaling,chen2024errorinterferencequantumsimulation} that the error bounds from the triangle inequality can be considerably improved upon using the ansatz $\ordnung{h^n} + \ordnung{\tau h^n}$.
Importantly, the coefficient in front of the asymptotically leading term with $\tau$-dependence can be very small.
Following the argument by Chen~\cite{chen2024trottererrortimescaling}, the idea behind this ansatz is that the error at each step can be split into a `parallel' and an `orthogonal' part.
The parallel part accumulates over time and should ideally be small.
The orthogonal part, on the other hand, only leads to an error oscillation and becomes negligible at long times.

More specifically, the parallel operator space $H_\parallel$ contains all operators (e.g.\ from \cref{eq:ord_1_errors,eq:ord_3_errors,eq:ord_5_errors}) that commute with the Hamiltonian.
Now any error operator $\mathcal{O}_\parallel\in H_\parallel$ will add up linearly with every time step since its contribution does not change over time.
In contrast, operators from the orthogonal complement $\mathcal{O}_\perp\in H_\perp$ do not commute with the Hamiltonian $[\mathcal{O}_\perp,H]\neq0$ and change `direction' averaging out over time.
This is reminiscent of the notion of a shadow Hamiltonian $\tilde H$ coined in the context of symplectic integrators (sec.~\ref{sec:symplectic})~\cite{Kennedy:2012gk}.
The Trotterized time evolution is equivalent to the exact time evolution with $\tilde H=H+\ordnung{h^{n}}$ defined by the sum of all error operators via the BCH formula~\eqref{eq:BCH}.
Operators $\mathcal{O}_\parallel$ that commute with the Hamiltonian define conserved quantities (in particular the energy directly related to $H$), so they are physically the most important to get right.
In other words, it is natural that error contributions from $\tilde H-H\in H_\parallel$ accumulate.

The resulting effect can be observed in the right panel of figure~\ref{fig:Schubert_comm_1d_error} where the long-time error (in Frobenius norm) is monitored.
First of all we confirm that all the different Suzuki-Trotter decompositions converge to a constant-in-time ratio $\text{error}/\tau$ and thus the error follows the expected asymptotic order $\ordnung{\tau}$.
On shorter time scales, however, there are clearly visible oscillations that can be explained by significant contributions from the orthogonal error in $H_\perp$.
Larger oscillations indicate smaller parallel coefficients and consequently smaller asymptotic errors.
This is particularly vividly demonstrated by the non-unitary scheme from Ref.~\cite{Ostmeyer:2022} that appears to have extremely small parallel $H_\parallel$ error contributions.
The fourth order scheme was constructed so that all complex-valued coefficients have the same real part $\real(c_i)=\real(d_i)=\frac{1}{10}$.

The way the error splits into parallel and orthogonal contributions strongly depends on the Trotterization.
Based on the observations from figure~\ref{fig:Schubert_comm_1d_error} it has been conjectured in Ref.~\cite{Ostmeyer:2022} that the most favourable error accumulation is obtained for maximally uniform (real parts of the) coefficients.
This has been proven by Maležič et al.~\cite{Malezic:2026bds} for second order product formulae using the framework derived by Chen~\cite{chen2024trottererrortimescaling}.
Ref.~\cite{Malezic:2026bds} also provides additional empirical evidence strengthening the conjecture for higher orders (see fig.~\ref{fig:origin_err}, right).
In fact, the most efficient 4th and 6th order schemes in \cref{tab:recommened6,tab:recommened14} already were selected by Maležič et al.\ as a compromise of small local errors and favourable error accumulation.

\subsection{Upper error bounds for symmetric schemes}\label{sec:symmetric_errors}

Throughout this review our main focus lies on symmetric schemes.
This property is not required for the error estimates in \cref{th:Schubert_error_bounds}, but it allows for a significantly simpler yet relatively tight error bound summarised in \cref{th:symmetric_error_bounds}.

\begin{theorem}[Symmetric Trotterization error bounds~\cite{Casas:2026krd}]\label{th:symmetric_error_bounds}
	Let $S_n(h)$ be a symmetric product formula of order $n$ as in equation~\eqref{eq:general_scheme} with $q$ cycles defined by the coefficients $c_i=d_{q+1-i}$.
	Then for $H=\sum_{k=1}^\Lambda A_k$ the error of the product formula is bounded by
	\begin{align}
		\norm*{S_n(h) - \eto{H h}} &\le \frac{\abs{h}^{n+1}}{2^{n}\br{n+1}!} \br{1+2\sum_{i=1}^q \abs*{c_i}}^{n+1} \br{\sum_{k=1}^\Lambda \norm*{A_k}}^{n+1}
	\end{align}
	in any given norm $\norm{\cdot}$.
\end{theorem}

\begin{remark}
	This \cref{th:symmetric_error_bounds} corresponds to Theorem~2.7 in Ref.~\cite{Casas:2026krd}.
	It presents a good compromise between simplicity and tightness.
	A generalisation that allows tighter error estimates tailored for symmetric schemes has also been derived in Ref.~\cite{Casas:2026krd}, see Theorem~3.3.
\end{remark}

\subsection{Lower error bounds}\label{sec:lower_error_bounds}

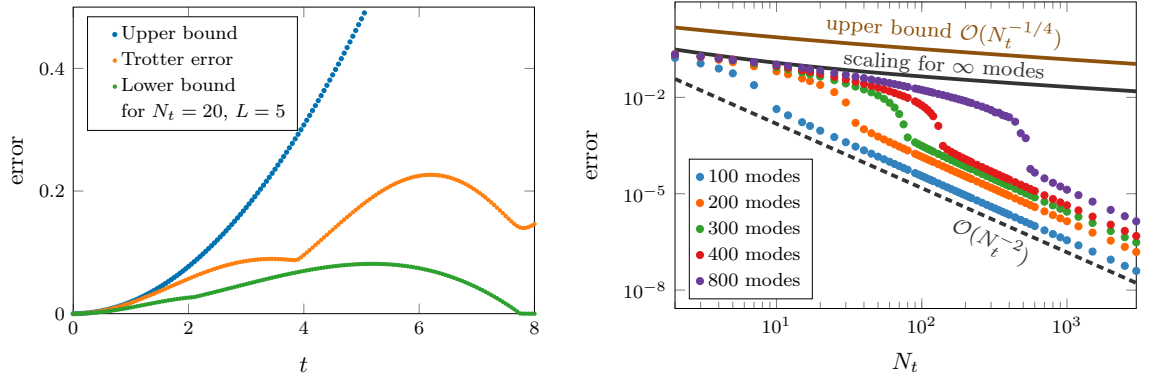
\begin{figure}
	\centering
\begin{tabular}{cc}
	\begin{tikzpicture}[mark size={0.7}, scale=.89]
		\begin{axis}[
			xlabel={$t$},
			ylabel={error},
			tick label style={font=\footnotesize},
			x post scale=1.0,
			y post scale=0.8,
			legend cell align={left},
			legend pos=north west,
			ylabel near ticks,
			yticklabel pos=left,
			xmin=0,
			xmax=8,
			ymin=0,
			ymax=0.5,
			]
			\addplot[color=plotblue, thick, mark=*, only marks] table[x=t, y=error, col sep=comma] {paper-source/Hahn-2410.03059v1/XY_Norm_Time_Upper_Bound.csv};
			\addplot[color=plotorange, mark=*, only marks] table[x=t, y=error, col sep=comma] {paper-source/Hahn-2410.03059v1/XY_Norm_Time_Error.csv};
			\addplot[color=plotgreen, mark=*, only marks] table[x=t, y=error, col sep=comma] {paper-source/Hahn-2410.03059v1/XY_Norm_Time_Lower_Bound.csv};
			\addlegendimage{empty legend};
			\legend{
				\footnotesize Upper bound,
				\footnotesize Trotter error,
				\footnotesize Lower bound,
				\footnotesize {for $N_t=20$, $L=5$}
			};
		\end{axis}
	\end{tikzpicture}
	&
	\begin{tikzpicture}[mark size={1.5}, scale=.89]
		\begin{axis}[
			xmode=log,
			ymode=log,
			xlabel={$N_t$},
			ylabel={error},
			tick label style={font=\footnotesize},
			x post scale=1,
			y post scale=0.8,
			legend pos=south west,
			legend columns=1,
			legend cell align={left},
			xmin=2, xmax=3000,
			]
			\addplot[color=1, mark=*, only marks] table[x=N, y=error, col sep=comma] {paper-source/Burgarth-2312.08044v2/Trotter_Error_2nd_order_groundstate_100.csv};
			\addplot[color=2, mark=*, only marks] table[x=N, y=error, col sep=comma] {paper-source/Burgarth-2312.08044v2/Trotter_Error_2nd_order_groundstate_200.csv};
			\addplot[color=3, mark=*, only marks] table[x=N, y=error, col sep=comma] {paper-source/Burgarth-2312.08044v2/Trotter_Error_2nd_order_groundstate_300.csv};
			\addplot[color=4, mark=*, only marks] table[x=N, y=error, col sep=comma] {paper-source/Burgarth-2312.08044v2/Trotter_Error_2nd_order_groundstate_400.csv};
			\addplot[color=5, mark=*, only marks] table[x=N, y=error, col sep=comma] {paper-source/Burgarth-2312.08044v2/Trotter_Error_2nd_order_groundstate_800.csv};
			\addplot[color=0, ultra thick, densely dashed, domain=2:3000]{0.15*x^(-2)};
			\addplot[color=6, ultra thick, domain=2:3000]{0.5*(2.04665*x^(-1/2) + 0.832107/x + 1.34365/x^(1/4))};
			\addplot[color=0, ultra thick, domain=2:3000]{0.31/x + 0.1/x^(1/2) + 0.1/x^(1/4)};
			\legend{
				\footnotesize $100$ modes,
				\footnotesize $200$ modes,
				\footnotesize $300$ modes,
				\footnotesize $400$ modes,
				\footnotesize $800$ modes,
			};
		\end{axis}
		\node[color=6,rotate=-4] at (4,4.15) {\footnotesize upper bound $\ord(N_t^{-1/4})$};
		\node[color=0,rotate=-4] at (4,3.6) {\footnotesize scaling for $\infty$ modes};
		\node[color=0,rotate=-25] at (4.7,1.) {\footnotesize $\ord(N_t^{-2})$};
	\end{tikzpicture}
\end{tabular} 	\caption{Trotter errors.
		Left: First order operator error $\norm*{S_1(-\im t) - \eto{-\im H t}}_2$ for the $XX$ model with a spin chain of length $L=5$.
		Upper and lower bounds (sec.~\ref{sec:lower_error_bounds}) are compared to the true error as a function of final time $t$ for a fixed number of steps $N_t=20$.
		Right: Second order error $\norm*{\brr{S_2(-\im t) - \eto{-\im H t}} \Psi_{100}}$ for the ground state $\Psi_{100}$ of the hydrogen atom after the fixed final time $t=1$ at different discretization levels ($\infty$ modes means no discretization).
		The analytic bound (sec.~\ref{sec:non-dff_errors}) for the singular Coulomb potential scales as $\ordnung{N_t^{-1/4}}$, not with the na\"ive $\ordnung{N_t^{-2}}$ applicable to smooth potentials.
			Reprinted figures by the HB+ group from Ref.~\cite{hahn2024lowerboundstrottererror} (left) and Ref.~\cite{burgarth2023strong} (right) with adapted labels.
		}
		\label{fig:Hahn_lower_error_bounds}
\end{figure}

So far, we have only discussed upper error bounds which typically greatly overestimate the true error.
It is instructive to investigate lower bounds as well because they could be used to estimate the minimal resources required for a given simulation.
Hahn et al.\ (referred to as part of the HB+ group below) provide such lower bounds for the first order Trotter method~\eqref{eq:1st_order_Trotter} and a Hamiltonian composed of two operators~\cite{hahn2024lowerboundstrottererror}.
While the generalisation to a similar level as in \cref{th:Schubert_error_bounds} will still required a lot of work, these bounds already demonstrate a great potential empirically.
In the left panel figure~\ref{fig:Hahn_lower_error_bounds} the lower error bounds from Ref.~\cite{hahn2024lowerboundstrottererror} are compared to the true error and the upper bounds from \cref{th:Schubert_error_bounds} for a simple toy model.
Rather intriguingly, the lower bounds tend to follow the trend of the true error much more closely than the upper bounds.
This discrepancy is certainly warrants further investigations.

\subsection{Worst case errors for non-differentiable and unbounded problems}\label{sec:non-dff_errors}

The usual error estimation based on power law expansions breaks down if the operators under investigation are not sufficiently smooth.
It turns out, however, that even for exponentials of non-differentiable operators rigorous bounds can be derived.
The convergence of Trotter formulae can still be guaranteed in most cases, but the error scaling might be worse than expected and is not even necessarily a polynomial with an integer power~\cite{Wiebe_2010}.

The first essential step when dealing with non-differentiable and especially unbounded operators is to use state-dependent vector norms for the error bounds rather than operator norms.
Not only can the operator norms diverge while individual initial states are evolved stably~\cite{An2021timedependent,Burgarth:2022lib}.
It has also been shown that in the non-divergent case vector norms can be exponentially tighter than the usual operator norm estimator~\cite{Zylberman:2025amu}.

Members of the HB+ group~\cite{burgarth2023strong} have derived state-dependent Trotter error bounds applicable for singular potentials.
In Ref.~\cite{Hahn:2024fcz}, they provide applications of these bounds to an open quantum system with a bosonic environment.
The ground state $\Psi_{100}$ of the hydrogen atom (Coulomb potential) is found to be a particularly interesting case.
Its slowest convergence in step size is limited to $\ordnung{h^{\frac14}}$ independently of the theoretical order of the product formula~\cite{burgarth2023strong}.
This behaviour can be clearly observed in the right panel of figure~\ref{fig:Hahn_lower_error_bounds} where the error scaling of $\Psi_{100}$ is plotted for different levels of regularisation.
The number of modes corresponds to the number of discrete points in position space.
At any finite number of modes the singularity of the Coulomb potential is shielded and for small enough step sizes (large enough $N_t$) the asymptotic scaling $\ordnung{h^2}$ is recovered.
This transition occurs at higher $N_t$ for finer discretisations and the continuous case ($\infty$ modes) is expected to follow the worst case $\ordnung{h^{\frac14}}$ scenario.

The state-dependent bounds have been generalised by Becker et al.~\cite{becker2024convergenceratestrotterkatosplitting} for potentials of the form $V(x)\propto \abs{x}^{-\alpha}$ with $\alpha\le 1$.
When non-differentiable states are also allowed, the convergence in the Trotter step size can be arbitrarily slow~\cite{facchi2025slowconvergencetrotterdecomposition}.

\subsection{Low-energy sector error bounds}\label{sec:low-energy_errors}

State-dependent error bounds are not only useful when the operator norm fails.
They can also be relevant in practical simulations which are often primarily interested in the ground state of a given system or the first few excited states.
For instance, Mizuta and Kuwahara~\cite{mizuta2025trotterizationsubstantiallyefficientlowenergy} have derived rigorous bounds for the Trotter error of low-energy states.
This error turns out to be substantially reduced compared to those of generic states, including better scaling with the system size.
Similarly, Bay-Smidt et al.~\cite{Bay-Smidt:2026vcs} have found that the energy gap typically has a much smaller Trotter error than might have been expected from commutator-based bounds.

In applications like adiabatic quantum computing (sec.~\ref{sec:qc_limitations}) it is particularly important to quantify the error in the low-energy sector during evolution with a time-dependent Hamiltonian (sec.~\ref{sec:time-dependent}).
Improved bounds for this specific scenario have been provided in Ref.~\cite{Zhou:2026cvd}.

When focusing on the ground state, it is a natural approach in physics to treat excited states as perturbations.
The efficacy of this ansatz for the low-energy error estimation has been demonstrated in Ref.~\cite{martínezmartínez2023estimating}.
Therein, the perturbative error estimates have been shown to feature a considerably higher correlation with the true error than generic bounds like those from \cref{th:Schubert_error_bounds}.
Unfortunately, so far the applicability of these bounds is limited to the second order Strang splitting.

\subsection{Model-specific error bounds}\label{sec:model-specific_errors}

The more information is available, the better the results will be.
Of course, this general rule is also applicable to the error estimation in time evolution simulations.
In particular, model-specific error bounds can be much more accurate than general estimations.
Below, we provide a list of references for well-investigated models without going into too much detail about their results.

\subsubsection{Relevance of entanglement}

Before naming any model explicitly, let us note that the entanglement entropy can often be estimated or at least bounded for systems with local interactions.
Whenever the entropy saturates to some maximal value $S_\text{max}$, the Trotter error for large systems scales with $S_\text{max}$, that is it becomes asymptotically independent of the system size~\cite{Kulkarni:2026yly}.

Moreover, the error of local observables is found to spread with the so-called operator scrambling~\cite{feng2025trotterizationoperatorscramblingentanglement}.
It has also been observed in Ref.~\cite{Zhang:2026cbs} that the variance of the Trotter error over different initial states with equal entanglement will depend on this entanglement.
While the error itself remains non-zero, its distribution is found to become more sharply peaked with higher entanglement and non-stabilizerness (``magic'').
Some states, on the other hand, exhibit unexpectedly small errors.
These states are dubbed ``Trotter scars'' in Ref.~\cite{Zhou:2026mlc}.

\subsubsection{Coulomb and Yukawa potentials}

As mentioned in \cref{sec:non-dff_errors}, the Coulomb potential features an extremely slow $\ordnung{h^{\frac14}}$ worst case convergence in the step size~\cite{burgarth2023strong}.
The error in the particle number $N$ for many-body Coulomb interactions has also been found to be a polynomial $\ordnung{N^{4.5}}$ with fractional exponents~\cite{fang2025trottererrormanybodyquantum,Fang:2026css}.
The same error bounds have also been reproduced for the Yukawa potential~\cite{Fang:2026ckf} indicating that the nature of the (well-behaved) tail of the potential is irrelevant.

\subsubsection{Hubbard and t-J models}

For the fermionic Hubbard and t-J models there might be no need to re-evaluate the commutators from \cref{th:Schubert_error_bounds}.
Schubert and Mendl~\cite{schubert2023trotter} have already applied their theorem to the Hubbard model and these results can be used directly.
Even earlier similar bounds had been derived specifically for Hubbard and t-J models in Ref.~\cite{https://doi.org/10.48550/arxiv.2302.04698}.

\subsubsection{SYK model}

Explicit error bounds based on higher order commutators have also been derived for the Sachdev-Ye-Kitaev (SYK) model~\cite{chen2025trottererrorgatecomplexity}.

\subsubsection{Transverse Ising model}

A less conventional error estimation has been presented for the transverse field Ising model.
In Ref.~\cite{Sinha:2025tqg} some conserved quantities are shown to stay exactly constant during time evolution even at finite Trotter step sizes.

\subsubsection{Bosonic systems}

Up to this point we have been implicitly identifying operators with their matrix representation for the sake of simplicity and intuition.
This identification fails for infinite-dimensional representation like bosonic systems that have continuous degrees of freedom.
Norms in infinite dimensions are not necessarily meaningful.
Therefore, the HB+ group has proposed, based on Refs.~\cite{Shirokov:2017vmf,Becker:2021spk}, to estimate the distance between the true time evolution operator and its Trotterized counterpart via a metric induced by the underlying Lie group~\cite{vanLuijk:2022zle}.

Spin-Boson models, on the other hand, are considered in Ref.~\cite{Hahn:2024fcz}.

\subsection{Empirical error estimation}\label{sec:empirical_error}

Finally, the most reliable way to determine an error in any specific case is to measure it.
A simple test is to measure known conserved quantities like the energy.
Whenever these quantities are not conserved well enough, one can reduce the Trotter step size adaptively (ADA-Trotter)~\cite{adaptive_trotter}.
A possible generalisation of this idea is to monitor individual observables of interest in order to derive case-dependent optimised Trotterizations.
In Ref.~\cite{li2024principaltrotterobservationerror} this approach has been used to minimise the Trotter error for the measurement of given observables.
It turns out that the order in which the Trotter operators are applied can lead to different projections of the error onto said observable.
A specialised quantum algorithm for the numerical error extraction has been proposed in Ref.~\cite{Ronfaut:2026knp}.
 	
	\section{Generalisations}\label{sec:generalisations}

Suzuki-Trotter decompositions can be generalised in many ways.
First and foremost, in \cref{sec:time-dependent} we will explore time-dependent Hamiltonians and the necessary adjustments summarised in \cref{th:time-dependent}.
Even for constant Hamiltonians, it might make sense to use time-dependent operators in the interaction picture.
In \cref{sec:hierarchy} this ansatz is explained together with other methods to exploit hierarchy separations between operators with different orders of magnitude.
\Cref{sec:complex,sec:processed,sec:multi-product} explain three different approaches to increase the efficiency of splitting methods.
At the cost of exact unitarity we can introduce complex-valued coefficients (sec.~\ref{sec:complex}) or multi-product formulae (sec.~\ref{sec:multi-product}).
The latter cancel out leading order errors through linear combinations of several independent Trotterizations with different time steps.
Processed methods introduced in \cref{sec:processed}, on the other hand, elevate the first-same-as-last property to a whole new level cancelling out significant parts of the cost at each step.
Finally, randomised methods are discussed in \cref{sec:random}.
They allow to reduce circuit depths considerably without introducing a systematic bias at the cost of statistical errors.

\subsection{Time-dependent Trotterizations}\label{sec:time-dependent}

Throughout this review we have been focusing on Hamiltonians without explicit time dependence.
Let us now briefly investigate the subtle changes required if the variation of the Hamiltonian with time cannot be ignored.
This is relevant for physical systems that are externally driven (pumped) as well as for adiabatic quantum computing in ground state search (sec.~\ref{sec:quantum_comp}) and interesting as a generalisation in its own right.

The formal solution to the Schrödinger equation with a time-dependent Hamiltonian
\begin{align}
	\im \frac{\md}{\md t}\ket{\psi(t)} &= H(t)\ket{\psi(t)}\\
	\Rightarrow \ket{\psi(t_1)} &= U(t_1,t_0) \ket{\psi(t_0)}
\end{align}
can still be expressed through a time evolution operator
\begin{align}
	U(t_1,t_0) &\equiv \mathcal{T}\brc{\eto{-\im\int_{t_0}^{t_1} \md t H(t)}}\\
	&= \lim\limits_{N_t\rightarrow\infty}\br{\eto{-\im H(t_1-h)h}\cdots \eto{-\im H(t_0+h)h}\eto{-\im H(t_0)h}}\,,
\end{align}
but now $U(t_1,t_0)$ depends on both the initial and the final times rather than simply their difference $t_1-t_0= N_t h$.
The time ordering operator $\mathcal{T}\brc{\cdot}$ ensures an ordered exponential.
This ordering is necessary since the Hamiltonian does not commute with itself at a different time any more.

Remarkably, every time-independent Suzuki-Trotter method is directly applicable for time-dependent Hamiltonians at the same computational cost using the prescription of \cref{th:time-dependent}.

\begin{theorem}[Generalising Trotter formulae for time-dependent Hamiltonians~\cite{Suzuki:1993timedep,Hatano_2005}]\label{th:time-dependent}
	Let the coefficients $c_i$ and $d_i$ as in equation~\eqref{eq:general_scheme} define a Suzuki-Trotter decomposition of order $n$ and $H(t)=\sum_{k=1}^\Lambda A_k(t)$ be a time-dependent Hamiltonian.
	Then the decomposition 
	\begin{align}
		S_n(t,h) &= \prod_{i=1}^q\brr{\left(\prod_{k = 1}^{\Lambda} \eto{c_{i} A_{k}(t+e_i h) h} \right) \left( \prod_{k = \Lambda}^{1} \eto{d_{i} A_{k}(t+e_i h) h}\right) }\,,\\
		e_i &\coloneqq \sum_{j=i}^q d_j + \sum_{j=i+1}^q c_j
	\end{align}
	is an order $n$ approximation of the time evolution operator
	\begin{align}
		\mathcal{T}\brc{\eto{\int_{t}^{t+h} \md s H(s)}} &= S_n(t,h)+\ordnung{h^{n+1}}\,.
	\end{align}
\end{theorem}

\begin{remark}
	Here we use the order convention $\prod_{i=1}^q M_i=M_1M_2\cdots M_q$ explicitly.
	Note also that the time shifts $e_i$ are the same for the entire $i$-th cycle (see fig.~\ref{fig:ramp}) and that they are not symmetric in $c_i$ and $d_i$.
\end{remark}

\Cref{th:time-dependent} presents the results of Refs.~\cite{Suzuki:1993timedep,Hatano_2005} using a notation different from the original in order to be consistent with the rest of this review.
Let us therefore re-iterate the crucial steps of the derivation leading up to \cref{th:time-dependent}.
We start with the following trick:
\begin{align}
	\eto{-\im A_k(t)}\eto{-\im \mathcal{T}h} &\coloneqq \eto{-\im \mathcal{T}h}\eto{-\im A_k(t+h)}\\
	\Rightarrow\mathcal{T}\brc{\eto{-\im\int_{t}^{t+h} \md s H(s)}} &= \eto{-\im \br{H(t) + \mathcal{T}}h}\,.\label{eq:time-shift-operator}
\end{align}
Suzuki~\cite{Suzuki:1993timedep} introduced the so-called time-shift operator $\eto{-\im \mathcal{T}h}$ and showed that the identity~\eqref{eq:time-shift-operator} is fulfilled exactly.
An intuitive proof is provided by Hatano and Suzuki in sec.~4 of Ref.~\cite{Hatano_2005}.

Once the time ordering has been symbolically absorbed into the exponent, we can use any classical Trotterization for the effective Hamiltonian $H_\text{eff}(t)\equiv \sum_k A_k(t) + \mathcal{T}$.
A general product formula based on equation~\eqref{eq:general_scheme} then becomes
\begin{align}	
	\begin{split}
		S_n(t,h) &= \eto{c_1\mathcal{T}h} \left(\prod_{k = 1}^{\Lambda} \eto{c_{1} A_{k}(t) h} \right) \left( \prod_{k = \Lambda}^{1} \eto{d_{1} A_{k}(t) h}\right) \eto{d_1\mathcal{T}h} \cdots\\
		&\quad\left( \prod_{k = \Lambda}^{1} \eto{d_{q-1} A_{k}(t) h}\right) \eto{d_{q-1}\mathcal{T}h}\, \eto{c_q\mathcal{T}h} \left(\prod_{k = 1}^{\Lambda} \eto{c_{q} A_{k}(t) h} \right) \left( \prod_{k = \Lambda}^{1} \eto{d_{q} A_{k}(t) h}\right) \eto{d_q\mathcal{T}h}
	\end{split}\\
	\begin{split}
		&= \eto{\mathcal{T}h} \left(\prod_{k = 1}^{\Lambda} \eto{c_{1} A_{k}(t+(1-c_1) h) h} \right) \left( \prod_{k = \Lambda}^{1} \eto{d_{1} A_{k}(t+(1-c_1) h) h}\right)  \cdots\\
		&\quad\left( \prod_{k = \Lambda}^{1} \eto{d_{q-1} A_{k}(t+(d_q+c_{q}+d_{q-1})h) h}\right) \left(\prod_{k = 1}^{\Lambda} \eto{c_{q} A_{k}(t+d_q h) h} \right) \left( \prod_{k = \Lambda}^{1} \eto{d_{q} A_{k}(t+d_q h) h}\right).
	\end{split}
\end{align}
Collecting the time shift coefficients into a general prescription for $e_i$ immediately reproduces \cref{th:time-dependent}.

Ikeda et al.~\cite{Ikeda:2022tlb} have derived specialised splitting methods up to order $n\le 6$ for the special case that the time dependence of every operator is contained in a scalar prefactor $A_k(t) = x_k(t) A_k$.
While these methods outperform the classical Suzuki formulae (sec.~\ref{sec:suzuki}) in some cases, the schemes from \cref{tab:recommened6,tab:recommened14} by Maležič et al.~\cite{Malezic:2026bds} in combination with \cref{th:time-dependent} are expected to be more efficient.

\subsubsection{Magnus expansion}\label{sec:magnus_expansion}

For the sake of completeness we also mention the Magnus expansion~\cite{Magnus:1954zz} that utilises the ansatz
\begin{align}
	U(t+h,t) &= \eto{\Omega(t,h)}\,,
\end{align}
where $\Omega(t,h)=\sum_{k=0}^\infty \Omega_k(t,h)$ can be expanded in terms of nested integrals~\cite{2009PhR...470..151B}
\begin{align}
	\Omega_1(t,h) &= \int_{t}^{t+h}\md s\, H(s)\,,\\
	\Omega_2(t,h) &= \frac 12 \int_{t}^{t+h}\md s_1 \int_{t}^{s_1}\md s_2\, [H(s_1),H(s_2)] \,,
\end{align}
and so on, to arbitrary order $\Omega_k = \ordnung{h^k}$.
In practice, this ansatz is rather expensive and it is more common to resort to commutator-free quasi-Magnus (CFQM) methods.
Efficient higher order realisations of CFQM methods have been derived in Ref.~\cite{2017CoPhC.220..243B}.
Even these improved algorithms require the exponentiation of dense (or at least not very sparse) matrices which makes them impractical for the simulation of large systems.
We refer to Sec.~6.3 of the review~\cite{Blanes_Casas_Murua_2024} for a broader overview.

\subsection{Making use of operator magnitude hierarchies}\label{sec:hierarchy}

Let us assume a Hamiltonian
\begin{align}
	H&= H_0 + \epsilon \sum_k A_k~\label{eq:hamilton_hierarchy}
\end{align}
with $\epsilon\ll1$ and, in consequence, a clear separation of scale between the main contribution $H_0$ and perturbations $A_k$.
This is not a rare situation and it can be exploited to significantly reduce the simulation error.

Any regular Trotterization $S_n$ of order $n$ comes with an error scaling as $\ordnung{\epsilon h^n}$.
In particular, the leading error is always linear in $\epsilon$.

Omelyan et al.\ have employed their framework discussed in \cref{sec:construction} to address this limitation.
Rather than treating all the higher order nested commutators contributing to the error on an equal footing, errors with low powers of $\epsilon$ (equivalently, high powers of $H_0$) are prioritised.
This way, a method scaling as $\ordnung{\epsilon^2h^4}+\ordnung{\epsilon h^6}$ and one with the error of order $\ordnung{\epsilon^2h^2}+\ordnung{\epsilon h^8}$ are derived.
Their results are listed in Appendix~A of Ref.~\cite{Omelyan_2002}.
Note that these splitting methods were originally designed for exactly two operators, but their generalised efficacy with the help of \cref{th:2-opt-to-many} has been demonstrated in Refs.~\cite{bosse2024efficientpracticalhamiltoniansimulation,Moreno:2026inv}.

Hierarchies including not two but three different scales $H= \sum_k H_{0,k} + \epsilon_1 \sum_k A_{k} + \epsilon_2 \sum_k B_{k}$ with $\epsilon_2\ll\epsilon_1\ll1$ have been explored in Ref.~\cite{Casares:2026lyr}.
The so-called Symmetry-Protected Randomized near-Integrable Trotter (SPRINT) formulae presented therein use different order Trotter schemes (or even randomised methods, sec.~\ref{sec:random}) for the splitting of the individual contributions from $H_0$, $A$, and $B$, respectively.

\subsubsection{Separation of time scales}

An alternative common scenario is a Hamiltonian
\begin{align}
	H &= \sum_k A_k + \epsilon \sum_k B_k
\end{align}
with some dominating terms $A_k$ and small corrections $\epsilon B_k$.
In this case it makes sense to spend more computational effort on the important parts $A_k$ by separating the simulation time scales between the different contributions.
Let $S_n^A(h)=\eto{h \sum_k A_k}+\ordnung{h^{n+1}}$ be some product formula as in equation~\eqref{eq:general_scheme} that evolves only the dominant contribution to the Hamiltonian.
Then we can construct the full Trotter scheme for a larger time step~\cite{Urbach:2005ji,Krieg:2018pqh}
\begin{align}
	S_n^{A,B}(N_B h) &= \prod_{i=1}^q\brr{\left( \prod_{k = 1}^{\Lambda} \eto{N_B c_{i} \epsilon B_{k} h} \right) S_n^A((c_i+d_i)h)^{N_B} \left( \prod_{k = \Lambda}^{1} \eto{N_B d_{i} \epsilon B_{k} h} \right)}
\end{align}
that reduces the computational effort of the $B$-part by a factor $N_B$ (and increases the cost of the $A$-part by a factor $q$).
Since the error contributed by the $B$-part scales as $\ordnung{\br{N_B\epsilon h}^n}$, a good and natural choice for the time scale separation factor is $N_B\approx \epsilon^{-1}$.

In this construction $S_n^A$ and $S_n^{A,B}$ rely on the same coefficients and, in consequence, have the same order $n$.
This is not strictly necessary and for very small $\epsilon$ it might make sense to choose some alternative scheme $S_{n'}^{A,B}$ with $n'\le n$ for the outer integration.
The dominant error in $A$ remains of order $n$.

This concept is readily generalisable to an arbitrary number of time scales using a recursive procedure~\cite{Urbach:2005ji}.

\subsubsection{Interaction picture}

Again assuming a Hamiltonian as in equation~\eqref{eq:hamilton_hierarchy} and allowing for the additional limitation that the operator exponential $\eto{(H_0+\epsilon A_k)h}$ can be calculated for each $k$, we can express the problem in the so-called interaction picture~\cite{Low:2018pte,Rajput:2021khs,ciavarella2024quantumsimulationsu3lattice}
\begin{align}
	S_1^\text{int}(h) &= \eto{H_0 h}\prod_k \left(\eto{-H_0 h} \eto{(H_0+\epsilon A_k) h}\right)\label{eq:1st_order_THRIFT}
\end{align}
with the local error~\cite{bosse2024efficientpracticalhamiltoniansimulation}
\begin{align}
	\norm*{U(h)-S_1^\text{int}(h)}_2 &\leq \epsilon^2\int_{0}^{h}dv\int_{0}^{v}ds\sum_{k_1<k_2}\norm*{\brr{A_{k_1}(s),A_{k_2}(v)}}_2 =\ordnung{\epsilon^2 h ^2}\,.
\end{align}
Crucially, there is no error contribution linear in $\epsilon$ to all orders in $h$.
If this condition is dropped and only the leading $\ordnung{\epsilon}$ error is required to vanish, then the factors $\eto{(H_0+\epsilon A_k)h}$ can be Trotterized.
This recovers the more generally applicable methods by Omelyan~\cite{Omelyan_2002} discussed above~\cite{Moreno:2026inv}.

Bosse et al.~\cite{bosse2024efficientpracticalhamiltoniansimulation} discuss the method based on equation~\eqref{eq:1st_order_THRIFT} and its generalisations in great detail.
In Ref.~\cite{bosse2024efficientpracticalhamiltoniansimulation} it is called Trotter Heuristic Resource Improved Formulas for Time-dynamics (THRIFT).
The extensive benchmarks and comparisons with alternative methods presented in Ref.~\cite{bosse2024efficientpracticalhamiltoniansimulation} (including Omelyan's~\cite{Omelyan_2002} method) are highly educational.
It is also found in Ref.~\cite{bosse2024efficientpracticalhamiltoniansimulation} that higher order THRIFT methods can be constructed from the first order building block~\eqref{eq:1st_order_THRIFT} in much the same way one would construct standard product formulae starting with the first order Trotterization~\eqref{eq:1st_order_Trotter}.
Randomisation (sec.~\ref{sec:random}) can be used to further elevate the error order to $\ordnung{\epsilon^2}$ for general Trotter schemes and to $\ordnung{\epsilon^3}$ for THRIFT~\cite{Kim:2026vbh}.

As a rule, more specialised algorithms are also more efficient and all the available analytic knowledge should be utilised.
This is why operator magnitude hierarchies~\eqref{eq:hamilton_hierarchy} are worth exploiting, be it with Omelyan's specialised schemes~\cite{Omelyan_2002}, via time scale separation~\cite{Urbach:2005ji}, or by means of the interaction picture~\cite{bosse2024efficientpracticalhamiltoniansimulation}.

\subsection{Complex coefficients}\label{sec:complex}

The coefficients $c_i,d_i$ in the product formula~\eqref{eq:general_scheme} are typically limited to real values in order to guarantee exact unitarity $S_n(\im h)^\dagger=S_n(\im h)^{-1}$ (preservation of phase space in symplectic integrators).
Exact unitarity is not always required, though, and unitarity violations to the same order $\ordnung{h^n}$ as the `usual' error might be acceptable.
In this case complex-valued coefficients $c_i,d_i\in \mathds{C}$ introduce additional degrees of freedom for the efficiency optimisation described in \cref{sec:construction}.

Some highly optimised complex-valued 4th order Trotter schemes have been derived and benchmarked in Ref.~\cite{Ostmeyer:2022}.
This includes a 4th order method with constant real parts $\real(c_i)=\real(d_i)=\frac{1}{10}$ with remarkably favourable long-time error accumulation (sec.~\ref{sec:long-time_errors}).
Real coefficients require at least one negative element for order $n\ge 4$ decompositions.
A larger variety of product formulae with complex coefficients~\cite{Chambers_2003,CASAS2022126700,Auzinger:2017,blanes2022symmetric,Blanes:2022} has been derived for higher order symplectic integrators (sec.~\ref{sec:symplectic}).

Interestingly, there are two intuitive yet incompatible choices generalising the real symmetry of the coefficients.
Either the symmetry $c_i=d_{q+1-i}$ is simply carried over preserving reversibility $S(\im h)^{-1}=S(-\im h)$, dubbed \textit{palindromic} in Ref.~\cite{Blanes:2022}.
Alternatively, \textit{symmetric-conjugate} schemes obeying $c_i=d_{q+1-i}^*$ allow to identify complex conjugation with time reversal $S_n(\im h)^\dagger=S_n(-\im h)$.
Symmetric-conjugate Trotterizations are found to violate unitarity less over long times~\cite{Blanes:2022} and are thus preferable.
Either version can easily be converted into the other by using the complex conjugates $c_i^*,d_i^*$ in every second step.
Such composite methods are found to be particularly stable for long evolution times~\cite{Bernier:2025}.

The apparent efficiency of splitting methods with complex coefficients can be deceptive.
Each multiplication of two complex numbers on a classical computer requires four real number multiplications and this overhead needs to be taken into account.

\subsection{Processed methods}\label{sec:processed}

A time step of a Suzuki-Trotter decompositions can also be written as
\begin{align}
	S(h) &= P(h)\, \hat S(h)\, P(h)^{-1}\,,
\end{align}
where $\hat S$ denotes the \textit{kernel} and $P$ the \textit{processor}.
Up to this point, we have been working with trivial processors (i.e.\ $S=\hat S$ and $P=\id$), but truly processed methods are readily available in the literature~\cite{morales2022greatly,Blanes:2024foel} including those optimised for symplectic integrators~\cite{10.1063/1.2203609} (sec.~\ref{sec:symplectic}).
The idea to generalise product formulae in this way was first mentioned by Butcher~\cite{Butcher:1969} and refined by Blanes et al.~\cite{Blanes:2006cmfd}.
While processed methods are rather cumbersome and inefficient for individual time steps, repeated applications can be contracted
\begin{align}
	S(h)^{N_\tau} &= P(h)\, \hat S(h)^{N_\tau}\, P(h)^{-1}\,,\label{eq:processed_advantage}
\end{align}
potentially saving a significant part of the computational cost.
To reach maximal efficiency, the kernel $\hat S$ should be as cheap as possible.
It does not even have to be of the same order $n$ as the full processed method $S$.

Since the computation of intermediate results would require additional insertions of the processor $P$, it would negate the advantage gained in equation~\eqref{eq:processed_advantage}.
Therefore, processed formulae are viable primarily for long-time evolution without interest in the full trajectory.

\subsection{Multi-product formulae}\label{sec:multi-product}

Starting with a Trotter scheme $S_{n_0}$ of even order $n_0$, we can build a \textit{multi-product formula} (MPF) or \textit{Richardson extrapolation}
\begin{align}
	S^{(s)}_{n,n_0}(h) &= \sum_{j=1}^m a_j\, S_{n_0}\br{\frac{h}{k_j}}^{k_j}\label{eq:mpf}
\end{align}
that approximates the time evolution operator to order
\begin{align}
	n &= n_0 + 2(m-1)
\end{align}
using $m$ linear combinations of the base scheme $S_{n_0}$ with different step sizes $\frac{h}{k_j}$.
Blanes et al.~\cite{Blanes:1999CeMDA} have shown that the unitarity (or symplectic nature) of such a method is preserved at least up to order
\begin{align}
	s &= \max\br{2n_0+1,n+3}\,.
\end{align}
MPFs can, for instance, be calculated during the post processing~\cite{CarreraVazquez2023wellconditioned,Park:2026ozj} of a quantum computation (sec.~\ref{sec:quantum_comp}).
In the context of quantum computing MPFs are also often referred to as \textit{linear combinations of unitary} (LCU) formulae~\cite{Childs:2012gwh}.

The coefficients $a_j$ have to be determined from a matrix equation~\cite{Blanes:1999CeMDA} and solutions exist for any tuple of distinct positive integers $k_j$.
In the special case of $n_0=2$ Chin~\cite{Chin:2008} has derived a closed form solution
\begin{align}
	a_j &= \prod_{l\neq j} \frac{k_j^2}{k_j^2-k_l^2}\,.
\end{align}
Particularly efficient combinations of $k_j$ tuples and coefficients $a_j$ have been derived by Low et al.\ and tabulated at the very end of Ref.~\cite{low2019wellconditionedmultiproducthamiltoniansimulation}.
In this case, a high efficiency is achieved when the total number of Trotter steps $\sum_j k_j$ is small and at the same time the coefficients $a_j$ are well conditioned, i.e.\ $\sum_j|a_j|$ is also small.
Crucially, only $\ordnung{n^2\log n}$ steps are required for a `well-conditioned' MPF (meaning $\sum_j|a_j|=\ordnung{\log n}$) of order $n$ which is an exponential improvement over classical (single) product formulae~\cite{low2019wellconditionedmultiproducthamiltoniansimulation,watson2024exponentiallyreducedcircuitdepths}.
More precise error bounds for MPFs similar to those of \cref{th:Schubert_error_bounds} have been derived by Zhuk et al.~\cite{Zhuk:2023zeh}, generalised by Aftab et al.~\cite{aftab2024multiproducthamiltoniansimulationexplicit} and refined by Mizuta~\cite{mizuta2025commutatorscalinghamiltoniansimulation}.

An alternative constant-depth MPF~\cite{Bark:2026znk} of order $n=4$ can be constructed using the ansatz
\begin{align}
	S_{4,2}(2h) &= \sum_{j=1}^2 a_j\, S_1^\uparrow\br{\kappa_j h}S_1^\downarrow\br{(1-\kappa_j) h}S_1^\uparrow\br{(1-\kappa_j) h}S_1^\downarrow\br{\kappa_j h}\,.
\end{align}
While less general than equation~\eqref{eq:mpf}, appropriately chosen coefficients $a_j$, $\kappa_j$ allow to keep the number of steps (circuit depth) constant for both constituent time evolutions.

MPFs do not preserve unitarity exactly, but the deviation is often negligible since $s>n$.
For this reason, they can be a good alternative when very high precision is required and polynomial expansions (sec.~\ref{sec:polynomials}) are not possible.

\subsection{Randomised methods}\label{sec:random}

There are several conceptually distinct ways to randomise Suzuki-Trotter decompositions.
For starters, the order of the operators $H=\sum_k A_k$ within each step can be permuted randomly and to some extent the step size can be varied.
This is found to reduce the 1st and 2nd order Trotter errors~\cite{Childs_2019,david2024fasterquantumsimulationmarkovian}, but the effect becomes negligible at higher orders~\cite{Childs_2019}.

Trotterization does not preserve all the physical symmetries of a given system exactly.
As a consequence, theoretically conserved quantities (e.g.\ the total magnetisation in a spin system) might experience an unphysical drift.
This drift can be alleviated by means of taking the average over several randomly chosen representatives of the same symmetry group~\cite{Lee:2026qao}.
The observable of interest does not have to be conserved (which would be rather boring), it only needs to share some symmetries with the Hamiltonian for the averaging to be applicable.

\subsubsection{Reducing the number of operators}

Alternatively, some of the operators (or gates) can be dropped entirely according to a given probability.
In the occasions when they are sampled, they need to be applied with correspondingly larger steps.
Such an approach trades sample-wise accuracy for shallower time evolution circuits.
The accuracy can be recovered via averaging over multiple random realisations.
This leads to an exponential multiplicative sampling overhead.
More specifically, the number of repetitions (shots) $\Gamma$ required for a constant error rate scales as
\begin{align}
	\Gamma &= \eto{\ordnung{\frac{\norm{H}_1^2\tau^2}{N_g}}}\,,
\end{align}
where $N_g$ denotes the (average) total number of gates in one shot~\cite{kiumi2024tepaiexacttimeevolution}.

In quantum computations shallower circuits reduce the hardware noise while the large number of independent simulations is prone to efficient parallelisation in classical simulations~\cite{Hasselgren:2026trv}.
There are several algorithms realising this idea in different ways.

The first such algorithm, dubbed quantum stochastic drift protocol (qDRIFT), was proposed by Campbell~\cite{PhysRevLett.123.070503} and generalised in Ref.~\cite{Kiss:2022lja}.
In qDRIFT at each Trotter step exactly one of the operators is sampled according to its weight in the Hamiltonian.
More flexibility is introduced in the Time Evolution via Probabilistic Angle Interpolation (TE-PAI) algorithm~\cite{kiumi2024tepaiexacttimeevolution}:
every operator is either applied at a given time or not, completely independently of the number of other operators applied at this same time.
Note that the TE-PAI algorithm is not randomising the usual Trotter sequence, but it is built from a limited set of specified angle operations PAI~\cite{Koczor:2023shj}.
Based on qDRIFT, the higher order methods qSHIFT~\cite{Lee:2026nqh} and pathwise random Hamiltonian simulation (PRHS)~\cite{Cugini:2026nra} have been put forward.
Adjusting the gate sampling according to properly chosen conditional probabilities exactly cancel lower order Trotter errors on average.

Combinations of Trotterization and qDRIFT are applicable to imaginary time evolution and, in consequence, quantum Monte Carlo (QMC) algorithms~\cite{Pocrnic:2023lrz}.
In Ref.~\cite{Martyn:2026iep} QMC simulations of systems with weak long-range interactions have been argued to benefit the most from randomisation on the operator level.

Randomised Trotter schemes can also be combined with multi-product formulae (MPF, sec.~\ref{sec:multi-product}), further increasing their effective order~\cite{Faehrmann2022randomizingmulti,Cho:2022gxr}.
MPFs can be exploited even more efficiently by sampling only the remainder error after a Trotter step.
This ansatz has been used in Ref.~\cite{Wang:2026lqk}.
Based on the earlier work~\cite{Zeng:2022pim}, Wang et al.\ have derived an algorithm with poly-logarithmic error scaling that moreover allows to elevate an order $n$ Trotter formula to order $2n+1$.

\subsubsection{Partially random methods}

As discussed in \cref{sec:hierarchy}, hierarchies in the magnitude of Hamiltonian terms should be exploited whenever possible.
Hybrid methods are a natural approach to this problem.
Refs.~\cite{Ouyang:2020jzz,Li:2021koi} independently advocate very similar partially randomised product formulae.
The idea is to use classical Trotterization on the dominating terms and add in the remaining terms at random using qDRIFT or similar.
This technique has subsequently been refined~\cite{Hagan:2022vyw} and demonstrated to be efficient in quantum chemistry simulations~\cite{Gunther:2025txp}.

\subsubsection{Eliminating Trotter errors}

Rather than just increasing the Trotter order, with randomised time evolution it is possible to eliminate the discretisation entirely.
This has been achieved by Granet and Dreyer~\cite{granet2023continuous} by applying the operators at different times distributed uniformly between $0$ and the final time $T$.
The number of times each operator is applied has to be chosen from a Poisson distribution depending on the weight of the respective operator.
Since the times are chosen from a continuum, no discretisation error is introduced.
Ref.~\cite{granet2023continuous} has also served as an inspiration for a continuous formulation of TE-PAI~\cite{Hayata:2026qcz,Dai:2026oqj}.

An alternative way to remove the Trotterization error is to add an operator after each time step that exactly cancels the error, very similarly to the approach in Refs.~\cite{Zeng:2022pim,Wang:2026lqk}.
While the additional operator would be infeasible to calculate exactly, it can be sampled without bias for instance using the continuous time TE-PAI algorithm.
This bias correction has been dubbed probabilistic Trotter error reversal (PTER)~\cite{Murota:2026cyh}.
It has a relatively small overhead with respect to the underlying Trotter scheme and introduces low noise levels because the main time evolution is still performed deterministically.
Both, the computational overhead and the noise level, can be systematically reduced with efficient higher order splitting methods.

Randomised time evolution methods like the one by Granet and Dreyer~\cite{granet2023continuous}, the continuous TE-PAI algorithm~\cite{Hayata:2026qcz}, or hybrid approaches~\cite{Gunther:2025txp,Murota:2026cyh} can be useful tools to avoid Trotter errors and to reduce the circuit depth, especially on noisy quantum computers.
It is, however, crucial to keep in mind that these methods introduce statistical errors leading to a substantial increase in the required number of shots.
It is strongly case- and hardware-dependent whether this trade-off is worthwhile.
 	
	\section{Alternatives and related methods}\label{sec:alternatives}

In the beginning of \cref{sec:practice} we have cautioned against the use of Trotterization without having considered potential alternatives.
These alternative methods for time evolution are discussed in the following.
There is certainly no guarantee that the list below is complete, but it covers most topics discussed in the recent literature on quantum time evolution.

We start with symplectic integrators of classical equations of motion in \cref{sec:symplectic}.
They are a special case of Suzuki-Trotter decompositions and allow some simplifications.
In particular, one of the third order nested commutators appearing in the BCH expansion~\eqref{eq:BCH} can be evaluated explicitly.
The resulting force-gradient term discussed in \cref{sec:force-gradient} can be included into the integrator.
This is itself a special case of splitting methods based on the Zassenhaus formula (sec.~\ref{sec:zassenhaus}) that incorporate the maximal amount of analytic knowledge about higher order commutators.

From \cref{sec:polynomials} on we leave Trotterization behind entirely.
Therein, we discuss polynomial expansions of the matrix exponential and argue that they are the first choice in any classical computation.
Theoretically exact unitarity is a small price to pay for the superior efficiency and precision achievable through Chebyshev or Taylor approximations.
In \cref{sec:tdvp} the two primary time evolution methods TEBD and TDVP employed in matrix product state and, more generally, tensor network simulations are briefly introduced.
Yet another approach has recently gained popularity for simulations on (noisy) quantum computers: quantum signal processing and circuit optimisation.
Minimal depth quantum circuits that are compiled case-by-case are reviewed in \cref{sec:qsp}.

Finally, the Crouch-Grossman and Munthe-Kaas methods for the numerical integration on manifolds (sec.~\ref{sec:integration_on_manifolds}) allow to solve the generalised Schrödinger equation with a time- and state-dependent Hamiltonian.
These methods, however, start with the premise that the exponential $\eto{h H}$ can be calculated for any given time and state.
Thus, they solve a very different problem and can be decoupled entirely from the Trotterization itself.

\subsection{Symplectic integrators}\label{sec:symplectic}

The applicability of Suzuki-Trotter decompositions is not limited to the quantum world.
Very prominently, the equations of motion (EOM) in classical mechanics can also be solved numerically using Trotter schemes.
In this context they are referred to as \textit{symplectic integrators} and the exact unitarity of splitting methods translates to exact phase space volume conservation~\cite{Hairer:2006gni,Hairer:1993}.
Among other advantages, this property can be used for sampling algorithms like the hybrid Monte Carlo (HMC)~\cite{Duane:1987de} where it guarantees detailed balance.
In particular, the energy does not experience any drift over time.
This is the main distinguishing feature from standard \textit{Runge-Kutta} (RK) methods~\cite{Runge:1895hdo,Kutta} which will typically see the energy diverge or approach zero in the long run.

In their original work~\cite{OMELYAN2003272} that also forms the basis of \cref{sec:omelyan} Omelyan et al.\ have catalogued all (efficient) 2nd and 4th order symplectic integrators.
See especially table~2 of Ref.~\cite{OMELYAN2003272}.
A few sporadic higher order integrators are provided in Ref.~\cite{OMELYAN2003272} as well.
Unless very high accuracy is required, these are still the best schemes available.

\subsubsection{Energy conservation}

We start with a classical Hamiltonian (identified as the energy)
\begin{align}
	\mathcal{H} &= T(p) + V(x)\,,
\end{align}
where $T$ denotes the kinetic energy solely depending on the momentum $p$ and $V$ is a potential that depends on the position $x$.
The absence of mixing terms containing both $p$ and $x$ is essential for the efficacy of symplectic integrators.
Now the EOM read
\begin{align}
	\begin{split}
		\dot p &= -\del{\mathcal H}{x} = -V'(x)\,,\\
		\dot x &= \hphantom{-}\del{\mathcal H}{p} = \hphantom{-} T'(p)\,,
	\end{split}\label{eq:classical_EOM}
\end{align}
where $\dot x\equiv \del{x}{\tau}$ is used as the standard physicists' notation for the time derivative.

Numerical symplectic integration does not lead to an exact conservation of $\mathcal{H}$ at finite step sizes.
However, any order $n$ symplectic integrator conserves a so-called shadow Hamiltonian $\tilde{\mathcal{H}} = \mathcal{H} + \ordnung{h^n}$ exactly~\cite{Clark:2008gh}.
The specific form of $\tilde{\mathcal{H}}$ depends on the chosen integrator and it can be derived in terms of nested commutators (or Poisson brackets) of the operators involved~\cite{Kennedy:2012gk}.
Since $\tilde{\mathcal H}$ is constant in time, the true Hamiltonian $\mathcal H$ can never deviate by more than $\ordnung{h^n}$ from the conserved energy.

\subsubsection{Comparison to Runge-Kutta methods}\label{sec:runge-kutta}

There are three conceptually different ways to perform a minimal time step for these EOM numerically.
They are listed in \cref{tab:rk_vs_symplectic} and they differ in the amount of information from the future time $\tau+h$ used in the update at time $\tau$.
In RK methods $p$ and $x$ are updated at the same time while symplectic integrators update them alternately.

\begin{table}
	\centering
	\begin{tabular}{l|l|l}
		Explicit Runge-Kutta & Symplectic & Implicit Runge-Kutta\\\hline
		$x(\tau+h) = x(\tau) + h T'(p(\tau))$ & $x(\tau+h) = x(\tau) + h T'(p(\tau))$ & $x(\tau+h) = x(\tau) + h T'(p(\tau+h))$\\
		$p(\tau+h) = p(\tau) - h V'(x(\tau))$ & $p(\tau+h) = p(\tau) - h V'(x(\tau+h))$ & $p(\tau+h) = p(\tau) - h V'(x(\tau+h))$\\
	\end{tabular}
	\caption{Different update schemes for the time step $\tau \rightarrow \tau+h$ for classical EOM~\eqref{eq:classical_EOM}.
		These steps define a valid first order (Euler) method or, more generally, the building blocks (ramps, see fig.~\ref{fig:ramp}) for higher order methods.
		Of course, the symplectic integration can also start with a momentum update.}
		\label{tab:rk_vs_symplectic}
\end{table}

For the sake of completeness let it be noted that RK methods are designed to solve ODEs of the form
\begin{align}
	\dot z &= f(t,z)
\end{align}
and, in the given setting, one would interpret the tuple $z=(x,p)^\trans$ as a vector.
This more general approach ignores some of the physical information at our disposal resulting in worse performance.
An RK method with $s$ stages is defined via a so-called \textit{Butcher tableau}~\cite{Butcher_1964}

\begin{equation}
	\begin{tabular}{l|lcl}
		$\tilde{c}_1$ & $\tilde{a}_{11}$ & $\cdots$ & $\tilde{a}_{1s}$ \\
		$\vdots$ & $\vdots$ & $\ddots$ & $\vdots$ \\
		$\tilde{c}_s$ & $\tilde{a}_{s1}$ & $\cdots$ & $\tilde{a}_{ss}$ \\\hline
		& $\tilde{b}_1$ & $\cdots$ & $\tilde{b}_s$\rule{0pt}{2.6ex}
	\end{tabular}\label{eq:butcher-tableau}
\end{equation}

\noindent
listing the coefficients $\tilde{a}_{ij}$, $\tilde{b}_i$, $\tilde{c}_i$ with $\tilde{c}_i=\sum_{j=1}^s \tilde{a}_{ij}$ and $\sum_{i=1}^s\tilde{b}_i=1$ for the update step
\begin{align}
	z(\tau + h) &= z(\tau) + h \sum_{i=1}^{s}\tilde{b}_i k_i + \ordnung{h^{n+1}}\,,\\
	k_i &= f\br{\tau + \tilde{c}_i h, z(\tau) + h\sum_{j=1}^s \tilde{a}_{ij} k_j}\,.\label{eq:RK_ki}
\end{align}
Here, we use the tilde to distinguish these coefficients from those in Suzuki-Trotter decompositions.
Butcher tableaus for efficient $n$th order RK methods are readily available in the literature~\cite{Hairer:1996,dormand_prince,matlab_ode}.
We can now generalise the notion of explicit and implicit RK methods introduced in \cref{tab:rk_vs_symplectic}.
Explicit methods only use information available from previous steps, that is $\tilde{a}_{ij}=0$ for all $j\ge i$.
All RK methods that are not explicit are implicit.

\subsubsection{Relation to Trotterization}

We write the EOM~\eqref{eq:classical_EOM} in the compact form
\begin{align}
	\dot z &= D_\mathcal{H} z\,,
\end{align}
with the linear operator $D_\mathcal{H} = D_T+D_V$ defined via the Poisson brackets $D_A z\coloneqq \brc{z,A}$, i.e.\ $D_T = T'(p)\partial_x$ and $D_V = V'(x)\partial_p$.
The formal solution of the EOM is then simply given by
\begin{align}
	z(\tau) &= \eto{\tau D_\mathcal{H}} z(0)
\end{align}
and $\eto{\tau D_\mathcal{H}}$ can be interpreted as a time evolution operator that needs to be Trotterized.
Crucially,
\begin{alignat}{4}
	\eto{h D_T} &= 1 + h D_T \quad && \text{and} \quad & \eto{h D_V} &= 1 + h D_V\\
	\Rightarrow \eto{h D_T} \matr{x\\p} &=  \matr{x\\p} + h\matr{T'(p)\\0} \quad && \text{and} \quad & \eto{h D_V} \matr{x\\p} &=  \matr{x\\p} + h\matr{0\\-V'(x)}
\end{alignat}
can be evaluated exactly because $D_T^2=D_V^2=0$ due to the absence of $x$,$p$-mixing terms in $\mathcal{H}$.
Therefore, any update scheme that alternates between $x$ and $p$ steps with coefficients from a Suzuki-Trotter decomposition is also a valid symplectic integrator.

In classical (non-relativistic) mechanics the kinetic energy is quadratic in the momentum $T(p)\propto p^2$ which allows the additional simplification
\begin{align}
	[D_V,[D_V,[D_V,D_T]]] &= 0\,.\label{eq:symplectic_zero}
\end{align}
Thus, for orders $n\ge4$ some of the nested commutators that make up the leading order error vanish.
For instance, in equation~\eqref{eq:ord_5_errors} the $\gamma_4$ and $\gamma_6$ terms drop out assuming $A=D_T$ and $B=D_V$.
The resulting simplified error functions can be minimised in order to obtain maximally efficient symplectic integrators at a given order.
Be aware that efficient symplectic integrators are not necessarily efficient general splitting methods.
In fact, the exploitation of the identity~\eqref{eq:symplectic_zero} even means that some 6th order symplectic integrators are only 4th order Trotter schemes.

For lower-order polynomial potentials even more order conditions can be dropped in a similar way.
This has been exploited in Ref.~\cite{escorihuelatomàs2026efficientsymplecticintegratorscubic} where efficient symplectic integrators are identified for cubic and quartic potentials.

\subsubsection{Force-gradient integrators}\label{sec:force-gradient}

Again assuming a quadratic kinetic energy, i.e.\ constant $T''\equiv T''(p)$, we can also explicitly derive the so-called force-gradient term
\begin{align}
	D_F \coloneqq [D_V,[D_T,D_V]] &= 2V'(x) \cdot T'' \cdot V''(x)\cdot \partial_p \label{eq:force-grad-term}\\
	\Rightarrow \eto{h^3 D_F}  \matr{x\\p} &=  \matr{x\\p} + h^3\matr{0 \\ 2 V''(x) \cdot T'' \cdot V'(x)}.
\end{align}
This 3rd order update commutes with the 1st order momentum update $D_V$ and both updates can be applied simultaneously.
A full order $n$ update step
\begin{align}
	\eto{h D_\mathcal{H} + \ordnung{h^{n+1}}} &= \eto{ a_1h D_T}\eto{b_1hD_V + c_1 h^3 D_F}\eto{ a_2hD_T}\cdots\eto{b_qhD_V + c_q h^3 D_F}\eto{ a_{q+1}hD_T}\label{eq:force-grad}
\end{align}
is defined by the coefficients $a_i$, $b_i$, $c_i$ where the $a_i$ and $b_i$ correspond to position $D_T$ and momentum $D_V$ updates, respectively, and the $c_i$ are used for the force-gradient $D_F$.
The catalogue in Ref.~\cite{OMELYAN2003272} includes highly efficient force-gradient methods.
Updates are abbreviated by $A=\eto{a_i h D_T}$, $B=\eto{b_i h D_V}$ and $C=\eto{b_i h D_V + c_i h^3 D_F}$ therein.

Note that the second derivatives $T''$ and $V''$ are Hessian matrices, so their exact evaluation and subsequent matrix-vector multiplications can be computationally very expensive.
The application of $T''$ cannot be avoided and is very rarely the bottleneck of a simulation.
The Hessian of the potential $V''$, on the other hand, does not have to be calculated explicitly.
Instead, the force-gradient update can be replaced by a two-step procedure~\cite{Yin:2011np,schaefers2024hessianfree}
\begin{align}
	\eto{b_i h D_V + c_i h^3 D_F}  \matr{x\\p} &=  \matr{x\\p} + h\matr{0 \\ -b_i V'(\tilde x)} + \ordnung{h^5}\,,\\
	\tilde x &= x - 2 h^2\,\frac{c_i}{b_i}\, T'' \cdot V'(x)\,.
\end{align}
That is, the momentum update is simply performed using the gradient $V'(\tilde x)$ at a slightly modified position $\tilde x$.
The position itself remains unchanged, i.e.\ the next position update starts at $x$ (not $\tilde x$).

Since the two-step procedure is only equivalent to the true force-gradient update up to $\ordnung{h^5}$, it requires different coefficients $a_i$, $b_i$, $c_i$ for valid higher order schemes and maximal efficiency.
Schäfers et al.~\cite{schaefers2024hessianfree} have derived and catalogued these coefficients for all Hessian-free force-gradient methods up to $q\le5$ cycles in a very similar style to Ref.~\cite{OMELYAN2003272}.
Their table~1 includes all 2nd, most 4th and several 6th order symplectic integrators with the abbreviations $A$,$B$,$C$ as in Ref.~\cite{OMELYAN2003272} and $D$ denoting the two-step procedure.
In a subsequent study~\cite{Schafers:2025mgt} the same authors have verified that the stability of two-step Hessian-free methods coincides with their full force-gradient counterparts.
Moreover, symplectic integrators with reduced efficiency but improved stability are presented in Ref.~\cite{Schafers:2025mgt}.

\subsection{Zassenhaus formula}\label{sec:zassenhaus}

The idea behind the force-gradient integrator can be generalised beyond symplectic integration.
It boils down to including higher orders terms in the \textit{Zassenhaus formula} (the dual of the BCH formula~\eqref{eq:BCH})
\begin{align}
	\eto{A+B} &= \eto{A}\eto{B}\eto{-\frac12 [A,B]}\eto{\frac16\br{2[B,[A,B]]+[A,[A,B]]}}\cdots
\end{align}
explicitly into the time evolution.
For instance, the force-gradient term~\eqref{eq:force-grad-term} corresponds to the third order nested commutator $[B,[A,B]]$.

While the coefficients within the Zassenhaus formula can be computed efficiently using a recursive procedure~\cite{Casas:2012nqk}, the commutators themselves are often very hard to derive and exponentiate.
This is why the Zassenhaus formula is rarely used for time evolution in practice.
In some special cases, however, the higher order nested commutators and their exponentials can be constructed explicitly.
This additional analytic insight (e.g.\ in-depth understanding of the Cartan subalgebra~\cite{Chaudhary:2026ksu}) can then be exploited to increase the efficiency of the splitting method.

More specifically, Zassenhaus-based formulae up to 11th order have been derived for the transverse-field Ising model~\cite{peetz2025hamiltoniansimulationstochasticzassenhaus}.
Broader classes of spin-type models are tackled in Ref.~\cite{nguyen2025zassenhausexpansionsolvingschrodinger}, be it only to second order, based on a Cartan decomposition of the underlying Lie algebra~\cite{Kokcu:2021ctj}.

The applicability of the Zassenhaus formula has been generalised considerably by Jourdan and Cassam-Chenaï~\cite{Jourdan:2025gqs}.
They have shown that assuming the so-called no-mixed adjoint property
\begin{align}
	\forall k\in \mathds{N}: \; \ad_B^{\vphantom{l}} \ad_A^k B &= 0\,,
\end{align}
with the $l$-th adjoint as in equation~\eqref{eq:adjoint}, the Zassenhaus formula simplifies to
\begin{align}
	\eto{A+B} &= \eto{A} \exp\br{\sum_{k=0}^{\infty} \frac{(-1)^k}{(k+1)!} \ad_A^k B}.
\end{align}
In particular, this decomposition allows highly efficient unitary coupled cluster simulations for some strongly correlated electron systems.
As demonstrated in Ref.~\cite{burdine2024trotterlesssimulationopenquantum}, some open systems governed by the Lindblad equation can also be time evolved efficiently under the condition $[A,B] = \xi_1 B + \xi_2$ with $\xi_{1,2}\ge 0$ which implies the no-mixed adjoint property.
In this case an expansion in Kraus operators is advocated as an alternative to Trotterization.

Applications of the Zassenhaus formula are an extreme case of model-dependent optimisation.
The necessary higher order commutators can rarely be derived with reasonable effort.
But when they are available, their knowledge can and should be used to accelerate simulations.

\subsection{Polynomial expansions}\label{sec:polynomials}

As a rule, the most efficient way to approximate the exponential of a sparse matrix on a classical computer is to use a polynomial expansion.
This is not always possible and comes at the price of exact unitarity.
Note that the latter obstacle is primarily conceptual since polynomial expansions readily allow to reach machine precision and thus restore unitarity for any practical purpose.
The most natural choices are the Taylor series
\begin{align}
	\eto{H h} &= \sum_{k=0}^n\frac{\left(H h\right)^k}{k!} + \ordnung{h^{n+1}}\,,\label{eq:taylor_expansion}
\end{align}
and the Chebyshev polynomials
\begin{align}
	T_0(H) &= \id\,, \quad
	T_1(H) = H\,, \quad 
	T_{k+1}(H) = 2H\cdot T_k(H) - T_{k-1}(H) \\
	\Rightarrow \eto{H h} &= I_0\br{\Gamma h} + 2\sum_{k=1}^n I_k\br{\Gamma h} T_k\left(\frac{H}{\Gamma}\right) + \ordnung{h^{n+1}}\label{eq:chebyshev_expansion}
\end{align}
with a cut-off scale $\Gamma\ge\norm*{H}_2$.
The $I_k$ denote the modified Bessel functions of first kind.

For real time evolution using the optimal Chebyshev expansion~\eqref{eq:chebyshev_expansion} the polynomial order $n$ and thus cost per time step $h$ scales as
\begin{align}
	n &= \Gamma h + \ordnung{\log\br{\frac1\varepsilon}}
\end{align}
with the global error $\varepsilon$.
The minimal total cost is achieved using $\Gamma = \norm*{H}_2$ and performing just a single time step $h=\im t$ for the entire real time $t$.
Multiple shorter time steps $N_t$ might be easier to realise, but they also come with the overhead $\ordnung{N_t \log\br{\nicefrac1\varepsilon}}$.
In either case the cost stays linear in $\norm*{H}_2 t$ and logarithmic in $\varepsilon$ which is in stark contrast to the $\ordnung{\br{\norm*{H}_2 t}^{1+\nicefrac1n} \varepsilon^{-\nicefrac1n}}$ cost of Trotterization.

Taylor should be used whenever an approximation in the entire complex plane is required, while Chebyshev expansions provide the most accurate approximations on purely real (or imaginary intervals)~\cite{Ostmeyer:2023xju}.
Of course, in either case the matrix exponential should never be calculated explicitly.
Instead, higher order polynomials should be constructed by repeated application of the sparse (Hamiltonian) matrix onto a vector (sec.~\ref{sec:sparsity}).

\begin{figure}
	\centering
	\includegraphics[width=.4\textwidth]{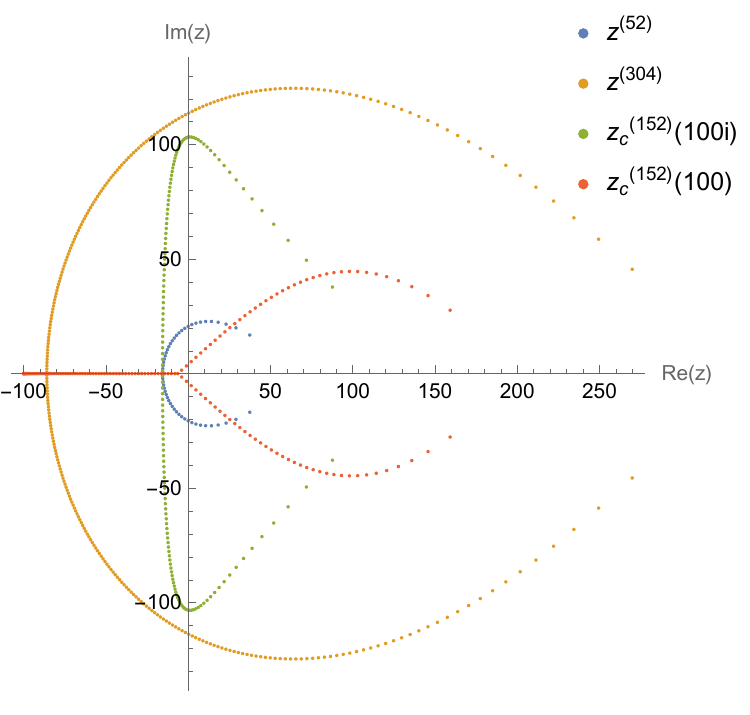}%
	\hfill%
	\resizebox{.55\textwidth}{!}{{\large\input{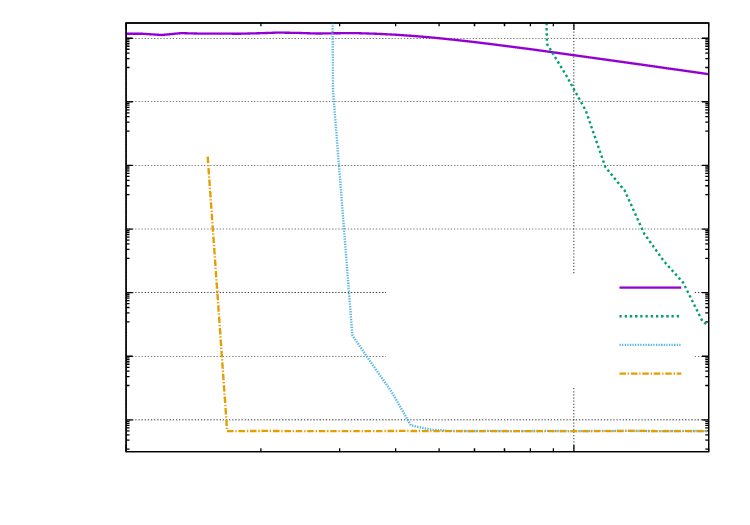}}}
	\caption{Time evolution methods based on polynomial expansions to order $n$.
	Left: zeros $z^{(n)}$ of the truncated Taylor series~\eqref{eq:taylor_expansion} and zeros $z_c^{(n)}(\Gamma h)$ of the Chebyshev polynomial~\eqref{eq:chebyshev_expansion} best describing the interval $[-\Gamma h,\Gamma h]$, respectively.
	Right: error $\norm*{S_n(-\im t) - \eto{-\im H t}}_F$ for different step sizes $h$ after the fixed time $t$ as a function of the approximate computational cost $n/h$ (we chose $n=4$ for the Strang splitting, certainly underestimating the cost).
	The Taylor expansion either uses the na\"ive sum `$\sum$' as in eq.~\eqref{eq:taylor_expansion} or the product `$\prod$' as in eq.~\eqref{eq:exp_factorisation}.
	Adapted figure from~\cite{Ostmeyer:2023xju}.}\label{fig:zeros-coeffs}
\end{figure}

In the literature (e.g.~\cite{PhysRevLett.114.090502,Chundury:2026wpt}) Taylor polynomials are often chosen for Hamiltonian simulations due to their simplicity.
However, some additional care is advised for this choice.
As opposed to the recursively constructed Chebyshev polynomials, a na\"ive construction of the Taylor series is numerically highly unstable.
This can be clearly observed in the right panel of figure~\ref{fig:zeros-coeffs} where the direct evaluation of the sum `$\sum$' requires substantially smaller steps $h$ to reach the same precision as the factorised version `$\prod$'
\begin{align}
	\eto{Hh} &= \prod_{i=1}^{n}\left(1-\frac{h}{z_i^{(n)}}H\right) + \ordnung{h^{n+1}}\label{eq:exp_factorisation}
\end{align}
of the very same polynomial.
The roots $z^{(n)}$ of the truncated Taylor series are very well behaved and follow the Szeg\H{o} curve~\cite{VARGA2010298} asymptotically for large $n$.
They rather intuitively enclose the region of the complex plain for with the polynomial approximation is valid (see fig.~\ref{fig:zeros-coeffs}, left) since the exponential function does not have any zeros.

We assume that the cost of the polynomials in figure~\ref{fig:zeros-coeffs} for each time step is dominated by the number of Hamiltonian applications $n$.
The Strang splitting is provided primarily to demonstrate the vastly superior accuracy of the polynomial methods.
For its cost estimation $n=4$ has been used as a very conservative estimate.
In realistic classical computations Trotter formulae are typically even less favourable.

An alternative method to ensure numerical stability of the Taylor expansion (or, in fact, any polynomial) is Horner's scheme.
For instance, the $n$-th order Taylor approximation of the time evolution operator would be applied to a state $\ket{\psi}$ as follows
\begin{align}
	\eto{h H} \ket{\psi} &= \ket{\psi} + \br{h H \br{\ket{\psi} + \cdots + \frac{h}{n-1} H\br{\ket{\psi} + \frac{h}{n} H\ket{\psi}}\cdots}} + \ordnung{h^{n+1}}\,.\label{eq:horner}
\end{align}
It is also worth noting that Chebyshev expansions suffer from rapid oscillations.
They can be smoothened with the help of a so-called kernel $g_k$ that interpolates smoothly between $g_0=1$ and $g_{n+1}=0$.
Ref.~\cite{RevModPhys.78.275} (especially Tab.~1 therein) provides a great overview over useful kernels.

In a nutshell, the Chebyshev expansion~\eqref{eq:chebyshev_expansion} should be the first choice for classical time evolution or, more generally, the application of a sparse matrix exponential with purely real (or imaginary) eigenvalues to a vector.
For sparse matrices with eigenvalues spread out in the complex plane, one should use the factorised Taylor series~\eqref{eq:exp_factorisation} or Horner's method~\eqref{eq:horner}.
Only if these polynomial approximations are ruled out, for instance because exact unitarity is required, Trotterization becomes a viable alternative.

\subsection{Quantum signal processing}\label{sec:qsp}

In the context of quantum computing (sec.~\ref{sec:quantum_comp}) polynomial expansions cannot be used directly for time evolution since they do not preserve exact unitarity.
They can, however, be encoded in the so-called \textit{quantum signal processing} (QSP) framework~\cite{Low:2016sck,Motlagh:2023oqc}.
QSP promises to inherit the superior convergence properties of polynomial expansions (total cost scaling as $\ordnung{\norm*{H}_2 t} + \ordnung{\log\br{\nicefrac1\varepsilon}}$).
For a detailed explanation of all the steps leading up to Hamiltonian time evolution with QSP we refer once more to the lecture notes~\cite{Lin:2022vrd}.
A concise summary of different time evolution algorithms including QSP can be found in Ref.~\cite{Hariprakash:2023tla}, with their respective computational complexities listed explicitly.
Ref.~\cite{Kane:2025ybw}, on the other hand, provides a more practical guide line for simulations of lattice field theories approaching continuous time.

Time evolution with QSP is typically realised via block encoding (BE)
\begin{align}
	U_A &= \matr{A & * \\ * & *}
\end{align}
of the target matrix $A$ as part of the unitary $U_A$.
For a physical volume $V$ the BE will typically come at the cost of an additional factor $\ordnung{V}$ in runtime or gate count (see e.g.\ Ref.~\cite{Hariprakash:2023tla}, Tab.~V).
The derivation of BEs is not straight forward, but some realisations are known for selected physical systems including lattice gauge theories~\cite{Draper:2026bcj,Rhodes:2024zbr}.
Qubitization~\cite{Low:2016znh} and quantum singular value transformation~\cite{Gilyen:2018khw} then allow unitary representations of polynomials, e.g.\ the Chebyshev expansion of the Hamiltonian.
Alternatively, QSP can be related to non-linear Fourier transform~\cite{Laneve:2025nvj}.

The BE causes a significant overhead in the dimension of the Hamiltonian so that Trotterization can be preferable for large systems even though QSP features favourable long-time scaling.
Moreover, at the time of writing we could not identify a full and concise `recipe-style` description of QSP that would allow to implement time evolution with reasonable effort.
For these reasons we believe that QSP is not yet useful in practice despite its great theoretical potential.
It remains an open question whether QSP, Trotterization or something else entirely will prove the best alternative for quantum computing in the long run.

\subsubsection{Circuit optimisation}\label{sec:circuit_opt}

While exact QSP can still be prohibitively expensive and complicated, some empirical approximations are feasible.
There has been a great effort in the last few years to construct quantum circuits for time evolution using variational optimisation.
The general idea is that asymptotic convergence is not essential as long as the approximation is more accurate than the hardware noise.
Thus, in the noisy intermediate-scale quantum (NISQ) era (sec.~\ref{sec:nisq}) deep circuits based on Trotterization or QSP are sometimes replaced by shallower circuits tailored directly for the problem at hand~\cite{Kikuchi:2023qbb}.
Initially, variational quantum simulations have been introduced for the evolution of specific states~\cite{Li:2016vmf}, but their development has largely been abandoned in favour of direct approximations of the full time evolution operator~\cite{Assi:2025ifp,Shoji:2026kyr}

The first operator-based ansatz was the so-called ``brickwall circuit'' formed by quantum gates for a fixed short evolution time~\cite{classical_opt_circuits,optimised_circuits}.
A number of variational parameters needs to be optimised classically, fixing the number of layers and the Hamiltonian up front.
The optimisation of the resulting time evolution approximation in Frobenius norm leads to very accurate results, but it comes at a high classical optimisation cost.

Subsequently, a multitude of related approaches has been developed with the goal to improve the circuit compression and the classical scalability.
In Refs.~\cite{classical_opt_circuits,gibbs2024deepcircuitcompressionquantum} a matrix product operator (MPO) ansatz has been utilised.
Ref.~\cite{Luiz:2026ter}, in contrast, starts with the Trotter circuit as an initial guess and subsequently optimises the circuit to reduce depth.

An alternative ansatz has been pursued in Ref.~\cite{variational_circuits} for translationally invariant systems.
Therein scalability is achieved through optimisation on small subsystems which are extrapolated to the full circuit.
The very opposite approach has been followed in Ref.~\cite{le2025riemannianquantumcircuitoptimization} where symmetries were excluded from the parametrisation deliberately and scalability is expected from a tensor network representation.

Perhaps surprisingly, variational quantum circuits have also been used for imaginary time evolution and ground state search~\cite{PRXQuantum.2.010342,Wolf:2025mjo}.
There appears to be a very smooth transition from circuit optimisation for time evolution to variational quantum algorithms.

Overall quantum circuit optimisation has emerged as a promising alternative to Suzuki-Trotter methods on noisy quantum devices since their shallower circuits allow to minimise hardware noise.
They do, however, require substantial case-by-case tuning and it is expected that they will not scale well beyond the NISQ era.

\subsection{Time-dependent variational principle}\label{sec:tdvp}

\textit{Tensor networks} (TNs) are the standard way to represent quantum systems on a classical computer nowadays.
The idea is to approximate the exponentially many degrees of freedom by polynomially many states with a structure inspired by the geometry of the underlying problem.
Typically, systems with low levels of entanglement can be represented well with this ansatz.
In particular, the ground state of a gapped local Hamiltonian has an accurate TN approximation~\cite{Hastings_2006}.	

Initially, the so-called matrix product states (MPS) have been introduced for systems in 1 spatial dimension~\cite{Verstraete:2004gdw,Perez-Garcia:2006nqo}.
MPS and their relation to the density-matrix renormalization group method (DMRG) are explained in Ref.~\cite{Schollwoeck:2010uqf} in great detail.
In higher dimensions there is a large variety of different TNs and we deliberately restrain here from naming any representatives or reviewing them.
Instead, Orús's introduction~\cite{tensor_intro} and concise review~\cite{Orus:2018dya} are highly recommended.
The particularly beginner-friendly lecture notes~\cite{lami2025beginnerslecturenotesquantum} introduce quantum spin chains, exact diagonalisation and TNs, those by Misguich~\cite{Misguich:2026mow} go into more details about MPS algorithms, while the notes~\cite{Waintal:2026tsc} focus on a comparison with quantum computing.
Finally, time evolution is ubiquitous in TN simulations.
It is performed in real (see e.g.~\cite{Hemery:2019} for a concise overview), imaginary (e.g.\ for ground state search~\cite{my_tensor_networks}) and, recently, even complex tilted time (e.g.\ for Lindbladian spectra~\cite{Westhoff:2025mxp}).
Multiple time evolution methods have been developed specifically for the simulation of TNs~\cite{White:2018kbw,Rakovszky:2022hss,Roy:2024irm}.
Paeckel et al.~\cite{Paeckel:2019yjf} have reviewed the time evolution of MPS in much greater detail than we could cover here.

Nevertheless, let us provide an overview of the two predominant methods.
In the context of TNs, time evolution using Suzuki-Trotter methods is referred to as \textit{time-evolving block decimation} (TEBD).
The main alternative algorithm is called \textit{time-dependent variational principle} (TDVP) and based on the eponymous concept in quantum mechanics~\cite{LANGHOFF:1972hex,Kramer:1981}.
The TDVP for MPS was introduced~\cite{Haegeman:2011zz} and refined~\cite{Haegeman:2015ezw} by Haegeman et al.\ in order to overcome the main shortcomings of TEBD.
In addition to the usual time discretisation errors, classical splitting methods suffer from large truncation effects through the polynomial-size constraint of TNs.
Trotterized time evolution tends to leave the manifold currently parametrising the MPS.
The necessary subsequent truncation does not even conserve constants like the energy in general.
These problems are almost entirely eliminated in TDVP at the cost of a considerably more complex algorithm.

The central idea in the TDVP is to time evolve the variables $X$ in the TN (more generally, some parametrised state~\cite{Hackl:2020viw}) directly
\begin{align}
	\Ket{\nabla_X\psi(X)} \cdot \dot X &= -\im H \Ket{\psi(X)}\,,
\end{align}
rather than approximately evolving the quantum state $\ket{\psi(X)}$ parametrised by said variables with the gradient $\nabla_X$.
This is achieved through a projection
\begin{align}
	\Braket{\nabla_{\bar X}\psi(\bar X)|\nabla_X\psi(X)} \cdot \dot X &= -\im \Bra{\nabla_{\bar X}\psi(\bar X)} H \Ket{\vphantom{\bar X}\psi(X)}
\end{align}
onto the tangent space $T_X \mathcal{M}_V$ of the manifold $\mathcal{M}_V=\brc{\ket{\psi(X)} | X\in V}$ at position $X$.
Here, $\mathcal{M}_V$ is defined by all the states parametrisable via $X$ with the complex vector space $V$ that encodes the structure of the given TN (including the current bond dimension) and $\bar X$ denotes the complex conjugate.
Note that $\braket{\nabla_{\bar X}\psi(\bar X)|\nabla_X\psi(X)}$ is the so-called Gram matrix acting on $V\rightarrow V$ that formally needs to be inverted.
In practice, the inversion of the full matrix can be avoided and replaced by approximations based on local operations, see e.g.\ Algorithms~5 and~6 in Ref.~\cite{Paeckel:2019yjf}.

In short, the TDVP is a powerful and accurate algorithm for the time evolution of TN states, in particular MPS.
It should always be considered as a more than viable alternative to the more straight forward Trotterized TEBD.

\subsection{Numerical integration on manifolds}\label{sec:integration_on_manifolds}

The numerical integration on manifolds can be viewed as a further generalisation of the Schrödinger equation
\begin{align}
	\dot \psi &= H(\tau,\psi) \cdot \psi\,.\label{eq:ODE_on_manifold}
\end{align}
Not only is the Hamiltonian $H(\tau,\psi)$ time-dependent (sec.~\ref{sec:time-dependent}), it can also depend on the state $\psi$ itself.
Moreover, one can drop the normalisation condition in favour of $\psi$ living on a manifold whose structure is preserved by the time evolution.
These are the premises Crouch and Grossman set out to derive numerical solutions for~\cite{Crouch:1993}.

\subsubsection{Crouch-Grossman methods}\label{sec:crouch-grossman}

As opposed to Suzuki-Trotter decompositions, the key assumption of \textit{Crouch-Grossman methods} is that the ``frozen'' solution $\eto{h H(\tau,\psi_1)}\cdot \psi_2$ can be calculated without difficulties for any $\tau$ and $\psi_{1,2}$.
In addition, this solution is required to remain on the manifold exactly, even at a finite step size $h$.
Of course, a standard RK update $\br{1+h H(\tau,\psi)}\cdot \psi$ does not stay on the manifold.
Higher order Crouch-Grossman methods are conceptually very similar to RK integrators with the linear update replaced by an exponential map.
The specific coefficients, however, need to be modified as the order conditions differ slightly~\cite{Owren:1999}.

\subsubsection{Munthe-Kaas methods}\label{sec:munthe-kaas}

These deviations can be corrected for by a small modification introduced by Munthe-Kaas~\cite{MUNTHEKAAS1999115}.
Rather than using the $k_i=H(\tau+h\tilde{c}_i, \dots)$ defined by an order $n$ RK method~\eqref{eq:RK_ki} directly, Munthe-Kaas proposed to use
\begin{align}
	\tilde{k}_i &\coloneqq \sum_{l=1}^{n-1} \frac{\mathcal{B}_l}{l!} \ad_{u_i}^l k_i\,,
\end{align}
where $\mathcal{B}_l$ denotes the $l$-th Bernoulli number, equation~\eqref{eq:adjoint} defines the $l$-th adjoint action, and the $u_i$ are given below.
Given the Butcher tableau~\eqref{eq:butcher-tableau} coefficients for an $s$-stage RK method, the full \textit{Runge-Kutta-Munthe-Kaas} (RKMK) method then reads
\begin{align}
	\psi(\tau + h) &= \eto{h \sum_{i=1}^{s}\tilde{b}_i \tilde{k}_i} \cdot \psi(\tau) + \ordnung{h^{n+1}}\,,\\
	u_i &= h\sum_{j=1}^s \tilde{a}_{ij} \tilde{k}_j\,,\\
	k_i &= H\br{\tau + \tilde{c}_i h, \eto{u_i} \cdot \psi(\tau)}\,.
\end{align}
Such an RKMK update is of the same order $n$ as the underlying RK method and guarantees that $\psi$ stays on the desired manifold.
Note that Algorithm~1 in Ref.~\cite{MUNTHEKAAS1999115} covers an even more general case and the procedure described here specifically tackles equation~\eqref{eq:ODE_on_manifold}.
 	
	\section{Quantum computing for time evolution}\label{sec:quantum_comp}

At the time of writing, the development of improved time evolution methods like Trotterization is mainly driven by their potential use in quantum computing.
Let us, therefore, briefly discuss the goals and requirements that differ between quantum and classical hardware.
We emphasise that this is not a review on quantum computing.
Recent reviews on the topic are, however, readily available.
For instance, Ref.~\cite{https://doi.org/10.48550/arxiv.2303.04850} also cover aspects of quantum time evolution in the context of many-body physics.
A reader not yet familiar with quantum computing will find that the book by Nielsen and Chuang~\cite{Nielsen_Chuang_2010} is a great starting point. 
More recent algorithms for scientific quantum computing are covered by Ref.~\cite{Lin:2022vrd} in a beginner-friendly way.

Simply speaking, quantum computers are naturally appealing for real time evolution because they rely exclusively on unitary operations.
These unitary operations need to be encoded in local gates, very reminiscent of the task phased by splitting methods.
Of course, there is also the analogy argument that ``quantum systems are best simulated on quantum hardware''.
For these reasons the first quantum advantage in physics applications is often expected to be demonstrated in simulations involving time evolution (see e.g.~\cite{Jordan:2012xnu,Simon:2025aie,Babbush:2025eqc}).

\subsection{Theoretical potential and limitations}\label{sec:qc_limitations}

Besides the obvious use for Hamiltonian dynamics, time evolution is also a building block for quantum phase estimation~\cite{Kitaev:1995qy,Cleve:1997dh,Pelofske:2026yor}, adiabatic quantum computing~\cite{Farhi:2000ikn,Cunningham:2026jgq} and other algorithms~\cite{Epperly:2021ugt,Shen:2022lmk}.
In fact, quantum signal processing (QSP)~\cite{Low:2016sck} relying only on real-time evolution $\eto{\im t H}$ oracles can be used to approximate any function on a quantum computer~\cite{silva2022fourierbased,Motlagh:2023oqc}.

\subsubsection{Computational complexities}

That said, quantum computers certainly have their limitations as well.
Any problem in the classical polynomial complexity class \texttt{P} is also in \texttt{BQP}, that is easily solvable on a quantum computer.
It is, however, unknown how much larger \texttt{BQP} is than \texttt{P} (they might even be equal) and what the relationship is between \texttt{BQP} and \texttt{NP}.
It is generally assumed that \texttt{NP}-hard problems are not within \texttt{BQP}~\cite{Nielsen_Chuang_2010}.
This implies that for instance the fermionic sign problem~\cite{Troyer:2004ge} is likely not solvable on quantum computer with polynomial resources either.
It might well be though that the hardness of the problem will be less obvious.
Rather than having non-positive sampling weight, the simulation might suffer from prohibitively expensive state preparation, noisy measurements resulting in the need of exponentially many repetitions (shots), an exponential grows of the circuit depth, or some other obstacle.

\subsubsection{No fast-forward theorem}

For time evolution such limitations become very explicit in the so-called ``no fast-forward theorem''~\cite{Berry:2005yrf,Atia:2016sax}.
It states\footnote{The no fast-forward theorem relies on the very plausible assumption that \texttt{BQP} is strictly contained in \texttt{PSPACE} (recall that both $\texttt{BQP}\subset\texttt{PSPACE}$ and $\texttt{NP}\subset\texttt{PSPACE}$).}
that the real time evolution of a generic Hamiltonian requires a quantum circuit depth at least proportional to the evolution time $t$.
To resolve exponentially small features in the spectrum one would need an exponentially long evolution time and, in consequence, exponentially long circuits.
This means that exponential runtime cannot be avoided in general, even on quantum computers, because the spectra of generic Hamiltonians become exponentially dense in system size.

\subsubsection{Tackling physical systems}

On the other hand, some more specific problems in physics might be more amenable to quantum simulations.
In quantum field theory (QFT) quantum algorithms (relying on time evolution) have been derived~\cite{Jordan:2012xnu} that achieve exponential speed up over their known classical counterparts.
Even without explicitly aiming for quantum advantage, it is still a non-trivial task to saturate the bounds dictated by the no-fast-forward theorem.
The QSP method (sec.~\ref{sec:qsp}) is one of the general ansätze promising optimal asymptotic runtime.

For some specific physical problems specialised algorithms have been developed.
For instance sparse, non-local fermionic models can be simulated on quantum hardware with asymptotically linear runtime using Trotterization~\cite{Irmejs:2025vuh} and the circuit depth per Trotter step for the simulation of electrons with atomic orbitals has recently been decreased to sub-linear in the number of orbitals~\cite{Gunther:2026pae}.

Mayhap surprisingly, Trotter schemes can not only be used to evolve a state but also to stabilise it.
If a state is measured often enough, then it is effectively frozen and can only change with a probability decreasing with the inverse number of measurements $\ordnung{N_\text{meas}^{-1}}$.
This phenomenon and some generalisation thereof are called Zeno effect~\cite{Misra:1976by,Facchi:2002jgu}.
It has now been shown in Ref.~\cite{Dizaji:2025xyu} that any order $n$ Trotterization can be used to define an improved projective measurement sequence with a Zeno error scaling as $\ordnung{N_\text{meas}^{-n}}$.

\subsubsection{Ground state search}

Another important application of time evolution is the ground state search.
On a quantum computer the probably most natural way to approach a ground state is via adiabatic quantum computing, smoothly evolving a simple ground state into the desired one.
By the adiabatic theorem the evolution speed and correspondingly computational cost are limited by the minimal gap to the first excited state.
In Ref.~\cite{An:2025yud} the discrete version of the same theorem has been used to show that Trotter time steps of $\ordnung{1}$ can be chosen independently of the target error for bounded Hamiltonians.
The convergence is then exponential in the evolution time.

On classical computers, ground states are often determined through imaginary time evolution.
While not natural, these updates can also be realised on a quantum device by embedding (block encoding) the non-unitary operator into a larger unitary one~\cite{PRXQuantum.2.010342}.
It is observed in Ref.~\cite{Gluza:2024lqq} that imaginary time evolution can be written as
\begin{align}
	\frac{\md}{\md \tau}\ket{\psi(\tau)} &= \brr{\ket{\psi(\tau)}\bra{\psi(\tau)},H} \ket{\psi(\tau)}\,.\label{eq:double_bracket}
\end{align}
The commutator of the density matrix and the Hamiltonian is anti-hermitian by construction, so equation~\eqref{eq:double_bracket} defines a unitary time evolution.
This gives rise to efficient ground state search algorithms~\cite{Gluza:2024lqq,Benavides-Riveros:2026hyf} via imaginary time evolution that can potentially be realised on quantum computers.

As a side note, we remark that there is yet another well-known quantum minimisation algorithm containing the term Trotter.
However, the Adaptive Derivative-Assembled Pseudo Trotter Variational Quantum Eigensolver (ADAPT-VQE)~\cite{Grimsley:2018wnd} does not use time evolution.
Although initially motivated by Trotterization, it is a purely variational method.

\subsubsection*{}
For more details on the theoretical and current potential of quantum computers for physics applications we refer to Refs.~\cite{Babbush:2025eqc,Zimboras:2025unr}.
Ref.~\cite{Babbush:2025eqc} lists multiple algorithms and physical systems on the road to quantum advantage which is also sketched.
A more approachable overview is provided in Ref.~\cite{Zimboras:2025unr} featuring a collection of myths surrounding quantum computing, mostly focused on the NISQ era.

\subsection{The NISQ era}\label{sec:nisq}

The year 2026 is clearly part of the \textit{noisy intermediate-scale quantum} (NISQ) era~\cite{Yin2026,Doyle:2026xcd}.
This means that the main problem in quantum computations is hardware noise.
Since noise accumulates with the number of gates, the reliable (high-fidelity) circuit depth is very limited.
As long as deep quantum circuits cannot be realised due to noise, there is little hope for useful quantum simulations of physical systems.
However, the noise level does not have to drop to zero to permit quantum advantage.
Noise-resilient quantum algorithms are steadily being improved in order to reduce hardware requirements and utilise near-term quantum computers to their full potential.
See Ref.~\cite{Endo:2020kro} for a review on error mitigation techniques and quantum-classical hybrid algorithms in the NISQ era.

Specifically for time evolution, it has been found that Trotterized simulations on NISQ devices are often more robust against noise than one might expect.
In Ref.~\cite{chertkov2024robustnessnearthermaldynamicsdigital} this has been demonstrated for systems near thermal equilibrium.
A similar effect has been observed for adiabatic evolution where the total 1st order Trotter error is identified to scale with $\ordnung{t^{-1}h}=\ordnung{N_t^{-1}}$ instead of $\ordnung{th}=\ordnung{t^2 N_t^{-1}}$ (Ref.~\cite{PhysRevLett.131.060602} quotes the squares of these results because they refer to the infidelity which is a quadratic quantity).
That is, the duration $t$ of the auxiliary time introduced to interpolate between the initial and target Hamiltonians is irrelevant for the final state fidelity.
This so-called self-healing has even been found to persist in some non-adiabatic cases~\cite{Vizzuso:2025isl}.

\subsubsection{Circuit depth reduction}

Of course, there are limits to the robustness acclaimed above and the best way to avoid irrecoverable hardware noise is to reduce the circuit depth in the first place.
All the methods presented throughout this review have the goal to reduce the computational cost in one way or another.
Hence, the starting point on the quest for the shortest circuit should be the practical guidelines in \cref{sec:practice}.
Noisy quantum computing does not introduce an entirely new hierarchy of algorithm efficiencies (e.g.\ contrary to some claims, 1st order Trotter should still never be used).
Quite the opposite, most of the hierarchies are strengthened.
What might be a negligible speed up on a classical computer might be the crucial reduction in circuit length that allows to extract a signal.
A notable exception is the use of multi-product formulae or linear combinations of unitaries (sec.~\ref{sec:multi-product}) as they allow to split a single long run into multiple short ones.

There are some additional techniques, though, that are unique to quantum computers and might reduce circuit depths beyond those obtained from classical methods.
For instance, according to Ref.~\cite{arseniev2024distributedquantumlogicalgorithm} the circuit depth can often be reduced considerably at the cost of introducing sufficiently many ancillary qubits.
Sometimes neighbouring gates can be combined in a more efficient circuit compilation.
To facilitate such synergies, it might pay off to consider different orderings of the operators in the Hamiltonian~\cite{thanh2025hamiltonianreorderingshallowertrotterization}.
This introduces an additional complexity on top of the Trotter error dependence on the ordering (sec.~\ref{sec:best_operators}).
An even more extreme ansatz, the so-called partial Trotterization~\cite{decker2025kernpilercompileroptimizationquantum}, is to group several operators together into a single one that is exponentiated completely.

\subsubsection{Error mitigation}

Having decided on the best possible noisy quantum simulation, one can apply error mitigation techniques in order to enhance the signal~\cite{Vovrosh:2021ocf,Urbanek:2021oej}.
While full error correction certainly remains the long-term goal on the way to fault-tolerant quantum computing, error mitigation will likely remain useful in the NISQ era~\cite{aharonov2025importanceerrormitigationquantum}.
Rahman et al.~\cite{Rahman:2022tkr} have introduced a simple yet powerful error mitigation technique for time evolution with symmetric Trotter schemes.
In addition to the main simulation for the physical measurement, the same evolution is repeated, first going forward for half the time then backward.
Without noise this second mitigation run should return to exactly the initial state.
In the presence of noise, however, part of the signal is lost as an effect of decoherence.
Since the circuits for the physical and the mitigation evolutions are identical up to (half) the time direction, the signal strength is expected to deteriorate equally in both runs.
The physical signal can now be restored to good approximation by rescaling with the decoherence factor obtained from the mitigation run.

\subsubsection{Quantum-classical hybrid methods}

So far we have only mentioned classical post-processing methods to reduce quantum noise.
The last few years have seen an alternative paradigm emerging that tries to prevent the hardware noise in the first place by performing only the most complicated tasks on the quantum device.
As much computational effort as possible is performed by a classical computer instead.
Such quantum-classical hybrids can for instance incorporate tensor networks (TNs) (sec.~\ref{sec:tdvp}) on the classical side.
Multiple algorithms developed for TNs can be adapted for quantum computing~\cite{Barratt_2021}.
Moreover, in hybrid TN~\cite{Yuan:2020xmq,Schuhmacher:2024spy} simulations the bulk of a highly entangled state can be time evolved using a quantum computer while boundaries are encoded with classical TNs~\cite{Bauer:2026niu}.
Yet another method has been presented in Ref.~\cite{gentinetta2025correctingextendingtrotterizedquantum} where a quantum computer evolves a simplified Hamiltonian using Trotterization while the classical computer corrects for the mistakes.

\subsubsection*{}
Only time will tell which if any of the algorithms developed specifically for time evolution on NISQ era quantum devices will prove useful in the long run.
It is clear, however, that the current unreliable state of quantum hardware has spurred the development of efficient time evolution methods in general.

	\section{Conclusion}\label{sec:conclusion}

In this review on quantum time evolution methods we pursue two main directions.
The introduction to Suzuki-Trotter decompositions in \cref{sec:intro} and practical guidelines for their usage in \cref{sec:practice} as well as the collections of generalisations (sec.~\ref{sec:generalisations}) and alternatives (sec.~\ref{sec:alternatives}) to Trotterization can be understood as a handbook.
Conversely, \cref{sec:construction,sec:errors,sec:quantum_comp} summarise the recent development on the construction of efficient splitting methods, their error estimation and the relevance of time evolution in the context of quantum computing.
A historical approach is avoided deliberately in favour of a strong focus on the current state of the art.

If you are primarily interested in the hands-on implementation of an efficient method for the time evolution of a quantum system, \cref{sec:practice} is just what you have been looking for.
The currently best Trotter schemes of orders $n=4,6$ by Maležič et al.~\cite{Malezic:2026bds} (tabs.~\ref{tab:recommened6},\ref{tab:recommened14}) and $n=8,10$ by Morales et al.~\cite{morales2022greatly} are highlighted.
Besides augmenting these methods with a number of tips and tricks in \cref{sec:tips_and_tricks}, we list several generalisations and alternatives to consider instead of Trotterization.
The detailed explanations of the corresponding algorithms are postponed to \cref{sec:generalisations,sec:alternatives}, respectively, and can be looked up as needed.

The potentially most important generalisation of Suzuki-Trotter decompositions is their extension to time-dependent Hamiltonians condensed into \cref{th:time-dependent} within \cref{sec:time-dependent}.

Specialised approaches are called for when dealing with either of the following:
classical equations of motion (EOM) should be solved with symplectic integrators (sec.~\ref{sec:symplectic}), a specialised subset of splitting methods;
tensor networks might be evolved in time more efficiently by means of the time-dependent variational principle (TDVP, sec.~\ref{sec:tdvp});
quantum computers achieve optimal theoretical complexity scaling using quantum signal processing (QSP, sec.~\ref{sec:qsp}), while their circuit depth can be reduced with randomised methods (sec.~\ref{sec:random}) at the cost of more shots.
Moreover, operator hierarchies, i.e.\ if one term in the Hamiltonian is much larger than another, can often be exploited as laid out in \cref{sec:hierarchy}.
Another common scenario is that strict unitarity of the time evolution operator is not required.
In such a case polynomial expansion (sec.~\ref{sec:polynomials}) or multi-product formulae (sec.~\ref{sec:multi-product}) lead to greatly superior approximations.

Over the last few years the reliable estimation of maximally tight error bounds has been of particular interest to the community.
After reiterating the construction of higher order Trotter schemes in \cref{sec:construction} and introducing the framework required for their in-depth understanding (sec.~\ref{sec:omelyan}), we report on the progress regarding error estimation in \cref{sec:errors}.
The most generally applicable upper Trotter error bounds are quoted in \cref{th:Schubert_error_bounds} (sec.~\ref{sec:general_erros}) while \cref{th:symmetric_error_bounds} (sec.~\ref{sec:symmetric_errors}) provides relatively simple bounds for symmetric schemes.

Last but not least, quantum computing has been a central motivating factor behind the development of improved time evolution methods as well as their error estimation.
The scope of quantum computing for time evolution is discussed in \cref{sec:qc_limitations}.
While they are a promising tool in the long run, there are still many challenges to overcome in the current NISQ era (sec.~\ref{sec:nisq}).

\subsection{Open problems}

The research of quantum time evolution methods in general and Suzuki-Trotter decompositions in particular is anything but concluded.
Hopefully, this review has conveyed a feeling for the ongoing research activity and the areas of particular interest.
In addition, a certainly incomplete list of prominent challenges and open questions is outlined below.

The search for better Trotter schemes, especially in higher orders, continues.
At the same time, the understanding of what even makes a good scheme still has significant holes.
How can the performance in practice be anticipated beyond the theoretical efficiency introduced in \cref{sec:omelyan}?
Is there a way to predict the error accumulation over time (sec.~\ref{sec:long-time_errors}) beyond the observation that more uniform coefficients appear to be preferable?

Of course, these questions go hand in hand with the quest for ever tighter error bounds.
In particular, the derivation of lower error bounds applicable to order $n>1$ Trotterizations is still an open problem.

Given a Hamiltonian, it is often unclear how best to decompose it into a sum of operators $H=\sum_k A_k$.
Other than the rule of thumb that fewer operators are typically better (sec.~\ref{sec:best_operators}), we have little insight to guide us.

We have learned that in many cases there are specialised methods that outperform product formulae by a wide margin.
Are there any more cases that offer a similar potential?

Let me conclude this review with an appeal to the community.
Yes, this includes you.
It is my firm belief that we should make the vast font of knowledge we collectively posses more easily accessible.
To name an example, someone better qualified than me is called for to write a complete recipe-style summary of the QSP time evolution algorithm so that it can be implemented easily.
Also, why is there still no Wikipedia article on Suzuki-Trotter decompositions? 	
	\ack{
		Great thanks to Marko Maležič for the highly fruitful and enjoyable ongoing collaboration on Trotterization as well as for carefully proof-reading this review.
		Afterwards, there remained a single typo for Claude\footnote{AI-assisted proofreading tool \href{https://www.claude.ai}{claude.ai} by \href{https://www.anthropic.com/}{Anthropic}, model Sonnet~5, set up with the help of Aniket Sen.} to find.
		Christopher Kane also deserves special credit for educating me on the topic of QSP.
		Moreover, I would like to thank
		Christian Bauer, Evan Berkowitz, Luca Dellantonio, Bartosz Kostrzewa, Paul Ludwig, Tom Luu, Emanuele Mendicelli, Pranay Naredi, Enrico Rinaldi, Alessandro Roggero, Hersh Singh, and Jinzhao Sun
		for insightful discussions and literature recommendations.
		I am deeply grateful to the authors of Refs.~\cite{hahn2024lowerboundstrottererror,burgarth2023strong} and those of Ref.~\cite{schubert2023trotter} for kindly granting me permission to reuse their figures.
		This work was funded in part by the Deutsche Forschungsgemeinschaft (DFG, German Research Foundation) as part of the CRC 1639 NuMeriQS -- project no.\ 511713970.
	}

	\FloatBarrier
	\addcontentsline{toc}{section}{References}
	\printbibliography

\end{document}

%% file: plots/2-stage_fix-dt.tex
% GNUPLOT: LaTeX picture with Postscript
\begingroup
  % Encoding inside the plot.  In the header of your document, this encoding
  % should to defined, e.g., by using
  % \usepackage[latin1,<other encodings>]{inputenc}
  \inputencoding{latin1}%
  \makeatletter
  \providecommand\color[2][]{%
    \GenericError{(gnuplot) \space\space\space\@spaces}{%
      Package color not loaded in conjunction with
      terminal option `colourtext'%
    }{See the gnuplot documentation for explanation.%
    }{Either use 'blacktext' in gnuplot or load the package
      color.sty in LaTeX.}%
    \renewcommand\color[2][]{}%
  }%
  \providecommand\includegraphics[2][]{%
    \GenericError{(gnuplot) \space\space\space\@spaces}{%
      Package graphicx or graphics not loaded%
    }{See the gnuplot documentation for explanation.%
    }{The gnuplot epslatex terminal needs graphicx.sty or graphics.sty.}%
    \renewcommand\includegraphics[2][]{}%
  }%
  \providecommand\rotatebox[2]{#2}%
  \@ifundefined{ifGPcolor}{%
    \newif\ifGPcolor
    \GPcolortrue
  }{}%
  \@ifundefined{ifGPblacktext}{%
    \newif\ifGPblacktext
    \GPblacktexttrue
  }{}%
  % define a \g@addto@macro without @ in the name:
  \let\gplgaddtomacro\g@addto@macro
  % define empty templates for all commands taking text:
  \gdef\gplbacktext{}%
  \gdef\gplfronttext{}%
  \makeatother
  \ifGPblacktext
    % no textcolor at all
    \def\colorrgb#1{}%
    \def\colorgray#1{}%
  \else
    % gray or color?
    \ifGPcolor
      \def\colorrgb#1{\color[rgb]{#1}}%
      \def\colorgray#1{\color[gray]{#1}}%
      \expandafter\def\csname LTw\endcsname{\color{white}}%
      \expandafter\def\csname LTb\endcsname{\color{black}}%
      \expandafter\def\csname LTa\endcsname{\color{black}}%
      \expandafter\def\csname LT0\endcsname{\color[rgb]{1,0,0}}%
      \expandafter\def\csname LT1\endcsname{\color[rgb]{0,1,0}}%
      \expandafter\def\csname LT2\endcsname{\color[rgb]{0,0,1}}%
      \expandafter\def\csname LT3\endcsname{\color[rgb]{1,0,1}}%
      \expandafter\def\csname LT4\endcsname{\color[rgb]{0,1,1}}%
      \expandafter\def\csname LT5\endcsname{\color[rgb]{1,1,0}}%
      \expandafter\def\csname LT6\endcsname{\color[rgb]{0,0,0}}%
      \expandafter\def\csname LT7\endcsname{\color[rgb]{1,0.3,0}}%
      \expandafter\def\csname LT8\endcsname{\color[rgb]{0.5,0.5,0.5}}%
    \else
      % gray
      \def\colorrgb#1{\color{black}}%
      \def\colorgray#1{\color[gray]{#1}}%
      \expandafter\def\csname LTw\endcsname{\color{white}}%
      \expandafter\def\csname LTb\endcsname{\color{black}}%
      \expandafter\def\csname LTa\endcsname{\color{black}}%
      \expandafter\def\csname LT0\endcsname{\color{black}}%
      \expandafter\def\csname LT1\endcsname{\color{black}}%
      \expandafter\def\csname LT2\endcsname{\color{black}}%
      \expandafter\def\csname LT3\endcsname{\color{black}}%
      \expandafter\def\csname LT4\endcsname{\color{black}}%
      \expandafter\def\csname LT5\endcsname{\color{black}}%
      \expandafter\def\csname LT6\endcsname{\color{black}}%
      \expandafter\def\csname LT7\endcsname{\color{black}}%
      \expandafter\def\csname LT8\endcsname{\color{black}}%
    \fi
  \fi
    \setlength{\unitlength}{0.0500bp}%
    \ifx\gptboxheight\undefined%
      \newlength{\gptboxheight}%
      \newlength{\gptboxwidth}%
      \newsavebox{\gptboxtext}%
    \fi%
    \setlength{\fboxrule}{0.5pt}%
    \setlength{\fboxsep}{1pt}%
    \definecolor{tbcol}{rgb}{1,1,1}%
\begin{picture}(7200.00,5040.00)%
    \gplgaddtomacro\gplbacktext{%
      \csname LTb\endcsname%%
      \put(946,1382){\makebox(0,0)[r]{\strut{}$10^{-7}$}}%
      \csname LTb\endcsname%%
      \put(946,2352){\makebox(0,0)[r]{\strut{}$10^{-6}$}}%
      \csname LTb\endcsname%%
      \put(946,3322){\makebox(0,0)[r]{\strut{}$10^{-5}$}}%
      \csname LTb\endcsname%%
      \put(946,4291){\makebox(0,0)[r]{\strut{}$10^{-4}$}}%
      \csname LTb\endcsname%%
      \put(1078,484){\makebox(0,0){\strut{}$0$}}%
      \csname LTb\endcsname%%
      \put(2223,484){\makebox(0,0){\strut{}$2$}}%
      \csname LTb\endcsname%%
      \put(3368,484){\makebox(0,0){\strut{}$4$}}%
      \csname LTb\endcsname%%
      \put(4513,484){\makebox(0,0){\strut{}$6$}}%
      \csname LTb\endcsname%%
      \put(5658,484){\makebox(0,0){\strut{}$8$}}%
      \csname LTb\endcsname%%
      \put(6803,484){\makebox(0,0){\strut{}$10$}}%
    }%
    \gplgaddtomacro\gplfronttext{%
      \csname LTb\endcsname%%
      \put(5816,3091){\makebox(0,0)[r]{\strut{}Strang~\eqref{eq:leapfrog}}}%
      \csname LTb\endcsname%%
      \put(5816,2871){\makebox(0,0)[r]{\strut{}Suzuki~\cite{suzuki_original}}}%
      \csname LTb\endcsname%%
      \put(5816,2651){\makebox(0,0)[r]{\strut{}Unif.~Non-Unitary~\cite{Ostmeyer:2022}}}%
      \csname LTb\endcsname%%
      \put(5816,2431){\makebox(0,0)[r]{\strut{}Blanes~\&~Moan~\cite{BLANES2002313}}}%
      \csname LTb\endcsname%%
      \put(209,2761){\rotatebox{-270.00}{\makebox(0,0){\strut{}$\mathrm{error}/t$}}}%
      \put(3940,154){\makebox(0,0){\strut{}$t$}}%
    }%
    \gplbacktext
    \put(0,0){\includegraphics[width={360.00bp},height={252.00bp}]{2-stage_fix-dt}}%
    \gplfronttext
  \end{picture}%
\endgroup

%% file: plots/3-stage_fix-t_ordRest.tex
% GNUPLOT: LaTeX picture with Postscript
\begingroup
  % Encoding inside the plot.  In the header of your document, this encoding
  % should to defined, e.g., by using
  % \usepackage[latin1,<other encodings>]{inputenc}
  \inputencoding{latin1}%
  \makeatletter
  \providecommand\color[2][]{%
    \GenericError{(gnuplot) \space\space\space\@spaces}{%
      Package color not loaded in conjunction with
      terminal option `colourtext'%
    }{See the gnuplot documentation for explanation.%
    }{Either use 'blacktext' in gnuplot or load the package
      color.sty in LaTeX.}%
    \renewcommand\color[2][]{}%
  }%
  \providecommand\includegraphics[2][]{%
    \GenericError{(gnuplot) \space\space\space\@spaces}{%
      Package graphicx or graphics not loaded%
    }{See the gnuplot documentation for explanation.%
    }{The gnuplot epslatex terminal needs graphicx.sty or graphics.sty.}%
    \renewcommand\includegraphics[2][]{}%
  }%
  \providecommand\rotatebox[2]{#2}%
  \@ifundefined{ifGPcolor}{%
    \newif\ifGPcolor
    \GPcolortrue
  }{}%
  \@ifundefined{ifGPblacktext}{%
    \newif\ifGPblacktext
    \GPblacktexttrue
  }{}%
  % define a \g@addto@macro without @ in the name:
  \let\gplgaddtomacro\g@addto@macro
  % define empty templates for all commands taking text:
  \gdef\gplbacktext{}%
  \gdef\gplfronttext{}%
  \makeatother
  \ifGPblacktext
    % no textcolor at all
    \def\colorrgb#1{}%
    \def\colorgray#1{}%
  \else
    % gray or color?
    \ifGPcolor
      \def\colorrgb#1{\color[rgb]{#1}}%
      \def\colorgray#1{\color[gray]{#1}}%
      \expandafter\def\csname LTw\endcsname{\color{white}}%
      \expandafter\def\csname LTb\endcsname{\color{black}}%
      \expandafter\def\csname LTa\endcsname{\color{black}}%
      \expandafter\def\csname LT0\endcsname{\color[rgb]{1,0,0}}%
      \expandafter\def\csname LT1\endcsname{\color[rgb]{0,1,0}}%
      \expandafter\def\csname LT2\endcsname{\color[rgb]{0,0,1}}%
      \expandafter\def\csname LT3\endcsname{\color[rgb]{1,0,1}}%
      \expandafter\def\csname LT4\endcsname{\color[rgb]{0,1,1}}%
      \expandafter\def\csname LT5\endcsname{\color[rgb]{1,1,0}}%
      \expandafter\def\csname LT6\endcsname{\color[rgb]{0,0,0}}%
      \expandafter\def\csname LT7\endcsname{\color[rgb]{1,0.3,0}}%
      \expandafter\def\csname LT8\endcsname{\color[rgb]{0.5,0.5,0.5}}%
    \else
      % gray
      \def\colorrgb#1{\color{black}}%
      \def\colorgray#1{\color[gray]{#1}}%
      \expandafter\def\csname LTw\endcsname{\color{white}}%
      \expandafter\def\csname LTb\endcsname{\color{black}}%
      \expandafter\def\csname LTa\endcsname{\color{black}}%
      \expandafter\def\csname LT0\endcsname{\color{black}}%
      \expandafter\def\csname LT1\endcsname{\color{black}}%
      \expandafter\def\csname LT2\endcsname{\color{black}}%
      \expandafter\def\csname LT3\endcsname{\color{black}}%
      \expandafter\def\csname LT4\endcsname{\color{black}}%
      \expandafter\def\csname LT5\endcsname{\color{black}}%
      \expandafter\def\csname LT6\endcsname{\color{black}}%
      \expandafter\def\csname LT7\endcsname{\color{black}}%
      \expandafter\def\csname LT8\endcsname{\color{black}}%
    \fi
  \fi
    \setlength{\unitlength}{0.0500bp}%
    \ifx\gptboxheight\undefined%
      \newlength{\gptboxheight}%
      \newlength{\gptboxwidth}%
      \newsavebox{\gptboxtext}%
    \fi%
    \setlength{\fboxrule}{0.5pt}%
    \setlength{\fboxsep}{1pt}%
    \definecolor{tbcol}{rgb}{1,1,1}%
\begin{picture}(7200.00,5040.00)%
    \gplgaddtomacro\gplbacktext{%
      \csname LTb\endcsname%%
      \put(1078,1009){\makebox(0,0)[r]{\strut{}$10^{-12}$}}%
      \csname LTb\endcsname%%
      \put(1078,1620){\makebox(0,0)[r]{\strut{}$10^{-10}$}}%
      \csname LTb\endcsname%%
      \put(1078,2231){\makebox(0,0)[r]{\strut{}$10^{-8}$}}%
      \csname LTb\endcsname%%
      \put(1078,2841){\makebox(0,0)[r]{\strut{}$10^{-6}$}}%
      \csname LTb\endcsname%%
      \put(1078,3452){\makebox(0,0)[r]{\strut{}$10^{-4}$}}%
      \csname LTb\endcsname%%
      \put(1078,4063){\makebox(0,0)[r]{\strut{}$10^{-2}$}}%
      \csname LTb\endcsname%%
      \put(1078,4673){\makebox(0,0)[r]{\strut{}$10^{0}$}}%
      \csname LTb\endcsname%%
      \put(1210,484){\makebox(0,0){\strut{}$10$}}%
      \csname LTb\endcsname%%
      \put(5509,484){\makebox(0,0){\strut{}$100$}}%
    }%
    \gplgaddtomacro\gplfronttext{%
      \csname LTb\endcsname%%
      \put(5816,2279){\makebox(0,0)[r]{\strut{}Strang~\eqref{eq:leapfrog}}}%
      \csname LTb\endcsname%%
      \put(5816,2004){\makebox(0,0)[r]{\strut{}Taylor~$(n=304,\,\sum)$}}%
      \csname LTb\endcsname%%
      \put(5816,1729){\makebox(0,0)[r]{\strut{}Taylor~$(n=304,\,\prod)$}}%
      \csname LTb\endcsname%%
      \put(5816,1454){\makebox(0,0)[r]{\strut{}Chebyshev~$(n=152)$}}%
      \csname LTb\endcsname%%
      \put(209,2761){\rotatebox{-270.00}{\makebox(0,0){\strut{}error}}}%
      \put(4006,154){\makebox(0,0){\strut{}$n/h$}}%
    }%
    \gplbacktext
    \put(0,0){\includegraphics[width={360.00bp},height={252.00bp}]{3-stage_fix-t_ordRest}}%
    \gplfronttext
  \end{picture}%
\endgroup

%% file: bibliography.bib
@article{Malezic:2026bds,
	author = "Male{\v{z}}i{\v{c}}, Marko and Ostmeyer, Johann",
	title = "{Efficient Trotter-Suzuki Schemes for Long-time Quantum Dynamics}",
	eprint = "2601.18756",
	archivePrefix = "arXiv",
	primaryClass = "quant-ph",
	doi = "10.1088/1751-8121/ae919f",
	journal = "J. Phys. A",
	volume = "59",
	number = "34",
	pages = "345301",
	year = "2026",
	addendumx = {Staff Pick by \href{https://community.wolfram.com/groups/-/m/t/3665192}{Wolfram Community}}
}

@inproceedings{Malezic:2026xyi,
	author = "Male{\v{z}}i{\v{c}}, Marko and Ostmeyer, Johann",
	title = "{Reducing the Gate Count with Efficient Trotter-Suzuki Schemes}",
	booktitle = "{42th International Symposium on Lattice Field Theory}",
	eprint = "2602.21145",
	archivePrefix = "arXiv",
	primaryClass = "hep-lat",
	month = "2",
	doi = "10.22323/1.518.0131",
	journal = "PoS",
	volumex = "LATTICE2025",
	year = "2026"
}

@article{markomalezic_2026_18347430,
	author = {Marko Maležič},
	title = "{MarkoMalezic/efficient-trotterizations: Efficient Trotterizations}",
	month = jan,
	year = 2026,
	publisher = {Zenodo},
	version = {v1.0},
	doi = {10.5281/zenodo.18347430},
	addendum = {\url{https://github.com/MarkoMalezic/efficient-trotterizations}}
}

@article{my_tensor_networks,
	title="{Simulating both parity sectors of the Hubbard Model with Tensor Networks}", 
	author={Manuel Schneider and Johann Ostmeyer and Karl Jansen and Thomas Luu and Carsten Urbach},
	eprint = "2106.13583",
	archivePrefix = "arXiv",
	primaryClass = "physics.comp-ph",
	journal = {Phys. Rev. B},
	volume = {104},
	issue = {15},
	pages = {155118},
	numpages = {18},
	year = {2021},
	month = {Oct},
	publisher = {American Physical Society},
	doi = {10.1103/PhysRevB.104.155118},
	urlx = {https://link.aps.org/doi/10.1103/PhysRevB.104.155118}
}

@article{Ostmeyer:2022,
	author = "Ostmeyer, Johann",
	title = "{Optimised Trotter decompositions for classical and quantum computing}",
	eprint = "2211.02691",
	archivePrefix = "arXiv",
	primaryClass = "quant-ph",
	doi = "10.1088/1751-8121/acde7a",
	journal = "J. Phys. A",
	volume = "56",
	number = "28",
	pages = "285303",
	year = "2023",
	publisher = {{IOP} Publishing}
}

@inproceedings{Ostmeyer:2023xju,
	author = "Ostmeyer, Johann",
	title = "{Simple Ways to improve Discrete Time Evolution}",
	booktitle = "{40th International Symposium on Lattice Field Theory {\textemdash} PoS(LATTICE2023)}",
	eprint = "2309.03389",
	archivePrefix = "arXiv",
	primaryClass = "quant-ph",
	doi = "10.22323/1.453.0025",
	journal = "PoS",
	volumex = "LATTICE2023",
	pages = "025",
	year = "2024"
}

@article{Krieg:2018pqh,
	author = "Krieg, Stefan and Luu, Thomas and Ostmeyer, Johann and Papaphilippou, Philippos and Urbach, Carsten",
	title = "{Accelerating Hybrid Monte Carlo simulations of the Hubbard model on the hexagonal lattice}",
	eprint = "1804.07195",
	archivePrefix = "arXiv",
	primaryClass = "cond-mat.str-el",
	doi = "10.1016/j.cpc.2018.10.008",
	journal = "Comput. Phys. Commun.",
	volume = "236",
	pages = "15--25",
	year = "2019"
}

@book{NumericalRecipes:2007,
	author = {Press, William H. and Teukolsky, Saul A. and Vetterling, William T. and Flannery, Brian P.},
	title = "{Numerical Recipes 3rd Edition: The Art of Scientific Computing}",
	year = {2007},
	isbn = {0521880688},
	publisher = {Cambridge University Press},
	address = {USA},
	edition = {3},
	note = {\url{https://numerical.recipes/}},
	addendum = {see esp.\ chap.\ 17}
}

@article{trotter_original,
	ISSN = {00029939, 10886826},
	URL = {http://www.jstor.org/stable/2033649},
	author = {H. F. Trotter},
	journal = {Proceedings of the American Mathematical Society},
	number = {4},
	pages = {545--551},
	publisher = {American Mathematical Society},
	title = "{On the Product of Semi-Groups of Operators}",
	urldate = {2022-09-07},
	volume = {10},
	year = {1959}
}

@article{Verlet:1967,
	title = "{Computer "Experiments" on Classical Fluids. I. Thermodynamical Properties of Lennard-Jones Molecules}",
	author = {Verlet, Loup},
	journal = {Phys. Rev.},
	volume = {159},
	issue = {1},
	pages = {98--103},
	numpages = {0},
	year = {1967},
	month = {Jul},
	publisher = {American Physical Society},
	doi = {10.1103/PhysRev.159.98},
	urlx = {https://link.aps.org/doi/10.1103/PhysRev.159.98}
}

@article{Strang:1968,
	author = {Strang, Gilbert},
	title = {On the Construction and Comparison of Difference Schemes},
	journal = {SIAM Journal on Numerical Analysis},
	volume = {5},
	number = {3},
	pages = {506-517},
	year = {1968},
	doi = {10.1137/0705041},	
	URL = {https://doi.org/10.1137/0705041},
	eprint = {https://doi.org/10.1137/0705041}
}

@article{Suzuki:1976be,
	author = "Suzuki, M.",
	title = "{Generalized Trotter's Formula and Systematic Approximants of Exponential Operators and Inner Derivations with Applications to Many Body Problems}",
	doi = "10.1007/BF01609348",
	journal = "Commun. Math. Phys.",
	volume = "51",
	pages = "183--190",
	year = "1976"
}

@article{suzuki_original,
	author = {Suzuki,Masuo },
	title = "{Decomposition formulas of exponential operators and Lie exponentials with some applications to quantum mechanics and statistical physics}",
	journal = {Journal of Mathematical Physics},
	volume = {26},
	number = {4},
	pages = {601-612},
	year = {1985},
	doi = {10.1063/1.526596},
	urlx = {
	https://doi.org/10.1063/1.526596
	},
}

@incollection{Hatano_2005,
	doi = {10.1007/11526216_2},	
	urlx = {https://doi.org/10.1007%2F11526216_2},	
	eprint = "math-ph/0506007",
	archivePrefix = "arXiv",
	year = 2005,
	month = {nov},	
	publisher = {Springer Berlin Heidelberg},	
	pages = {37--68},	
	author = {Naomichi Hatano and Masuo Suzuki},	
	title = "{Finding Exponential Product Formulas of Higher Orders}",	
	booktitle = {Quantum Annealing and Other Optimization Methods}
}

@article{Hall1934,
	author = {Hall, P.},
	title = "{A Contribution to the Theory of Groups of Prime-Power Order}",
	journal = {Proceedings of the London Mathematical Society},
	volume = {s2-36},
	number = {1},
	pages = {29-95},
	doi = {https://doi.org/10.1112/plms/s2-36.1.29},
	url = {https://londmathsoc.onlinelibrary.wiley.com/doi/abs/10.1112/plms/s2-36.1.29},
	eprint = {https://londmathsoc.onlinelibrary.wiley.com/doi/pdf/10.1112/plms/s2-36.1.29},
	year = {1934}
}

@article{Lyndon1958,
	ISSN = {0003486X, 19398980},
	URL = {http://www.jstor.org/stable/1970044},
	author = {K. T. Chen and R. H. Fox and R. C. Lyndon},
	journal = {Annals of Mathematics},
	number = {1},
	pages = {81--95},
	publisher = {Annals of Mathematics},
	title = "{Free Differential Calculus, IV. The Quotient Groups of the Lower Central Series}",
	urldate = {2026-01-10},
	volume = {68},
	year = {1958}
}

@article{Casas:2009,
	author = {Casas, Fernando and Murua, Ander},
	title = "{An efficient algorithm for computing the Baker–Campbell–Hausdorff series and some of its applications}",
	journal = {Journal of Mathematical Physics},
	volume = {50},
	number = {3},
	pages = {033513},
	year = {2009},
	month = {03},
	issn = {0022-2488},
	doi = {10.1063/1.3078418},
	url = {https://doi.org/10.1063/1.3078418},
	eprint = {https://pubs.aip.org/aip/jmp/article-pdf/doi/10.1063/1.3078418/16002991/033513_1_online.pdf},
}

@article{Barratt_2021,
	author = "Barratt, Fergus and Dborin, James and Bal, Matthias and Stojevic, Vid and Pollmann, Frank and Green, Andrew G.",
	title = "{Parallel quantum simulation of large systems on small NISQ computers}",
	eprint = "2003.12087",
	archivePrefix = "arXiv",
	primaryClass = "quant-ph",
	doi = "10.1038/s41534-021-00420-3",
	journal = "npj Quantum Inf.",
	volume = "7",
	pages = "79",
	year = "2021"
}

@article{Omelyan_2002,
	doi = {10.1016/s0010-4655(02)00451-4},	
	urlx = {https://doi.org/10.1016%2Fs0010-4655%2802%2900451-4},	
	year = 2002,
	month = {jul},	
	publisher = {Elsevier {BV}	},	
	volume = {146},	
	number = {2},	
	pages = {188--202},	
	author = {I.P. Omelyan and I.M. Mryglod and R. Folk},	
	title = "{Optimized Forest{\textendash}Ruth- and Suzuki-like algorithms for integration of motion in many-body systems}",	
	journal = {Computer Physics Communications}
}

@article{OMELYAN2003272,
	title = "{Symplectic analytically integrable decomposition algorithms: classification, derivation, and application to molecular dynamics, quantum and celestial mechanics simulations}",
	journal = {Computer Physics Communications},
	volume = {151},
	number = {3},
	pages = {272-314},
	year = {2003},
	issn = {0010-4655},
	doi = {https://doi.org/10.1016/S0010-4655(02)00754-3},
	urlx = {https://www.sciencedirect.com/science/article/pii/S0010465502007543},
	author = {I.P. Omelyan and I.M. Mryglod and R. Folk},
}

@article{Wiebe_2010,
	doi = {10.1088/1751-8113/43/6/065203},
	urlx = {https://doi.org/10.1088/1751-8113/43/6/065203},
	eprint = "0812.0562",
	archivePrefix = "arXiv",
	primaryClass = "math-ph",
	year = 2010,
	month = {jan},
	publisher = {{IOP} Publishing},
	volume = {43},
	number = {6},
	pages = {065203},
	author = {Nathan Wiebe and Dominic Berry and Peter H{\o}yer and Barry C Sanders},
	title = "{Higher order decompositions of ordered operator exponentials}",
	journal = {Journal of Physics A: Mathematical and Theoretical},
}

@article{Childs_2019,
	doi = {10.22331/q-2019-09-02-182},	
	urlx = {https://doi.org/10.22331%2Fq-2019-09-02-182},	
	eprint = "1805.08385",
	archivePrefix = "arXiv",
	primaryClass = "quant-ph",
	year = 2019,
	month = {sep},	
	publisher = {Verein zur Forderung des Open Access Publizierens in den Quantenwissenschaften},	
	volume = {3},	
	pages = {182},	
	author = {Andrew M. Childs and Aaron Ostrander and Yuan Su},	
	title = {Faster quantum simulation by randomization},	
	journal = {Quantum}
}

@article{YOSHIDA1990262,
	title = {Construction of higher order symplectic integrators},
	journal = {Physics Letters A},
	volume = {150},
	number = {5},
	pages = {262-268},
	year = {1990},
	issn = {0375-9601},
	doi = {https://doi.org/10.1016/0375-9601(90)90092-3},
	urlx = {https://www.sciencedirect.com/science/article/pii/0375960190900923},
	author = {Haruo Yoshida},
}

@article{BLANES2002313,
	title = "{Practical symplectic partitioned Runge–Kutta and Runge–Kutta–Nyström methods}",
	journal = {Journal of Computational and Applied Mathematics},
	volume = {142},
	number = {2},
	pages = {313-330},
	year = {2002},
	issn = {0377-0427},
	doi = {https://doi.org/10.1016/S0377-0427(01)00492-7},
	urlx = {https://www.sciencedirect.com/science/article/pii/S0377042701004927},
	author = {S. Blanes and P.C. Moan},
}

@article{Blanes_2019,
	doi = {10.1016/j.apnum.2019.07.022},	
	urlx = {https://doi.org/10.1016%2Fj.apnum.2019.07.022},	
	year = 2019,
	month = {dec},	
	publisher = {Elsevier {BV}},	
	volume = {146},	
	pages = {400--415},	
	author = {Sergio Blanes and Fernando Casas and Mechthild Thalhammer},	
	title = {Splitting and composition methods with embedded error estimators},	
	journal = {Applied Numerical Mathematics}
}

@article{Blanes:2022,
	title = {Applying splitting methods with complex coefficients to the numerical integration of unitary problems},
	journal = {Journal of Computational Dynamics},
	volume = {9},number = {2},pages = {85-101},
	year = {2022},
	issn = {2158-2491},
	doi = {10.3934/jcd.2021022},
	urlx = {/article/id/61d39ce42d80b75f0b6d022f},
	eprint = "2104.02412",
	archivePrefix = "arXiv",
	primaryClass = "math.NA",
	author = {Sergio Blanes and Fernando Casas and Alejandro Escorihuela-Tomàs}
}

@article{blanes2022symmetric,
	title="{On symmetric-conjugate composition methods in the numerical integration of differential equations}",
	author={Blanes, Sergio and Casas, Fernando and Chartier, Philippe and Escorihuela-Tom{\`a}s, Alejandro},
	journal={Mathematics of Computation},
	volume={91},
	number={336},
	pages={1739--1761},
	year={2022},
	archivePrefix = {arXiv},
	eprint = {2101.04100},
	primaryClass = {math.NA},
	url={https://www.ams.org/journals/mcom/2022-91-336/S0025-5718-2021-03715-1/viewer/}
}

@article{CASAS2022126700,
	title = {High order integrators obtained by linear combinations of symmetric-conjugate compositions},
	journal = {Applied Mathematics and Computation},
	volume = {414},
	pages = {126700},
	year = {2022},
	issn = {0096-3003},
	doi = {https://doi.org/10.1016/j.amc.2021.126700},
	urlx = {https://www.sciencedirect.com/science/article/pii/S0096300321007840},
	archivePrefix = {arXiv},
	eprint = {2106.06503},
	primaryClass = {math.NA},
	author = {F. Casas and A. Escorihuela-Tomàs},
}

@article{Auzinger:2017,
	author = {Auzinger, Winfried and Hofst\"{a}tter, Harald and Ketcheson, David and Koch, Othmar},
	title = {Practical Splitting Methods for the Adaptive Integration of Nonlinear Evolution Equations. Part I: Construction of Optimized Schemes and Pairs of Schemes},
	year = {2017},
	issue_date = {Mar 2017},
	publisher = {BIT Computer Science and Numerical Mathematics},
	address = {USA},
	volume = {57},
	number = {1},
	issn = {0006-3835},
	urlx = {https://doi.org/10.1007/s10543-016-0626-9},
	doi = {10.1007/s10543-016-0626-9},
	journal = {BIT},
	month = {mar},
	pages = {55–74},
	numpages = {20}
}

@article{Chambers_2003,
	doi = {10.1086/376844},
	urlx = {https://dx.doi.org/10.1086/376844},
	year = {2003},
	month = {aug},
	publisher = {},
	volume = {126},
	number = {2},
	pages = {1119},
	author = {J. E. Chambers},
	title = {Symplectic Integrators with Complex Time Steps},
	journal = {The Astronomical Journal},
}

@article{adaptive_trotter,
	author = "Zhao, Hongzheng and Bukov, Marin and Heyl, Markus and Moessner, Roderich",
	title = "{Making Trotterization Adaptive and Energy-Self-Correcting for NISQ Devices and Beyond}",
	eprint = "2209.12653",
	archivePrefix = "arXiv",
	primaryClass = "quant-ph",
	doi = "10.1103/PRXQuantum.4.030319",
	journal = "PRX Quantum",
	volume = "4",
	number = "3",
	pages = "030319",
	year = "2023"
}

@article{PhysRevX.11.011020,
	title = "{Theory of Trotter Error with Commutator Scaling}",
	author = {Childs, Andrew M. and Su, Yuan and Tran, Minh C. and Wiebe, Nathan and Zhu, Shuchen},
	eprint = "1912.08854",
	archivePrefix = "arXiv",
	primaryClass = "quant-ph",
	journal = {Phys. Rev. X},
	volume = {11},
	issue = {1},
	pages = {011020},
	numpages = {49},
	year = {2021},
	month = {Feb},
	publisher = {American Physical Society},
	doi = {10.1103/PhysRevX.11.011020},
	urlx = {https://link.aps.org/doi/10.1103/PhysRevX.11.011020}
}

@article{Heyl_2019,
	doi = {10.1126/sciadv.aau8342},	
	urlx = {https://doi.org/10.1126%2Fsciadv.aau8342},	
	year = 2019,
	month = {apr},	
	publisher = {American Association for the Advancement of Science ({AAAS})},	
	volume = {5},	
	number = {4},	
	author = {Markus Heyl and Philipp Hauke and Peter Zoller},	
	title = "{Quantum localization bounds Trotter errors in digital quantum simulation}",	
	journal = {Science Advances}
}

@article{optimised_circuits,
	title="{{Optimal compression of quantum many-body time evolution operators into brickwall circuits}}",
	author={Maurits S. J. Tepaske and  Dominik Hahn and David J. Luitz},
	eprint = "2205.03445",
	archivePrefix = "arXiv",
	primaryClass = "cond-mat.str-el",
	journal={SciPost Phys.},
	volume={14},
	pages={073},
	year={2023},
	publisher={SciPost},
	doi={10.21468/SciPostPhys.14.4.073},
	url={https://scipost.org/10.21468/SciPostPhys.14.4.073},
}

@article{classical_opt_circuits,
	author = "Keever, Conor Mc and Lubasch, Michael",
	title = "{Classically optimized Hamiltonian simulation}",
	eprint = "2205.11427",
	archivePrefix = "arXiv",
	primaryClass = "quant-ph",
	doi = "10.1103/PhysRevResearch.5.023146",
	journal = "Phys. Rev. Res.",
	volume = "5",
	number = "2",
	pages = "023146",
	year = "2023"
}

@article{Wolf:2025mjo,
	author = "Wolf, Stefan and Eckstein, Martin and Hartmann, Michael J.",
	title = "{Variational Time Evolution Compression for Solving Impurity Models on Quantum Hardware}",
	eprint = "2508.10526",
	archivePrefix = "arXiv",
	primaryClass = "quant-ph",
	month = "8",
	year = "2025"
}

@article{variational_circuits,
	doi = {10.1088/2058-9565/acb1d0},
	urlx = {https://dx.doi.org/10.1088/2058-9565/acb1d0},
	year = {2023},
	month = {jan},
	publisher = {IOP Publishing},
	volume = {8},
	number = {2},
	pages = {025006},
	author = {Refik Mansuroglu and Timo Eckstein and Ludwig Nützel and Samuel A Wilkinson and Michael J Hartmann},
	title = "{Variational Hamiltonian simulation for translational invariant systems via classical pre-processing}",
	journal = {Quantum Science and Technology},
	eprint = "2106.03680",
	archivePrefix = "arXiv",
	primaryClass = "quant-ph",
}

@book{lie1888theorie,
	title="{Theorie der Transformationsgruppen}",
	author={Lie, Sophus},
	volume={1},
	year={1888},
	publisher={\href{https://gdz.sub.uni-goettingen.de/id/PPN585383391}{BG Teubner}},
	url={https://gdz.sub.uni-goettingen.de/id/PPN585383391},
	doi={10.1007/978-3-662-46211-9},
	isbnx={978-3-662-46211-9},
	addendum={\href{https://link.springer.com/book/10.1007/978-3-662-46211-9}{English version, \textsc{isbn}: 978-3-662-46211-9}}
}

@article{silva2022fourierbased,
	title="{Fourier-based quantum signal processing}", 
	author={Thais de Lima Silva and Lucas Borges and Leandro Aolita},
	year={2022},
	eprint={2206.02826},
	archivePrefix={arXiv},
	primaryClass={quant-ph}
}

@article{PhysRevLett.114.090502,
	title = "{Simulating Hamiltonian Dynamics with a Truncated Taylor Series}",
	author = {Berry, Dominic W. and Childs, Andrew M. and Cleve, Richard and Kothari, Robin and Somma, Rolando D.},
	eprint = "1412.4687",
	archivePrefix = "arXiv",
	primaryClass = "quant-ph",
	journal = {Phys. Rev. Lett.},
	volume = {114},
	issue = {9},
	pages = {090502},
	numpages = {5},
	year = {2015},
	month = {Mar},
	publisher = {American Physical Society},
	doi = {10.1103/PhysRevLett.114.090502},
	urlx = {https://link.aps.org/doi/10.1103/PhysRevLett.114.090502}
}

@article{RevModPhys.78.275,
	title = {The kernel polynomial method},
	author = {Wei\ss{}e, Alexander and Wellein, Gerhard and Alvermann, Andreas and Fehske, Holger},
	journal = {Rev. Mod. Phys.},
	volume = {78},
	issue = {1},
	pages = {275--306},
	numpages = {0},
	year = {2006},
	month = {Mar},
	publisher = {American Physical Society},
	doi = {10.1103/RevModPhys.78.275},
	url = {https://link.aps.org/doi/10.1103/RevModPhys.78.275},
	archivePrefix = {arXiv},
	eprint = {cond-mat/0504627},
	primaryClass = {cond-mat.other},
}

@article{morales2022greatly,
	author = "Morales, Mauro E. S. and Costa, Pedro C. S. and Pantaleoni, Giacomo and Burgarth, Daniel K. and Sanders, Yuval R. and Berry, Dominic W.",
	title = "{Selection and Improvement of Product Formulae for Best Performance of Quantum Simulation}",
	eprint = "2210.15817",
	archivePrefix = "arXiv",
	primaryClass = "quant-ph",
	doi = "10.2478/qic-2025-0001",
	journal = "Quant. Inf. Comput.",
	volume = "25",
	number = "1",
	pages = "1--35",
	year = "2025"
}

@article{Ikeda:2022tlb,
	author = "Ikeda, Tatsuhiko N. and Abrar, Asir and Chuang, Isaac L. and Sugiura, Sho",
	title = "{Minimum Trotterization Formulas for a Time-Dependent Hamiltonian}",
	eprint = "2212.06788",
	archivePrefix = "arXiv",
	primaryClass = "quant-ph",
	doi = "10.22331/q-2023-11-06-1168",
	journal = "Quantum",
	volume = "7",
	pages = "1168",
	year = "2023"
}

@article{Arnal:2020xpt,
	author = "Arnal, Ana and Casas, Fernando and Chiralt, Cristina",
	title = "{A Note on the Baker{\textendash}Campbell{\textendash}Hausdorff Series in Terms of Right-Nested Commutators}",
	eprint = "2006.15869",
	archivePrefix = "arXiv",
	primaryClass = "math-ph",
	doi = "10.1007/s00009-020-01681-6",
	journal = "Mediterranean J. Math.",
	volume = "18",
	number = "2",
	pages = "53",
	year = "2021"
}

@article{Blanes_Casas_Murua_2024,
	title="{Splitting methods for differential equations}",
	eprint={2401.01722},
	archivePrefix={arXiv},
	primaryClass={math.NA},
	volume={33},
	DOI={10.1017/S0962492923000077},
	journal={Acta Numerica},
	author={Blanes, Sergio and Casas, Fernando and Murua, Ander},
	year={2024},
	pages={1–161}}

@article{McLachlan:1995otni,
	author = {McLachlan, Robert I.},
	title = "{On the Numerical Integration of Ordinary Differential Equations by Symmetric Composition Methods}",
	journal = {SIAM Journal on Scientific Computing},
	volume = {16},
	number = {1},
	pages = {151-168},
	year = {1995},
	doi = {10.1137/0916010},
	URL = {https://doi.org/10.1137/0916010},
	eprint = {https://doi.org/10.1137/0916010}
}

@article{Campbell:2020wqh,
	author = "Campbell, Earl T.",
	title = "{Early fault-tolerant simulations of the Hubbard model}",
	eprint = "2012.09238",
	archivePrefix = "arXiv",
	primaryClass = "quant-ph",
	doi = "10.1088/2058-9565/ac3110",
	journal = "Quantum Sci. Technol.",
	volume = "7",
	number = "1",
	pages = "015007",
	year = "2022"
}

@ARTICLE{Childs:2018PNAS,
	author = {{Childs}, Andrew M. and {Maslov}, Dmitri and {Nam}, Yunseong and {Ross}, Neil J. and {Su}, Yuan},
	title = "{Toward the first quantum simulation with quantum speedup}",
	journal = {Proceedings of the National Academy of Science},
	year = 2018,
	month = sep,
	volume = {115},
	number = {38},
	pages = {9456-9461},
	doi = {10.1073/pnas.1801723115},
	archivePrefix = {arXiv},
	eprint = {1711.10980},
	primaryClass = {quant-ph},
}

@ARTICLE{Childs:2019PhRv,
	author = {{Childs}, Andrew M. and {Su}, Yuan},
	title = "{Nearly Optimal Lattice Simulation by Product Formulas}",
	journal = {Phys. Rev. Lett.},
	year = 2019,
	month = aug,
	volume = {123},
	number = {5},
	eid = {050503},
	pages = {050503},
	doi = {10.1103/PhysRevLett.123.050503},
	archivePrefix = {arXiv},
	eprint = {1901.00564},
	primaryClass = {quant-ph},
}

@book {Hairer:2006gni,
	AUTHOR = {Hairer, Ernst and Lubich, Christian and Wanner, Gerhard},
	TITLE = {Geometric numerical integration},
	SERIES = {Springer Series in Computational Mathematics},
	VOLUME = {31},
	EDITION = {Second},
	NOTE = {Structure-preserving algorithms for ordinary differential
	equations},
	PUBLISHER = {Springer-Verlag, Berlin},
	YEAR = {2006},
	PAGES = {xviii+644},
	ISBN = {\href{https://link.springer.com/book/10.1007/3-540-30666-8}{3-540-30663-3; 978-3-540-30663-4}},
	url = {https://link.springer.com/book/10.1007/3-540-30666-8},
	doi = {10.1007/3-540-30666-8}
}

@article{Zylberman:2025amu,
	author = "Zylberman, Julien and Fredon, Thibault and Loureiro, Nuno F. and Debbasch, Fabrice",
	title = "{Trotter-based quantum algorithm for solving transport equations with exponentially fewer time-steps}",
	eprint = "2508.15691",
	archivePrefix = "arXiv",
	primaryClass = "quant-ph",
	month = "8",
	doi = "10.1088/2058-9565/ae488e",
	journal = "Quantum Sci. Technol.",
	volume = "11",
	number = "2",
	pages = "025015",
	year = "2026"
}

@article{An:2025yud,
	author = "An, Dong and Costa, Pedro C. S. and Berry, Dominic W.",
	title = "{Large time-step discretisation of adiabatic quantum dynamics}",
	eprint = "2509.00171",
	archivePrefix = "arXiv",
	primaryClass = "quant-ph",
	month = "8",
	year = "2025"
}

@article{VARGA2010298,
	title = "{Zeros of the partial sums of cos(z) and sin(z). III}",
	journal = {Applied Numerical Mathematics},
	volume = {60},
	number = {4},
	pages = {298-313},
	year = {2010},
	note = {Special Issue: NUMAN 2008},
	issn = {0168-9274},
	doi = {https://doi.org/10.1016/j.apnum.2009.08.007},
	url = {https://www.sciencedirect.com/science/article/pii/S016892740900155X},
	author = {Richard S. Varga and Amos J. Carpenter},
}

@InProceedings{Butcher:1969,
	author="Butcher, J. C.",
	editor="Morris, J. Li.",
	title="{The effective order of Runge-Kutta methods}",
	booktitle="{Conference on the Numerical Solution of Differential Equations}",
	year="1969",
	publisher="Springer Berlin Heidelberg",
	address="Berlin, Heidelberg",
	pages="133--139",
	isbn="978-3-540-36158-9",
	doi={10.1007/BFb0060019}
}

@article{Blanes:2006cmfd,
	author = {Blanes, S. and Casas, F. and Murua, A.},
	title = {Composition Methods for Differential Equations with Processing},
	journal = {SIAM Journal on Scientific Computing},
	volume = {27},
	number = {6},
	pages = {1817-1843},
	year = {2006},
	doi = {10.1137/030601223},
	URL = {https://doi.org/10.1137/030601223},
	eprint = {https://doi.org/10.1137/030601223},
}

@article{Blanes:2024foel,
	title = {Families of efficient low order processed composition methods},
	journal = {Applied Numerical Mathematics},
	volume = {204},
	pages = {86-100},
	year = {2024},
	issn = {0168-9274},
	doi = {https://doi.org/10.1016/j.apnum.2024.06.002},
	url = {https://www.sciencedirect.com/science/article/pii/S0168927424001429},
	archivePrefix = {arXiv},
	eprint = {2404.04340},
	primaryClass = {math.NA},
	author = {S. Blanes and F. Casas and A. Escorihuela-Tomàs},
}

@ARTICLE{Blanes:1999CeMDA,
	author = {{Blanes}, S. and {Casas}, F. and {Ros}, J.},
	title = "{Extrapolation of Symplectic Integrators}",
	journal = {Celestial Mechanics and Dynamical Astronomy},
	year = 1999,
	month = oct,
	volume = {75},
	number = {2},
	pages = {149-161},
	doi = {10.1023/A:1008364504014},
}

@article{Chin:2008,
	author = {Chin, Siu},
	year = {2008},
	month = {09},
	pages = {391-406},
	title = "{Multi-product splitting and Runge-Kutta-Nyström integrators}",
	volume = {106},
	journal = {Celestial Mechanics and Dynamical Astronomy},
	doi = {10.1007/s10569-010-9255-9},
	archivePrefix = {arXiv},
	eprint = {0809.0914},
	primaryClass = {cs.NA},
}

@article{Childs:2012gwh,
	author = "Childs, Andrew M. and Wiebe, Nathan",
	title = "{Hamiltonian Simulation Using Linear Combinations of Unitary Operations}",
	eprint = "1202.5822",
	archivePrefix = "arXiv",
	primaryClass = "quant-ph",
	doi = "10.26421/QIC12.11-12-1",
	journal = "Quant. Inf. Comput.",
	volume = "12",
	number = "11&12",
	pages = "0901--0924",
	year = "2012"
}

@article{low2019wellconditionedmultiproducthamiltoniansimulation,
	title="{Well-conditioned multiproduct Hamiltonian simulation}", 
	author={Guang Hao Low and Vadym Kliuchnikov and Nathan Wiebe},
	year={2019},
	eprint={1907.11679},
	archivePrefix={arXiv},
	primaryClass={quant-ph},
	url={https://arxiv.org/abs/1907.11679}, 
}

@article{Faehrmann2022randomizingmulti,
	title = {Randomizing multi-product formulas for {H}amiltonian simulation},
	author = {Faehrmann, Paul K. and Steudtner, Mark and Kueng, Richard and Kieferova, Maria and Eisert, Jens},
	eprint = "2101.07808",
	archivePrefix = "arXiv",
	primaryClass = "quant-ph",
	doi = "10.22331/q-2022-09-19-806",
	journal = "Quantum",
	volume = "6",
	pages = "806",
	year = "2022"
}

@article{CarreraVazquez2023wellconditioned,
	doi = {10.22331/q-2023-07-25-1067},
	url = {https://doi.org/10.22331/q-2023-07-25-1067},
	title = {Well-conditioned multi-product formulas for hardware-friendly {H}amiltonian simulation},
	author = {Carrera Vazquez, Almudena and Egger, Daniel J. and Ochsner, David and Woerner, Stefan},
	eprint = "2207.11268",
	archivePrefix = "arXiv",
	primaryClass = "quant-ph",
	journal = {{Quantum}},
	issn = {2521-327X},
	publisher = {{Verein zur F{\"{o}}rderung des Open Access Publizierens in den Quantenwissenschaften}},
	volume = {7},
	pages = {1067},
	month = jul,
	year = {2023}
}

@article{Zhuk:2023zeh,
	author = "Zhuk, Sergiy and Robertson, Niall F. and Bravyi, Sergey",
	title = "{Trotter error bounds and dynamic multi-product formulas for Hamiltonian simulation}",
	eprint = "2306.12569",
	archivePrefix = "arXiv",
	primaryClass = "quant-ph",
	doi = "10.1103/PhysRevResearch.6.033309",
	journal = "Phys. Rev. Res.",
	volume = "6",
	number = "3",
	pages = "033309",
	year = "2024"
}

@article{aftab2024multiproducthamiltoniansimulationexplicit,
	title="{Multi-product Hamiltonian simulation with explicit commutator scaling}", 
	author={Junaid Aftab and Dong An and Konstantina Trivisa},
	year={2024},
	eprint={2403.08922},
	archivePrefix={arXiv},
	primaryClass={quant-ph},
	url={https://arxiv.org/abs/2403.08922}, 
}

@article{Yin:2011np,
	author = "Yin, Hantao and Mawhinney, Robert D.",
	editor = "Vranas, Pavlos",
	title = {{Improving DWF Simulations: the Force Gradient Integrator and the M{\"o}bius Accelerated DWF Solver}},
	eprint = "1111.5059",
	archivePrefix = "arXiv",
	primaryClass = "hep-lat",
	doi = "10.22323/1.139.0051",
	journal = "PoS",
	volume = "LATTICE2011",
	pages = "051",
	year = "2011"
}

@article{Schafers:2025mgt,
	author = {Sch{\"a}fers, Kevin and Finkenrath, Jacob and G{\"u}nther, Michael and Knechtli, Francesco},
	title = "{Numerical stability of force-gradient integrators and their Hessian-free variants in lattice QCD simulations}",
	eprint = "2506.08813",
	archivePrefix = "arXiv",
	primaryClass = "hep-lat",
	reportNumber = "CERN-TH-2025-093",
	doi = "10.1016/j.cpc.2026.110034",
	journal = "Comput. Phys. Commun.",
	volume = "321",
	pages = "110034",
	year = "2026"
}

@article{Clark:2008gh,
	author = "Clark, M. A. and Kennedy, A. D. and Silva, P. J.",
	editorx = "Aubin, Christopher and Cohen, Saul and Dawson, Chris and Dudek, Jozef and Edwards, Robert and Joo, Balint and Lin, Huey-Wen and Orginos, Kostas and Richards, David and Thacker, Hank",
	title = "{Tuning HMC using Poisson brackets}",
	eprint = "0810.1315",
	archivePrefix = "arXiv",
	primaryClass = "hep-lat",
	doi = "10.22323/1.066.0041",
	journal = "PoS",
	volume = "LATTICE2008",
	pages = "041",
	year = "2008"
}

@article{Kennedy:2012gk,
	author = "Kennedy, A. D. and Silva, P. J. and Clark, M. A.",
	title = "{Shadow Hamiltonians, Poisson Brackets, and Gauge Theories}",
	eprint = "1210.6600",
	archivePrefix = "arXiv",
	primaryClass = "hep-lat",
	doi = "10.1103/PhysRevD.87.034511",
	journal = "Phys. Rev. D",
	volume = "87",
	number = "3",
	pages = "034511",
	year = "2013"
}

@ARTICLE{Crouch:1993,
	author = {{Crouch}, P.~E. and {Grossman}, R.},
	title = "{Numerical integration of ordinary differential equations on manifolds}",
	journal = {Journal of NonLinear Science},
	year = 1993,
	month = dec,
	volume = {3},
	number = {1},
	pages = {1-33},
	doi = {10.1007/BF02429858},
}

@article{MUNTHEKAAS1999115,
	title = "{High order Runge-Kutta methods on manifolds}",
	journal = {Applied Numerical Mathematics},
	volume = {29},
	number = {1},
	pages = {115-127},
	year = {1999},
	note = {Proceedings of the NSF/CBMS Regional Conference on Numerical Analysis of Hamiltonian Differential Equations},
	issn = {0168-9274},
	doi = {https://doi.org/10.1016/S0168-9274(98)00030-0},
	url = {https://www.sciencedirect.com/science/article/pii/S0168927498000300},
	author = {Hans Munthe-Kaas},
}

@article{Owren:1999,
	author = {Owren, B. and Marthinsen, A.},
	title = "{Runge-Kutta Methods Adapted to Manifolds and Based on Rigid Frames}",
	year = {1999},
	issue_date = {Mar 1999},
	publisher = {BIT Computer Science and Numerical Mathematics},
	address = {USA},
	volume = {39},
	number = {1},
	issn = {0006-3835},
	url = {https://doi.org/10.1023/A:1022325426017},
	doi = {10.1023/A:1022325426017},
	journal = {BIT},
	month = mar,
	pages = {116–142},
	numpages = {27}
}

@book{Kramer:1981,
	year = {1981},
	author = {Kramer, Peter and Saraceno, Marcos},
	address = {Berlin ;},
	booktitle = {Geometry of the time-dependent variational principle in quantum mechanics},
	isbn = {\href{https://link.springer.com/book/10.1007/3-540-10579-4}{9783540105794}},
	language = {eng},
	publisher = {Springer-Verlag},
	series = {Lecture notes in physics ; 140.},
	title = {Geometry of the time-dependent variational principle in quantum mechanics },
}

@article{LANGHOFF:1972hex,
	author = "Langhoff, P. W. and Epstein, S. T. and Karplus, M.",
	title = "{Aspects of Time-Dependent Perturbation Theory}",
	doi = "10.1103/RevModPhys.44.602",
	journal = "Rev. Mod. Phys.",
	volume = "44",
	pages = "602--644",
	year = "1972"
}

@article{Hastings_2006,
	title="{Solving gapped Hamiltonians locally}",
	volume={73},
	ISSN={1550-235X},
	url={http://dx.doi.org/10.1103/PhysRevB.73.085115},
	DOI={10.1103/physrevb.73.085115},
	number={8},
	journal={Physical Review B},
	publisher={American Physical Society (APS)},
	author={Hastings, M. B.},
	year={2006},
	month=Feb,
	eprint = "cond-mat/0508554",
	archivePrefix = "arXiv",
}

@article{Verstraete:2004gdw,
	author = "Verstraete, F. and Garc{\'\i}a-Ripoll, J. J. and Cirac, J. I.",
	title = "{Matrix Product Density Operators: Simulation of Finite-Temperature and Dissipative Systems}",
	eprint = "cond-mat/0406426",
	archivePrefix = "arXiv",
	doi = "10.1103/PhysRevLett.93.207204",
	journal = "Phys. Rev. Lett.",
	volume = "93",
	number = "20",
	pages = "207204",
	year = "2004"
}

@article{Perez-Garcia:2006nqo,
	author = "Perez-Garcia, David and Verstraete, Frank and Wolf, Michael M. and Cirac, J. Ignacio",
	title = "{Matrix product state representations}",
	eprint = "quant-ph/0608197",
	archivePrefix = "arXiv",
	doi = "10.26421/QIC7.5-6-1",
	journal = "Quant. Inf. Comput.",
	volume = "7",
	number = "5-6",
	pages = "401--430",
	year = "2007"
}

@article{Schollwoeck:2010uqf,
	author = "Schollwoeck, Ulrich",
	title = "{The density-matrix renormalization group in the age of matrix product states}",
	eprint = "1008.3477",
	archivePrefix = "arXiv",
	primaryClass = "cond-mat.str-el",
	doi = "10.1016/j.aop.2010.09.012",
	journal = "Annals Phys.",
	volume = "326",
	pages = "96--192",
	year = "2011"
}

@article{Orus:2018dya,
	author = "Or{\'u}s, Rom{\'a}n",
	title = "{Tensor networks for complex quantum systems}",
	eprint = "1812.04011",
	archivePrefix = "arXiv",
	primaryClass = "cond-mat.str-el",
	doi = "10.1038/s42254-019-0086-7",
	journal = "Nature Rev. Phys.",
	volume = "1",
	pages = "538--550",
	year = "2019"
}

@article{Haegeman:2011zz,
	author = "Haegeman, Jutho and Cirac, J. Ignacio and Osborne, Tobias J. and Pizorn, Iztok and Verschelde, Henri and Verstraete, Frank",
	title = "{Time-Dependent Variational Principle for Quantum Lattices}",
	eprint = "1103.0936",
	archivePrefix = "arXiv",
	primaryClass = "cond-mat.str-el",
	doi = "10.1103/PhysRevLett.107.070601",
	journal = "Phys. Rev. Lett.",
	volume = "107",
	pages = "070601",
	year = "2011"
}

@article{Haegeman:2015ezw,
	author = "Haegeman, Jutho and Lubich, Christian and Oseledets, Ivan and Vandereycken, Bart and Verstraete, Frank",
	title = "{Unifying time evolution and optimization with matrix product states}",
	eprint = "1408.5056",
	archivePrefix = "arXiv",
	primaryClass = "quant-ph",
	doi = "10.1103/PhysRevB.94.165116",
	journal = "Phys. Rev. B",
	volume = "94",
	number = "16",
	pages = "165116",
	year = "2016"
}

@article{Hemery:2019,
	title = {Matrix product states approaches to operator spreading in ergodic quantum systems},
	author = {H\'emery, K\'evin and Pollmann, Frank and Luitz, David J.},
	archivePrefix = {arXiv},
	eprint = {1901.05793},
	primaryClass = {cond-mat.str-el},
	journal = {Phys. Rev. B},
	volume = {100},
	issue = {10},
	pages = {104303},
	numpages = {11},
	year = {2019},
	month = {Sep},
	publisher = {American Physical Society},
	doi = {10.1103/PhysRevB.100.104303},
	url = {https://link.aps.org/doi/10.1103/PhysRevB.100.104303}
}

@article{Paeckel:2019yjf,
	author = {Paeckel, Sebastian and K{\"o}hler, Thomas and Swoboda, Andreas and Manmana, Salvatore R. and Schollw{\"o}ck, Ulrich and Hubig, Claudius},
	title = "{Time-evolution methods for matrix-product states}",
	eprint = "1901.05824",
	archivePrefix = "arXiv",
	primaryClass = "cond-mat.str-el",
	doi = "10.1016/j.aop.2019.167998",
	journal = "Annals Phys.",
	volume = "411",
	pages = "167998",
	year = "2019"
}

@article{Westhoff:2025mxp,
	author = {Westhoff, Philipp and Moroder, Mattia and Schollw{\"o}ck, Ulrich and Paeckel, Sebastian},
	title = "{A Tensor Network Framework for Lindbladian Spectra and Steady States}",
	eprint = "2509.07709",
	archivePrefix = "arXiv",
	primaryClass = "quant-ph",
	month = "9",
	year = "2025"
}

@article{Hackl:2020viw,
	author = "Hackl, Lucas and Guaita, Tommaso and Shi, Tao and Haegeman, Jutho and Demler, Eugene and Cirac, Ignacio",
	title = "{Geometry of variational methods: dynamics of closed quantum systems}",
	eprint = "2004.01015",
	archivePrefix = "arXiv",
	primaryClass = "quant-ph",
	doi = "10.21468/SciPostPhys.9.4.048",
	journal = "SciPost Phys.",
	volume = "9",
	number = "4",
	pages = "048",
	year = "2020"
}

@article{White:2018kbw,
	author = "White, Christopher David and Zaletel, Michael and Mong, Roger S. K. and Refael, Gil",
	title = "{Quantum dynamics of thermalizing systems}",
	doi = "10.1103/PhysRevB.97.035127",
	journal = "Phys. Rev. B",
	volume = "97",
	number = "3",
	pages = "035127",
	year = "2018",
	archivePrefix = {arXiv},
	eprint = {1707.01506},
	primaryClass = {cond-mat.str-el},
}

@article{Rakovszky:2022hss,
	author = "Rakovszky, Tibor and von Keyserlingk, C. W. and Pollmann, Frank",
	title = "{Dissipation-assisted operator evolution method for capturing hydrodynamic transport}",
	doi = "10.1103/PhysRevB.105.075131",
	journal = "Phys. Rev. B",
	volume = "105",
	number = "7",
	pages = "075131",
	year = "2022",
	archivePrefix = {arXiv},
	eprint = {2004.05177},
	primaryClass = {cond-mat.str-el},
}

@article{Troyer:2004ge,
	author = "Troyer, Matthias and Wiese, Uwe-Jens",
	title = "{Computational complexity and fundamental limitations to fermionic quantum Monte Carlo simulations}",
	eprint = "cond-mat/0408370",
	archivePrefix = "arXiv",
	doi = "10.1103/PhysRevLett.94.170201",
	journal = "Phys. Rev. Lett.",
	volume = "94",
	pages = "170201",
	year = "2005"
}

@article{schubert2023trotter,
	author = "Schubert, Ansgar and Mendl, Christian B.",
	title = "{Trotter error with commutator scaling for the Fermi-Hubbard model}",
	eprint = "2306.10603",
	archivePrefix = "arXiv",
	primaryClass = "quant-ph",
	doi = "10.1103/PhysRevB.108.195105",
	journal = "Phys. Rev. B",
	volume = "108",
	number = "19",
	pages = "195105",
	year = "2023"
}

@article{Schubert:2023code,
	author = {Christian B. Mendl and Ansgar Schubert},
	title = "{Trotter error with commutator scaling for the Fermi-Hubbard model}",
	year = {2023},
	publisher = {GitHub},
	journal = {GitHub repository},
	note = {\url{https://github.com/qc-tum/fermi\_hubbard\_commutators.git}},
}

@article{https://doi.org/10.48550/arxiv.2302.04698,
	title = "{Quantum simulation costs for Suzuki-Trotter decomposition of quantum many-body lattice models}",
	author = {Myers, Nathan M. and Scott, Ryan and Park, Kwon and Scarola, Vito W.},
	eprint = "2302.04698",
	archivePrefix = "arXiv",
	primaryClass = "quant-ph",
	journal = {Phys. Rev. Res.},
	volume = {5},
	issue = {2},
	pages = {023199},
	numpages = {17},
	year = {2023},
	month = {Jun},
	publisher = {American Physical Society},
	doi = {10.1103/PhysRevResearch.5.023199},
	url = {https://link.aps.org/doi/10.1103/PhysRevResearch.5.023199}
}

@article{https://doi.org/10.48550/arxiv.2303.04850,
	author = "Ayral, Thomas and Besserve, Pauline and Lacroix, Denis and Ruiz Guzman, Edgar Andres",
	title = "{Quantum computing with and for many-body physics}",
	eprint = "2303.04850",
	archivePrefix = "arXiv",
	primaryClass = "quant-ph",
	doi = "10.1140/epja/s10050-023-01141-1",
	journal = "Eur. Phys. J. A",
	volume = "59",
	number = "10",
	pages = "227",
	year = "2023"
}

@article{granet2023continuous,
	author = "Granet, Etienne and Dreyer, Henrik",
	title = "{Hamiltonian dynamics on digital quantum computers without discretization error}",
	eprint = "2308.03694",
	archivePrefix = "arXiv",
	primaryClass = "quant-ph",
	doi = "10.1038/s41534-024-00877-y",
	journal = "npj Quantum Inf.",
	volume = "10",
	number = "1",
	pages = "82",
	year = "2024"
}

@article{martínezmartínez2023estimating,
	title="{Estimating Trotter Approximation Errors to Optimize Hamiltonian Partitioning for Lower Eigenvalue Errors}", 
	author={Shashank G. Mehendale and Luis A. Martínez-Martínez and Prathami Divakar Kamath and Artur F. Izmaylov},
	journal  ="Digital Discovery",
	year  ="2025",
	volume  ="4",
	issue  ="12",
	pages  ="3540-3551",
	publisher  ="RSC",
	doi  ="10.1039/D5DD00185D",
	url  ="http://dx.doi.org/10.1039/D5DD00185D",
	eprint={2312.13282},
	archivePrefix={arXiv},
	primaryClass={physics.chem-ph}
}

@article{burgarth2023strong,
	title = "{Strong error bounds for Trotter and strang-splittings and their implications for quantum chemistry}",
	author = {Burgarth, Daniel and Facchi, Paolo and Hahn, Alexander and Johnsson, Mattias and Yuasa, Kazuya},
	eprint = "2312.08044",
	archivePrefix = "arXiv",
	primaryClass = "quant-ph",
	journal = {Phys. Rev. Res.},
	volume = {6},
	issue = {4},
	pages = {043155},
	numpages = {42},
	year = {2024},
	month = {Nov},
	publisher = {American Physical Society},
	doi = {10.1103/PhysRevResearch.6.043155},
	url = {https://link.aps.org/doi/10.1103/PhysRevResearch.6.043155}
}

@article{Burgarth:2022lib,
	author = "Burgarth, Daniel and Galke, Niklas and Hahn, Alexander and van Luijk, Lauritz",
	title = "{State-dependent Trotter limits and their approximations}",
	eprint = "2209.14787",
	archivePrefix = "arXiv",
	primaryClass = "quant-ph",
	doi = "10.1103/PhysRevA.107.L040201",
	journal = "Phys. Rev. A",
	volume = "107",
	number = "4",
	pages = "L040201",
	year = "2023"
}

@article{Hahn:2024fcz,
	author = "Hahn, Alexander and Burgarth, Daniel and Lonigro, Davide",
	title = "{Efficiency of dynamical decoupling for (almost) any spin{\textendash}boson model}",
	eprint = "2409.15743",
	archivePrefix = "arXiv",
	primaryClass = "quant-ph",
	doi = "10.21468/SciPostPhys.19.2.035",
	journal = "SciPost Phys.",
	volume = "19",
	number = "2",
	pages = "035",
	year = "2025"
}

@article{vanLuijk:2022zle,
	author = "van Luijk, Lauritz and Galke, Niklas and Hahn, Alexander and Burgarth, Daniel",
	title = "{Error bounds for Lie group representations in quantum mechanics}",
	eprint = "2211.08582",
	archivePrefix = "arXiv",
	primaryClass = "quant-ph",
	doi = "10.1088/1751-8121/ad288b",
	journal = "J. Phys. A",
	volume = "57",
	number = "10",
	pages = "105301",
	year = "2024"
}

@article{schaefers2024hessianfree,
	author = {Sch{\"a}fers, Kevin and Finkenrath, Jacob and G{\"u}nther, Michael and Knechtli, Francesco},
	title = "{Hessian-free force-gradient integrators}",
	eprint = "2403.10370",
	archivePrefix = "arXiv",
	primaryClass = "math.NA",
	doi = "10.1016/j.cpc.2024.109478",
	journal = "Comput. Phys. Commun.",
	volume = "309",
	pages = "109478",
	year = "2025"
}

@article{10.1063/1.2203609,
	author = {Blanes, Sergio and Casas, Fernando and Murua, Ander},
	title = "{Symplectic splitting operator methods for the time-dependent Schrödinger equation}",
	journal = {The Journal of Chemical Physics},
	volume = {124},
	number = {23},
	pages = {234105},
	year = {2006},
	month = {06},
	issn = {0021-9606},
	doi = {10.1063/1.2203609},
	url = {https://doi.org/10.1063/1.2203609},
}

@article{li2024principaltrotterobservationerror,
	title = "{Principal Trotter observation error with truncated commutators}",
	author = {Li, Langyu},
	eprint={2408.03891},
	archivePrefix={arXiv},
	primaryClass={quant-ph},
	journal = {Phys. Rev. A},
	volume = {110},
	issue = {6},
	pages = {062614},
	numpages = {12},
	year = {2024},
	month = {Dec},
	publisher = {American Physical Society},
	doi = {10.1103/PhysRevA.110.062614},
	url = {https://link.aps.org/doi/10.1103/PhysRevA.110.062614}
}

@article{david2024fasterquantumsimulationmarkovian,
	title="{Faster Quantum Simulation Of Markovian Open Quantum Systems Via Randomisation}", 
	author={I. J. David and I. Sinayskiy and F. Petruccione},
	year={2024},
	eprint={2408.11683},
	archivePrefix={arXiv},
	primaryClass={quant-ph},
	url={https://arxiv.org/abs/2408.11683}, 
}

@article{watson2024exponentiallyreducedcircuitdepths,
	author = "Watson, James D. and Watkins, Jacob",
	title = "{Exponentially Reduced Circuit Depths Using Trotter Error Mitigation}",
	eprint = "2408.14385",
	archivePrefix = "arXiv",
	primaryClass = "quant-ph",
	doi = "10.1103/kw39-yxq5",
	journal = "PRX Quantum",
	volume = "6",
	number = "3",
	pages = "030325",
	year = "2025"
}

@article{chen2024trottererrortimescaling,
	author = "Chen, Yi-Hsiang",
	title = "{Trotter error timescaling separation via commutant decomposition}",
	eprint = "2409.16634",
	archivePrefix = "arXiv",
	primaryClass = "quant-ph",
	doi = "10.1103/PhysRevA.111.022612",
	journal = "Phys. Rev. A",
	volume = "111",
	number = "2",
	pages = "022612",
	year = "2025"
}

@article{gibbs2024deepcircuitcompressionquantum,
	author = "Gibbs, Joe and Cincio, Lukasz",
	title = "{Deep Circuit Compression for Quantum Dynamics via Tensor Networks}",
	eprint = "2409.16361",
	archivePrefix = "arXiv",
	primaryClass = "quant-ph",
	reportNumber = "LA-UR-24-30140",
	doi = "10.22331/q-2025-07-09-1789",
	journal = "Quantum",
	volume = "9",
	pages = "1789",
	year = "2025"
}

@article{hahn2024lowerboundstrottererror,
	author = "Hahn, Alexander and Hartung, Paul and Burgarth, Daniel and Facchi, Paolo and Yuasa, Kazuya",
	title = "{Lower bounds for the Trotter error}",
	eprint = "2410.03059",
	archivePrefix = "arXiv",
	primaryClass = "quant-ph",
	doi = "10.1103/PhysRevA.111.022417",
	journal = "Phys. Rev. A",
	volume = "111",
	number = "2",
	pages = "022417",
	year = "2025"
}

@article{burdine2024trotterlesssimulationopenquantum,
	author = {Burdine, Colin and Blair, Enrique P.},
	title = "{Trotterless Simulation of Open Quantum Systems for NISQ Quantum Devices}",
	eprint = "2410.03854",
	archivePrefix = "arXiv",
	primaryClass = "quant-ph",
	doi = "10.1002/qute.202400240",
	journal = "Adv. Quantum Technol.",
	volume = "8",
	number = "1",
	pages = "2400240",
	year = "2025"
}

@article{kiumi2024tepaiexacttimeevolution,
	title="{TE-PAI: Exact Time Evolution by Sampling Random Circuits}", 
	author={Chusei Kiumi and Bálint Koczor},
	eprint={2410.16850},
	archivePrefix={arXiv},
	primaryClass={quant-ph},
	url={https://arxiv.org/abs/2410.16850}, 
	doi = "10.1088/2058-9565/ae1160",
	journal = "Quantum Sci. Technol.",
	volume = "10",
	number = "4",
	pages = "045071",
	year = "2025"
}

@article{chertkov2024robustnessnearthermaldynamicsdigital,
	title="{Robustness of near-thermal dynamics on digital quantum computers}", 
	author={Eli Chertkov and Yi-Hsiang Chen and Michael Lubasch and David Hayes and Michael Foss-Feig},
	eprint={2410.10794},
	archivePrefix={arXiv},
	primaryClass={quant-ph},
	url={https://arxiv.org/abs/2410.10794}, 
	doi = "10.1103/nn2w-jxpf",
	journal = "Phys. Rev. Res.",
	volume = "8",
	number = "1",
	pages = "013255",
	year = "2026"
}

@article{chen2024errorinterferencequantumsimulation,
	title="{Error Interference in Quantum Simulation}", 
	author={Boyang Chen and Jue Xu and Qi Zhao and Xiao Yuan},
	eprint={2411.03255},
	archivePrefix={arXiv},
	primaryClass={quant-ph},
	journal = {Phys. Rev. Lett.},
	pages = {--},
	year = {2026},
	month = {Apr},
	publisher = {American Physical Society},
	doi = {10.1103/g2n5-qdxh},
	url = {https://link.aps.org/doi/10.1103/g2n5-qdxh}
}

@article{arseniev2024distributedquantumlogicalgorithm,
	title="{Distributed quantum logic algorithm}", 
	author={Boris Arseniev},
	year={2024},
	eprint={2411.11979},
	archivePrefix={arXiv},
	primaryClass={quant-ph},
	url={https://arxiv.org/abs/2411.11979}, 
}

@article{le2025riemannianquantumcircuitoptimization,
	author = "Le, Isabel Nha Minh and Sun, Shuo and Mendl, Christian B.",
	title = "{Riemannian quantum circuit optimization based on matrix product operators}",
	eprint = "2501.08872",
	archivePrefix = "arXiv",
	primaryClass = "quant-ph",
	doi = "10.22331/q-2025-08-27-1833",
	journal = "Quantum",
	volume = "9",
	pages = "1833",
	year = "2025"
}

@article{peetz2025hamiltoniansimulationstochasticzassenhaus,
	title="{Hamiltonian Simulation via Stochastic Zassenhaus Expansions}", 
	author={Joseph Peetz and Prineha Narang},
	year={2025},
	eprint={2501.13922},
	archivePrefix={arXiv},
	primaryClass={quant-ph},
	url={https://arxiv.org/abs/2501.13922}, 
}

@article{nguyen2025zassenhausexpansionsolvingschrodinger,
	title="{Zassenhaus Expansion in Solving the Schr\"odinger Equation}", 
	author={Molena Nguyen and Naihuan Jing},
	year={2025},
	eprint={2505.09441},
	archivePrefix={arXiv},
	primaryClass={quant-ph},
	url={https://arxiv.org/abs/2505.09441}, 
}

@article{bosse2024efficientpracticalhamiltoniansimulation,
	author = "Bosse, Jan Lukas and Childs, Andrew M. and Derby, Charles and Gambetta, Filippo Maria and Montanaro, Ashley and Santos, Raul A.",
	title = "{Efficient and practical Hamiltonian simulation from time-dependent product formulas}",
	eprint = "2403.08729",
	archivePrefix = "arXiv",
	primaryClass = "quant-ph",
	doi = "10.1038/s41467-025-57580-5",
	journal = "Nature Commun.",
	volume = "16",
	number = "1",
	pages = "2673",
	year = "2025"
}

@article{ciavarella2024quantumsimulationsu3lattice,
	author = "Ciavarella, Anthony N. and Bauer, Christian W.",
	title = "{Quantum Simulation of SU(3) Lattice Yang-Mills Theory at Leading Order in Large-Nc Expansion}",
	eprint = "2402.10265",
	archivePrefix = "arXiv",
	primaryClass = "hep-ph",
	doi = "10.1103/PhysRevLett.133.111901",
	journal = "Phys. Rev. Lett.",
	volume = "133",
	number = "11",
	pages = "111901",
	year = "2024"
}

@article{gentinetta2025correctingextendingtrotterizedquantum,
	title="{Correcting and extending Trotterized quantum many-body dynamics}", 
	author={Gian Gentinetta and Friederike Metz and Giuseppe Carleo},
	eprint={2502.13784},
	archivePrefix={arXiv},
	primaryClass={quant-ph},
	url={https://arxiv.org/abs/2502.13784}, 
	doi = "10.1103/1nqt-x5xh",
	journal = "PRX Quantum",
	volume = "6",
	number = "3",
	pages = "030361",
	year = "2025"
}

@article{chen2025trottererrorgatecomplexity,
	title="{Trotter error and gate complexity of the SYK and sparse SYK models}", 
	author={Yiyuan Chen and Jonas Helsen and Maris Ozols},
	eprint={2502.18420},
	archivePrefix={arXiv},
	primaryClass={quant-ph},
	url={https://arxiv.org/abs/2502.18420}, 
	doi = "10.22331/q-2026-02-09-1999",
	journal = "Quantum",
	volume = "10",
	pages = "1999",
	year = "2026"
}

@article{thanh2025hamiltonianreorderingshallowertrotterization,
	title="{Hamiltonian Reordering for Shallower Trotterization Circuits}", 
	author={Cédric Ho Thanh},
	year={2025},
	eprint={2503.11153},
	archivePrefix={arXiv},
	primaryClass={quant-ph},
	url={https://arxiv.org/abs/2503.11153}, 
}

@article{mizuta2025trotterizationsubstantiallyefficientlowenergy,
	title="{Trotterization is Substantially Efficient for Low-Energy States}", 
	author={Kaoru Mizuta and Tomotaka Kuwahara},
	eprint={2504.20746},
	archivePrefix={arXiv},
	primaryClass={quant-ph},
	url={https://arxiv.org/abs/2504.20746}, 
	journal = "Phys. Rev. Lett.",
	volume = "135",
	number = "13",
	pages = "130602",
	year = "2025"
}

@article{lami2025beginnerslecturenotesquantum,
	title="{Beginner's Lecture Notes on Quantum Spin Chains, Exact Diagonalization and Tensor Networks}", 
	author={Guglielmo Lami and Mario Collura and Nishan Ranabhat},
	year={2025},
	eprint={2503.03564},
	archivePrefix={arXiv},
	primaryClass={cond-mat.str-el},
	url={https://arxiv.org/abs/2503.03564}, 
}

@inproceedings{decker2025kernpilercompileroptimizationquantum,
	title="{Kernpiler: Compiler Optimization for Quantum Hamiltonian Simulation with Partial Trotterization}", 
	author={Ethan Decker and Lucas Goetz and Evan McKinney and Erik Gustafson and Junyu Zhou and Yuhao Liu and Alex K. Jones and Ang Li and Alexander Schuckert and Samuel Stein and Eleanor Crane and Gushu Li},
	booktitle="{2026 ACM/IEEE 53rd Annual International Symposium on Computer Architecture (ISCA)}",
	year={2026},
	volume={},
	number={},
	pages={905-920},
	eprint={2504.07214},
	archivePrefix={arXiv},
	primaryClass={quant-ph},
	doi = "10.1109/ISCA66397.2026.00073",
}

@article{feng2025trotterizationoperatorscramblingentanglement,
	title="{Trotterization, Operator Scrambling, and Entanglement}", 
	author={Tianfeng Feng and Yue Cao and Qi Zhao},
	year={2025},
	eprint={2506.23345},
	archivePrefix={arXiv},
	primaryClass={quant-ph},
	url={https://arxiv.org/abs/2506.23345}, 
}

@article{mizuta2025commutatorscalinghamiltoniansimulation,
	title="{On the commutator scaling in Hamiltonian simulation with multi-product formulas}", 
	author={Kaoru Mizuta},
	eprint={2507.06557},
	archivePrefix={arXiv},
	primaryClass={quant-ph},
	doi = "10.22331/q-2026-01-19-1974",
	journal = "Quantum",
	volume = "10",
	pages = "1974",
	year = "2026"
}

@article{facchi2025slowconvergencetrotterdecomposition,
	title="{Slow convergence of Trotter decomposition for rotations}", 
	author={Paolo Facchi and Francesco Perrini and Vito Viesti},
	eprint={2507.15421},
	archivePrefix={arXiv},
	primaryClass={quant-ph},
	url={https://arxiv.org/abs/2507.15421}, 
	doi = "10.1016/j.aop.2026.170539",
	journal = "Annals Phys.",
	volume = "491",
	pages = "170539",
	year = "2026"
}

@article{becker2024convergenceratestrotterkatosplitting,
	title="{Convergence Rates for the Trotter Splitting for Unbounded Operators}",
	ISSN={1615-3383},
	url={http://dx.doi.org/10.1007/s10208-025-09730-w},
	DOI={10.1007/s10208-025-09730-w},
	journal={Foundations of Computational Mathematics},
	publisher={Springer Science and Business Media LLC},
	author={Becker, Simon and Galke, Niklas and van Luijk, Lauritz and Salzmann, Robert},
	year={2025},
	eprint={2407.04045},
	archivePrefix={arXiv},
	primaryClass={math-ph},
}

@article{fang2025trottererrormanybodyquantum,
	title="{On the Trotter Error in Many-body Quantum Dynamics with Coulomb Potentials}", 
	author={Di Fang and Xiaoxu Wu and Avy Soffer},
	eprint={2507.22707},
	archivePrefix={arXiv},
	primaryClass={quant-ph},
	url={https://arxiv.org/abs/2507.22707}, 
	doi = "10.1007/s00220-026-05593-6",
	journal = "Commun. Math. Phys.",
	volume = "407",
	number = "4",
	pages = "83",
	year = "2026"
}

@article{aharonov2025importanceerrormitigationquantum,
	title="{On the Importance of Error Mitigation for Quantum Computation}", 
	author={Dorit Aharonov and Ori Alberton and Itai Arad and Yosi Atia and Eyal Bairey and Zvika Brakerski and Itsik Cohen and Omri Golan and Ilya Gurwich and Oded Kenneth and Eyal Leviatan and Netanel H. Lindner and Ron Aharon Melcer and Adiel Meyer and Gili Schul and Maor Shutman},
	year={2025},
	eprint={2503.17243},
	archivePrefix={arXiv},
	primaryClass={quant-ph},
	url={https://arxiv.org/abs/2503.17243}, 
}

@article{Blunt:2025zul,
	author = "Blunt, Nick S. and Ivanov, Aleksei V. and Bay-Smidt, Andreas Juul",
	title = "{A Monte Carlo approach to bound Trotter error}",
	eprint = "2510.11621",
	archivePrefix = "arXiv",
	primaryClass = "quant-ph",
	month = "10",
	year = "2025"
}

@article{Sinha:2025tqg,
	author = "Sinha, Akash and Padmanabhan, Pramod and Korepin, Vladimir",
	title = "{Non-invertible Kramers-Wannier duality-symmetry in the trotterized critical Ising chain}",
	eprint = "2511.03947",
	archivePrefix = "arXiv",
	primaryClass = "quant-ph",
	month = "11",
	year = "2025"
}

@article{Jourdan:2025gqs,
	author = {Jourdan, Louis and Cassam-Chena{\"\i}, Patrick},
	title = "{A Remarkable Application of Zassenhaus Formula to Strongly Correlated Electron Systems}",
	eprint = "2510.24364",
	archivePrefix = "arXiv",
	primaryClass = "math-ph",
	doi = "10.22331/q-2026-04-08-2057",
	journal = "Quantum",
	volume = "10",
	pages = "2057",
	year = "2026"
}

@article{Assi:2025ifp,
	author = "Assi, Ibsal and Vogl, Michael and Kumari, Meenu and LeBlanc, J. P. F.",
	title = "{Beyond Trotterization: Variational Product Formulas for Quantum Simulation}",
	eprint = "2511.15124",
	archivePrefix = "arXiv",
	primaryClass = "quant-ph",
	month = "11",
	year = "2025"
}

@article{Dizaji:2025xyu,
	author = "Dizaji, Kasra Rajabzadeh and Kim, Leeseok and Marvian, Milad and Arenz, Christian",
	title = "{Higher-order Zeno sequences}",
	eprint = "2511.20792",
	archivePrefix = "arXiv",
	primaryClass = "quant-ph",
	month = "11",
	year = "2025"
}

@article{Simon:2025aie,
	author = "Simon, William A. and Love, Peter J.",
	title = "{Quantum Advantage in Resource Estimation}",
	eprint = "2512.02131",
	archivePrefix = "arXiv",
	primaryClass = "quant-ph",
	month = "12",
	year = "2025"
}

@inproceedings{Waintal:2026tsc,
	author = "Waintal, Xavier and Huang, Chen-How and Groth, Christoph W.",
	title = "{Who can compete with quantum computers? Lecture notes on quantum inspired tensor networks computational techniques}",
	eprint = "2601.03035",
	archivePrefix = "arXiv",
	primaryClass = "quant-ph",
	month = "1",
	year = "2026"
}

@article{Irmejs:2025vuh,
	author = "Irmejs, Reinis and Cirac, J. Ignacio",
	title = "{Efficient Simulation of Sparse, Non-Local Fermion Models}",
	eprint = "2512.15843",
	archivePrefix = "arXiv",
	primaryClass = "quant-ph",
	month = "12",
	year = "2025"
}

@article{PhysRevLett.131.060602,
	title = {Self-Healing of Trotter Error in Digital Adiabatic State Preparation},
	author = {Kovalsky, Lucas K. and Calderon-Vargas, Fernando A. and Grace, Matthew D. and Magann, Alicia B. and Larsen, James B. and Baczewski, Andrew D. and Sarovar, Mohan},
	eprint = "2209.06242",
	archivePrefix = "arXiv",
	primaryClass = "quant-ph",
	journal = {Phys. Rev. Lett.},
	volume = {131},
	issue = {6},
	pages = {060602},
	numpages = {6},
	year = {2023},
	month = {Aug},
	publisher = {American Physical Society},
	doi = {10.1103/PhysRevLett.131.060602},
	url = {https://link.aps.org/doi/10.1103/PhysRevLett.131.060602}
}

@article{Vizzuso:2025isl,
	author = "Vizzuso, Mara and Passarelli, Gianluca and Cantele, Giovanni and Lucignano, Procolo and Chen, Xi and Paul, Koushik",
	title = "{Nonadiabatic Self-Healing of Trotter Errors in Digitized Counterdiabatic Dynamics}",
	eprint = "2512.22636",
	archivePrefix = "arXiv",
	primaryClass = "quant-ph",
	month = "12",
	year = "2025"
}

@article{Zhou:2026cvd,
	author = "Zhou, Shuo and Pan, Zhaokai and Gong, Weiyuan and Li, Tongyang",
	title = "{Time-Dependent Hamiltonian Simulation in the Low-Energy Subspace}",
	eprint = "2601.01550",
	archivePrefix = "arXiv",
	primaryClass = "quant-ph",
	month = "1",
	year = "2026"
}

@article{Kulkarni:2026yly,
	author = "Kulkarni, Prateek P.",
	title = "{Entanglement-Dependent Error Bounds for Hamiltonian Simulation}",
	eprint = "2602.00555",
	archivePrefix = "arXiv",
	primaryClass = "quant-ph",
	month = "1",
	year = "2026"
}

@article{Park:2026ozj,
	author = "Park, Seung and Lee, Sangjin and Baek, Kyunghyun",
	title = "{Dual channel multi-product formulas}",
	eprint = "2602.01713",
	archivePrefix = "arXiv",
	primaryClass = "quant-ph",
	month = "2",
	year = "2026"
}

@article{Ronfaut:2026knp,
	author = "Ronfaut, Clement and Ollive, Robin and Louise, Stephane",
	title = "{Numerical Error Extraction by Quantum Measurement Algorithm}",
	eprint = "2602.01927",
	archivePrefix = "arXiv",
	primaryClass = "quant-ph",
	month = "2",
	year = "2026"
}

@article{Zimboras:2025unr,
	author = "Zimbor{\'a}s, Zolt{\'a}n and others",
	title = "{Myths around quantum computation before full fault tolerance: What no-go theorems rule out and what they don't}",
	eprint = "2501.05694",
	archivePrefix = "arXiv",
	primaryClass = "quant-ph",
	month = "1",
	year = "2025"
}

@article{Wang:2026qyq,
	author = "Wang, Xinzhao and Zhou, Shuo and Wang, Xiaoyang and Zheng, Yi-Cong and Zhang, Shengyu and Li, Tongyang",
	title = "{Lindbladian Simulation with Commutator Bounds}",
	eprint = "2603.28602",
	archivePrefix = "arXiv",
	primaryClass = "quant-ph",
	reportNumber = "RIKEN-iTHEMS-Report-26",
	month = "3",
	year = "2026"
}

@article{Shoji:2026kyr,
	author = "Shoji, Satoru and Ito, Kosuke and Shimizu, Yukihiro and Fujii, Keisuke",
	title = "{Variational Quantum Operator Simulation}",
	eprint = "2603.06013",
	archivePrefix = "arXiv",
	primaryClass = "quant-ph",
	month = "3",
	year = "2026"
}

@article{Zhou:2026mlc,
	author = "Zhou, Bozhen and Zhao, Qi and Zhang, Pan",
	title = "{Trotter Scars: Trotter Error Suppression in Quantum Simulation}",
	eprint = "2603.29857",
	archivePrefix = "arXiv",
	primaryClass = "quant-ph",
	month = "3",
	year = "2026"
}

@article{Fang:2026css,
	author = "Fang, Di and Wu, Xiaoxu",
	title = "{Trotterization with Many-body Coulomb Interactions: Convergence for General Initial Conditions and State-Dependent Improvements}",
	eprint = "2604.07704",
	archivePrefix = "arXiv",
	primaryClass = "quant-ph",
	month = "4",
	year = "2026"
}

@article{Zhang:2026cbs,
	author = "Zhang, Xiangran and Xu, Jue and Zhao, Qi and Zhou, You",
	title = "{Taming Trotter Errors with Quantum Resources}",
	eprint = "2604.13486",
	archivePrefix = "arXiv",
	primaryClass = "quant-ph",
	month = "4",
	year = "2026"
}

@article{Hasselgren:2026trv,
	author = "Hasselgren, Fredrik and Koczor, B{\'a}lint",
	title = "{Quantum-inspired classical simulation through randomized time evolution}",
	eprint = "2604.13144",
	archivePrefix = "arXiv",
	primaryClass = "quant-ph",
	month = "4",
	year = "2026"
}

@article{Tate:2026inm,
	author = "Tate, Reuben and Aktar, Shamminuj and Eidenbenz, Stephan",
	title = "{An Analysis of Commutation-Based Trotter Ordering Strategies on Heisenberg-Style Hamiltonians}",
	eprint = "2604.23138",
	archivePrefix = "arXiv",
	primaryClass = "quant-ph",
	reportNumber = "LANL report LA-UR-26-23269",
	month = "4",
	year = "2026"
}

@article{PhysRevLett.123.070503,
	title = "{Random Compiler for Fast Hamiltonian Simulation}",
	author = {Campbell, Earl},
	eprint = "1811.08017",
	archivePrefix = "arXiv",
	primaryClass = "quant-ph",
	journal = {Phys. Rev. Lett.},
	volume = {123},
	issue = {7},
	pages = {070503},
	numpages = {5},
	year = {2019},
	month = {Aug},
	publisher = {American Physical Society},
	doi = {10.1103/PhysRevLett.123.070503},
	url = {https://link.aps.org/doi/10.1103/PhysRevLett.123.070503}
}

@article{Lee:2026nqh,
	author = "Lee, Sangjin and Cho, Sangkook",
	title = "{qSHIFT: An Adaptive Sampling Protocol for Higher-Order Quantum Simulation}",
	eprint = "2604.26263",
	archivePrefix = "arXiv",
	primaryClass = "quant-ph",
	month = "4",
	year = "2026"
}

@article{Bay-Smidt:2026vcs,
	author = "Bay-Smidt, Andreas Juul and Glaser, Nina and Fabian, Marcel D. and Campbell, Earl T. and Blunt, Nick S. and Solomon, Gemma C.",
	title = "{Quantum simulation of nanographenes and Trotter error cancellation}",
	eprint = "2605.00745",
	archivePrefix = "arXiv",
	primaryClass = "quant-ph",
	month = "5",
	year = "2026"
}

@article{Yan:2026jky,
	author = "Yan, WenBin",
	title = "{Physics Guided Generative Optimization for Trotter Suzuki Decomposition}",
	eprint = "2605.13268",
	archivePrefix = "arXiv",
	primaryClass = "quant-ph",
	month = "5",
	year = "2026"
}

@article{Negishi:2026ujz,
	author = "Negishi, Naoki and Yang, Bo",
	title = "{Beyond Commutativity: Redesigning Trotter Decomposition via Local Symmetry}",
	eprint = "2605.16016",
	archivePrefix = "arXiv",
	primaryClass = "quant-ph",
	month = "5",
	year = "2026"
}

@article{Koczor:2023shj,
	author = "Koczor, B{\'a}lint and Morton, John J. L. and Benjamin, Simon C.",
	title = "{Probabilistic Interpolation of Quantum Rotation Angles}",
	eprint = "2305.19881",
	archivePrefix = "arXiv",
	primaryClass = "quant-ph",
	doi = "10.1103/PhysRevLett.132.130602",
	journal = "Phys. Rev. Lett.",
	volume = "132",
	number = "13",
	pages = "130602",
	year = "2024"
}

@article{Chundury:2026wpt,
	author = "Chundury, Srikar and Burgstahler, Blake and Li, Jiajia and Suh, In-Saeng and Mueller, Frank",
	title = "{Diagonal-Budgeted Trotterization for Efficient Quantum Hamiltonian Simulation}",
	eprint = "2606.16959",
	archivePrefix = "arXiv",
	primaryClass = "quant-ph",
	month = "6",
	year = "2026"
}

@article{Lee:2026qao,
	author = "Lee, Sangjin and Choi, Sangkook",
	title = "{Mitigating Trotter Errors via Post-Processed Symmetry Restoration}",
	eprint = "2606.20242",
	archivePrefix = "arXiv",
	primaryClass = "quant-ph",
	month = "6",
	year = "2026"
}

@article{Casas:2012nqk,
	author = "Casas, Fernando and Murua, Ander and Nadinic, Mladen",
	title = "{Efficient computation of the Zassenhaus formula}",
	eprint = "1204.0389",
	archivePrefix = "arXiv",
	primaryClass = "math-ph",
	doi = "10.1016/j.cpc.2012.06.006",
	journal = "Comput. Phys. Commun.",
	volume = "183",
	pages = "2386--2391",
	year = "2012"
}

@article{Kokcu:2021ctj,
	author = {K{\"o}kc{\"u}, Efekan and Steckmann, Thomas and Wang, Yan and Freericks, J. K. and Dumitrescu, Eugene F. and Kemper, Alexander F.},
	title = "{Fixed Depth Hamiltonian Simulation via Cartan Decomposition}",
	eprint = "2104.00728",
	archivePrefix = "arXiv",
	primaryClass = "quant-ph",
	doi = "10.1103/PhysRevLett.129.070501",
	journal = "Phys. Rev. Lett.",
	volume = "129",
	number = "7",
	pages = "070501",
	year = "2022"
}

@article{Hayata:2026qcz,
	author = "Hayata, Tomoya and Kikuchi, Yuta",
	title = "{Continuous-time evolution via probabilistic angle interpolation and its applications}",
	eprint = "2604.02854",
	archivePrefix = "arXiv",
	primaryClass = "quant-ph",
	reportNumber = "RIKEN-iTHEMS-Report-26",
	month = "4",
	year = "2026"
}

@article{Dai:2026oqj,
	author = "Dai, Joshua W. and Hasselgren, Fredrik and Kiumi, Chusei",
	title = "{Structure-Aware Variance Reduction for Unbiased Randomized Hamiltonian Simulation}",
	eprint = "2606.23544",
	archivePrefix = "arXiv",
	primaryClass = "quant-ph",
	month = "6",
	year = "2026"
}

@article{Misguich:2026mow,
	author = "Misguich, Gr{\'e}goire",
	title = "{Introduction to matrix-product states and tensor networks}",
	eprint = "2606.24803",
	archivePrefix = "arXiv",
	primaryClass = "cond-mat.str-el",
	month = "6",
	year = "2026"
}

@article{Gluza:2024lqq,
	author = {Gluza, Marek and Son, Jeongrak and Tiang, Bi Hong and Zander, Ren{\'e} and Seidel, Raphael and Suzuki, Yudai and Holmes, Zo{\"e} and Ng, Nelly H. Y.},
	title = "{Double-Bracket Quantum Algorithms for Quantum Imaginary-Time Evolution}",
	eprint = "2412.04554",
	archivePrefix = "arXiv",
	primaryClass = "quant-ph",
	doi = "10.1103/rw81-k8vk",
	journal = "Phys. Rev. Lett.",
	volume = "136",
	number = "2",
	pages = "020601",
	year = "2026"
}

@article{Benavides-Riveros:2026hyf,
	author = "Benavides-Riveros, Carlos L. and Sharma, Prachi and {\v{S}}imkovic, Fedor",
	title = "{Augmenting Imaginary-Time Evolution with Local Geometric Information}",
	eprint = "2606.23934",
	archivePrefix = "arXiv",
	primaryClass = "quant-ph",
	month = "6",
	year = "2026"
}

@article{Becker:2021spk,
	author = "Becker, Simon and Datta, Nilanjana and Lami, Ludovico and Rouz{\'e}, Cambyse",
	title = "{Energy-constrained discrimination of unitaries, quantum speed limits and a Gaussian Solovay-Kitaev theorem}",
	eprint = "2006.06659",
	archivePrefix = "arXiv",
	primaryClass = "quant-ph",
	doi = "10.1103/PhysRevLett.126.190504",
	journal = "Phys. Rev. Lett.",
	volume = "126",
	pages = "190504",
	year = "2021"
}

@article{Shirokov:2017vmf,
	author = "Shirokov, M. E.",
	title = "{On the Energy-Constrained Diamond Norm and Its Application in Quantum Information Theory}",
	eprint = "1706.00361",
	archivePrefix = "arXiv",
	primaryClass = "quant-ph",
	doi = "10.1134/S0032946018010027",
	journal = "Probl. Info. Transm.",
	volume = "54",
	number = "1",
	pages = "20--33",
	year = "2018"
}

@ARTICLE{Bernier:2025,
	author = {{Bernier}, J. and {Blanes}, S. and {Casas}, F. and {Escorihuela-Tom{\`a}s}, A.},
	title = "{On alternating-conjugate splitting methods}",
	year = 2025,
	month = mar,
	archivePrefix = {arXiv},
	eprint = {2503.08453},
	primaryClass = {cs.NA},
}

@article{escorihuelatomàs2026efficientsymplecticintegratorscubic,
	title={Efficient symplectic integrators for cubic and quartic potentials}, 
	author={Alejandro Escorihuela-Tomàs},
	year={2026},
	eprint={2605.06975},
	archivePrefix={arXiv},
	primaryClass={math.NA},
	url={https://arxiv.org/abs/2605.06975}, 
}

@article{Bauer:2026niu,
	author = "Bauer, Tobias V. and Milbradt, Richard M. and Mendl, Christian B.",
	title = "{Time Evolution on Hybrid Tensor Networks -- A Novel and Parallelizable Algorithm}",
	eprint = "2606.28169",
	archivePrefix = "arXiv",
	primaryClass = "quant-ph",
	month = "6",
	year = "2026"
}

@article{Yuan:2020xmq,
	author = "Yuan, Xiao and Sun, Jinzhao and Liu, Junyu and Zhao, Qi and Zhou, You",
	title = "{Quantum Simulation with Hybrid Tensor Networks}",
	eprint = "2007.00958",
	archivePrefix = "arXiv",
	primaryClass = "quant-ph",
	doi = "10.1103/PhysRevLett.127.040501",
	journal = "Phys. Rev. Lett.",
	volume = "127",
	number = "4",
	pages = "040501",
	year = "2021"
}

@article{Schuhmacher:2024spy,
	author = "Schuhmacher, Julian and Ballarin, Marco and Baiardi, Alberto and Magnifico, Giuseppe and Tacchino, Francesco and Montangero, Simone and Tavernelli, Ivano",
	title = "{Hybrid Tree Tensor Networks for Quantum Simulation}",
	eprint = "2404.05784",
	archivePrefix = "arXiv",
	primaryClass = "quant-ph",
	doi = "10.1103/PRXQuantum.6.010320",
	journal = "PRX Quantum",
	volume = "6",
	number = "1",
	pages = "010320",
	year = "2025"
}

@article{Murota:2026cyh,
	author = "Murota, Keisuke and Kikuchi, Yuta and Rinaldi, Enrico and Sauvage, Fr{\'e}d{\'e}ric and Todo, Synge",
	title = "{Unbiased Hamiltonian Simulation by Reversing Trotter Error Dynamics}",
	eprint = "2606.29741",
	archivePrefix = "arXiv",
	primaryClass = "quant-ph",
	month = "6",
	year = "2026"
}

@article{Maxwell:2026ewi,
	author = "Maxwell, William and Casares, Pablo A. M. and Lang, Robert A. and Fomichev, Stepan and Arrazola, Juan Miguel and Jahangiri, Soran and Asadi, Ali and Meneses, Luis Alfredo Nunez and Germain, Thomas and Motlagh, Danial",
	title = "{Practical Estimation of Trotter Error for Hamiltonian Simulation}",
	eprint = "2606.30738",
	archivePrefix = "arXiv",
	primaryClass = "quant-ph",
	month = "6",
	year = "2026"
}

@article{Casares:2026lyr,
	author = "Casares, Pablo A. M. and Maxwell, William and Motlagh, Danial and Choubisa, Hitarth and Niu, Zy and Loaiza, Ignacio and Mueller, Jonathan E. and Voigt, Arne-Christian and Arrazola, Juan Miguel and Fomichev, Stepan",
	title = "{Theory and practice of Trotter product formulas for quantum chemistry}",
	eprint = "2606.30741",
	archivePrefix = "arXiv",
	primaryClass = "quant-ph",
	month = "6",
	year = "2026"
}

@article{Kiss:2022lja,
	author = "Kiss, Oriel and Grossi, Michele and Roggero, Alessandro",
	title = "{Importance sampling for stochastic quantum simulations}",
	eprint = "2212.05952",
	archivePrefix = "arXiv",
	primaryClass = "quant-ph",
	doi = "10.22331/q-2023-04-13-977",
	journal = "Quantum",
	volume = "7",
	pages = "977",
	year = "2023"
}

@article{Wang:2026lqk,
	author = "Wang, Xinzhao and Zhou, Shuo and Wang, Ziruo and Zeng, Pei and Sun, Jinzhao and Zhao, Qi and Gur, Tom and Li, Tongyang",
	title = "{Trotter error compensation with polylogarithmic precision and nested-commutator scaling without ancillas}",
	eprint = "2607.11856",
	archivePrefix = "arXiv",
	primaryClass = "quant-ph",
	month = "7",
	year = "2026"
}

@article{Bark:2026znk,
	author = "Bark, Chan Bin and Lee, Sangjin and Ahn, Kwang Jun and Cheon, Sangmo and Park, Moon Jip and Kim, Youngseok",
	title = "{Constant-Depth Multi-Product Formula for Trotter Error Mitigation in Near-Term Digital Quantum Simulation}",
	eprint = "2607.12225",
	archivePrefix = "arXiv",
	primaryClass = "quant-ph",
	month = "7",
	year = "2026"
}

@article{Zeng:2022pim,
	author = "Zeng, Pei and Sun, Jinzhao and Jiang, Liang and Zhao, Qi",
	title = "{Simple and High-Precision Hamiltonian Simulation by Compensating Trotter Error with Linear Combination of Unitary Operations}",
	eprint = "2212.04566",
	archivePrefix = "arXiv",
	primaryClass = "quant-ph",
	doi = "10.1103/PRXQuantum.6.010359",
	journal = "PRX Quantum",
	volume = "6",
	number = "1",
	pages = "010359",
	year = "2025"
}

@article{Simon:2026dtc,
	author = "Simon, Rick P. A. and Meth, Michael and Martini, Francesco and Tirler, Peter and Jena, Andrew and Ringbauer, Martin and Dellantonio, Luca",
	title = "{An Error-aware and Adaptive Method for the Estimation of Quantum Observables on Qudit-Based Quantum Computers}",
	eprint = "2605.00682",
	archivePrefix = "arXiv",
	primaryClass = "quant-ph",
	month = "5",
	year = "2026"
}

@article{Pillay:2026ojs,
	author = "Pillay, S. M. and David, I. J. and Sinayskiy, I. and Petruccione, F.",
	title = "{Optimising Trotter-Suzuki Simulations of Markovian Open Quantum Systems via Classical Search}",
	eprint = "2607.27060",
	archivePrefix = "arXiv",
	primaryClass = "quant-ph",
	doi = "10.1007/s11128-026-05267-1",
	journal = "Quant. Inf. Comput.",
	volume = "25",
	pages = "267",
	year = "2026"
}

@article{Kim:2026vbh,
	author = "Kim, Leeseok and Garc{\'\i}a-Pintos, Luis Pedro",
	title = "{Randomized product formulas beyond optimal deterministic scaling}",
	eprint = "2608.07720",
	archivePrefix = "arXiv",
	primaryClass = "quant-ph",
	month = "8",
	year = "2026"
}

@article{Cugini:2026nra,
	author = "Cugini, Davide",
	title = "{Pathwise Random Hamiltonian Simulation}",
	eprint = "2608.29756",
	archivePrefix = "arXiv",
	primaryClass = "quant-ph",
	month = "8",
	year = "2026"
}

@article{Pocrnic:2023lrz,
	author = "Pocrnic, Matthew and Hagan, Matthew and Carrasquilla, Juan and Segal, Dvira and Wiebe, Nathan",
	title = "{Composite Qdrift-product formulas for quantum and classical simulations in real and imaginary time}",
	eprint = "2306.16572",
	archivePrefix = "arXiv",
	primaryClass = "quant-ph",
	doi = "10.1103/PhysRevResearch.6.013224",
	journal = "Phys. Rev. Res.",
	volume = "6",
	number = "1",
	pages = "013224",
	year = "2024"
}

@article{Cho:2022gxr,
	author = "Cho, Chien Hung and Berry, Dominic W. and Hsieh, Min-Hsiu",
	title = "{Doubling the order of approximation via the randomized product formula}",
	eprint = "2210.11281",
	archivePrefix = "arXiv",
	primaryClass = "quant-ph",
	doi = "10.1103/PhysRevA.109.062431",
	journal = "Phys. Rev. A",
	volume = "109",
	number = "6",
	pages = "062431",
	year = "2024"
}

@article{Martyn:2026iep,
	author = "Martyn, John M. and Lin, Joshua and Warrington, Neill C. and Chuang, Isaac L. and Daley, Andrew J.",
	title = "{Faster Quantum Monte Carlo Simulation by Random Compilation}",
	eprint = "2609.10486",
	archivePrefix = "arXiv",
	primaryClass = "quant-ph",
	month = "9",
	year = "2026"
}

@article{Fang:2026ckf,
	author = "Fang, Di and Zhang, Jiaqi",
	title = "{Splitting Analysis for Yukawa Potential}",
	eprint = "2607.12153",
	archivePrefix = "arXiv",
	primaryClass = "math.NA",
	month = "7",
	year = "2026"
}

@article{Roy:2024irm,
	author = "Roy, Sayak Guha and Slagle, Kevin",
	title = "{Reweighted Time-Evolving Block Decimation for Improved Quantum Dynamics Simulations}",
	eprint = "2412.08730",
	archivePrefix = "arXiv",
	primaryClass = "quant-ph",
	doi = "10.21468/SciPostPhys.21.1.015",
	journal = "SciPost Phys.",
	volume = "21",
	pages = "015",
	year = "2026"
}

@article{Casas:2026krd,
	author = "Casas, Fernando and Murua, Ander",
	title = "{Error bounds for splitting methods in unitary problems}",
	eprint = "2604.01026",
	archivePrefix = "arXiv",
	primaryClass = "math.NA",
	month = "4",
	year = "2026"
}

@article{Gunther:2026pae,
	author = {G{\"u}nther, Jakob and Harrow, Aram W.},
	title = "{Sublinear-depth Quantum Simulation of Electrons with Atomic Orbitals}",
	eprint = "2610.00950",
	archivePrefix = "arXiv",
	primaryClass = "quant-ph",
	reportNumber = "MIT-CTP/6121",
	month = "10",
	year = "2026"
}

@article{Gunther:2025txp,
	author = {G{\"u}nther, Jakob and Witteveen, Freek and Schmidhuber, Alexander and Miller, Marek and Christandl, Matthias and Harrow, Aram W.},
	title = "{Phase Estimation with Partially Randomized Time Evolution}",
	eprint = "2503.05647",
	archivePrefix = "arXiv",
	primaryClass = "quant-ph",
	doi = "10.1103/ynxb-p2xq",
	journal = "PRX Quantum",
	volume = "7",
	number = "2",
	pages = "020332",
	year = "2026"
}

@article{Ouyang:2020jzz,
	author = "Ouyang, Yingkai and White, David R. and Campbell, Earl T.",
	title = "{Compilation by stochastic Hamiltonian sparsification}",
	eprint = "1910.06255",
	archivePrefix = "arXiv",
	primaryClass = "quant-ph",
	doi = "10.22331/q-2020-02-27-235",
	journal = "Quantum",
	volume = "4",
	pages = "235",
	year = "2020"
}

@article{Li:2021koi,
	author = "Li, Xiantao and Jin, Shi",
	title = "{A partially random Trotter algorithm for quantum Hamiltonian simulations}",
	eprint = "2109.07987",
	archivePrefix = "arXiv",
	primaryClass = "quant-ph",
	doi = "10.1007/s42967-023-00336-z",
	journal = "Commun. Appl. Math. Comput.",
	volume = "7",
	number = "2",
	pages = "442--469",
	year = "2025"
}

@article{Hagan:2022vyw,
	author = "Hagan, Matthew and Wiebe, Nathan",
	title = "{Composite Quantum Simulations}",
	eprint = "2206.06409",
	archivePrefix = "arXiv",
	primaryClass = "quant-ph",
	doi = "10.22331/q-2023-11-14-1181",
	journal = "Quantum",
	volume = "7",
	pages = "1181",
	year = "2023"
}

@article{Chaudhary:2026ksu,
	author = "Chaudhary, Manish and Byrnes, Tim",
	title = "{Efficient quantum simulation of unitary gates based on atomic ensembles}",
	eprint = "2610.22207",
	archivePrefix = "arXiv",
	primaryClass = "quant-ph",
	month = "10",
	year = "2026"
}

@article{Moreno:2026inv,
	author = "Moreno, Carlos A. Pi{\~n}eiro and Krause, Pascal and Bande, Annika",
	title = "{Norm-Aware Product Formulas for Hamiltonian Simulations}",
	eprint = "2610.31070",
	archivePrefix = "arXiv",
	primaryClass = "quant-ph",
	month = "10",
	year = "2026"
}

@book{Nielsen_Chuang_2010,
	place={Cambridge},
	title="{Quantum Computation and Quantum Information: 10th Anniversary Edition}",
	publisher={Cambridge University Press},
	author={Nielsen, Michael A. and Chuang, Isaac L.},
	year={2010},
url = {https://doi.org/10.1017/CBO9780511976667},
doi = {10.1017/CBO9780511976667},
isbn = {\href{https://doi.org/10.1017/CBO9780511976667}{9780511976667}}
}

@article{Atia:2016sax,
	author = "Atia, Yosi and Aharonov, Dorit",
	title = "{Fast-forwarding of Hamiltonians and Exponentially Precise Measurements}",
	eprint = "1610.09619",
	archivePrefix = "arXiv",
	primaryClass = "quant-ph",
	doi = "10.1038/s41467-017-01637-7",
	journal = "Nature Commun.",
	volume = "8",
	number = "1",
	pages = "1572",
	year = "2017"
}

@article{qiskit2024,
	title="{Quantum computing with {Q}iskit}",
	author={Javadi-Abhari, Ali and Treinish, Matthew and Krsulich, Kevin and Wood, Christopher J. and Lishman, Jake and Gacon, Julien and Martiel, Simon and Nation, Paul D. and Bishop, Lev S. and Cross, Andrew W. and Johnson, Blake R. and Gambetta, Jay M.},
	year={2024},
	doix={10.48550/arXiv.2405.08810},
	eprint={2405.08810},
	archivePrefix={arXiv},
	primaryClass={quant-ph},
	addendum = {\url{https://github.com/Qiskit/qiskit}},
}

@article{qiskit-omelyan,
	title="{qiskit-omelyan}",
	author={Marko Maležič},
	publisher    = {GitHub},
	journal = {GitHub repository},
	version      = {v0.2.1},
	year={2026},
	note={\url{https://github.com/MarkoMalezic/qiskit-omelyan}},
	addendum = {part of the \href{https://www.ibm.com/quantum/ecosystem}{Qiskit Ecosystem}},
}

@article{Rajput:2021khs,
	author = "Rajput, Abhishek and Roggero, Alessandro and Wiebe, Nathan",
	title = "{Hybridized Methods for Quantum Simulation in the Interaction Picture}",
	eprint = "2109.03308",
	archivePrefix = "arXiv",
	primaryClass = "quant-ph",
	doi = "10.22331/q-2022-08-17-780",
	journal = "Quantum",
	volume = "6",
	pages = "780",
	year = "2022"
}

@article{Luiz:2026ter,
	author = "Luiz, F. S. and Ferreira, P. N. and de Oliveira, M. C.",
	title = "{Hardware-Efficient Hamiltonian Simulation via Trotter-Initialized Variational Optimization with Native Placement}",
	eprint = "2604.26663",
	archivePrefix = "arXiv",
	primaryClass = "quant-ph",
	month = "4",
	year = "2026"
}

@article{Li:2016vmf,
	author = "Li, Ying and Benjamin, Simon C.",
	title = "{Efficient Variational Quantum Simulator Incorporating Active Error Minimization}",
	eprint = "1611.09301",
	archivePrefix = "arXiv",
	primaryClass = "quant-ph",
	doi = "10.1103/physrevx.7.021050",
	journal = "Phys. Rev. X",
	volume = "7",
	number = "2",
	pages = "021050",
	year = "2017"
}

@article{Pelofske:2026yor,
	author = "Pelofske, Elijah and Eidenbenz, Stephan",
	title = "{Numerical Experiments with Parameter Setting of Trotterized Quantum Phase Estimation for Quantum Hamiltonian Ground State Computation}",
	eprint = "2602.22349",
	archivePrefix = "arXiv",
	primaryClass = "quant-ph",
	reportNumber = "LA-UR-26-20498",
	month = "2",
	year = "2026"
}

@article{Cunningham:2026jgq,
	author = "Cunningham, Joseph and Roland, J{\'e}r{\'e}mie",
	title = "{Alternative adiabatic quantum dynamics with algorithmic applications}",
	eprint = "2605.30110",
	archivePrefix = "arXiv",
	primaryClass = "quant-ph",
	month = "5",
	year = "2026"
}

@article{Farhi:2000ikn,
	author = "Farhi, Edward and Goldstone, Jeffrey and Gutmann, Sam and Sipser, Michael",
	title = "{Quantum Computation by Adiabatic Evolution}",
	eprint = "quant-ph/0001106",
	archivePrefix = "arXiv",
	reportNumber = "MIT-CTP-2936, MIT-CTP-2936",
	month = "1",
	year = "2000"
}

@article{An2021timedependent,
	doi = {10.22331/q-2021-05-26-459},
	url = {https://doi.org/10.22331/q-2021-05-26-459},
	title = {Time-dependent unbounded {H}amiltonian simulation with vector norm scaling},
	author = {An, Dong and Fang, Di and Lin, Lin},
	journal = {{Quantum}},
	issn = {2521-327X},
	publisher = {{Verein zur F{\"{o}}rderung des Open Access Publizierens in den Quantenwissenschaften}},
	volume = {5},
	pages = {459},
	month = may,
	year = {2021}
}

@article{Low:2018pte,
	author = "Low, Guang Hao and Wiebe, Nathan",
	title = "{Hamiltonian Simulation in the Interaction Picture}",
	eprint = "1805.00675",
	archivePrefix = "arXiv",
	primaryClass = "quant-ph",
	month = "5",
	year = "2018"
}

@article{Suzuki:1993timedep,
	title="{General Decomposition Theory of Ordered Exponentials}",
	author={Masuo Suzuki},
	journal={Proceedings of the Japan Academy, Series B},
	volume={69},
	number={7},
	pages={161-166},
	year={1993},
	doi={10.2183/pjab.69.161}
}

@article{Misra:1976by,
	author = "Misra, B. and Sudarshan, E. C. G.",
	title = "{The Zeno's Paradox in Quantum Theory}",
	reportNumber = "ORO-3992-271",
	doi = "10.1063/1.523304",
	journal = "J. Math. Phys.",
	volume = "18",
	pages = "756",
	year = "1977"
}

@article{Facchi:2002jgu,
	author = "Facchi, P. and Pascazio, S.",
	title = "{Quantum Zeno Subspaces}",
	eprint = "quant-ph/0201115",
	archivePrefix = "arXiv",
	doi = "10.1103/PhysRevLett.89.080401",
	journal = "Phys. Rev. Lett.",
	volume = "89",
	number = "8",
	pages = "080401",
	year = "2002"
}

@article{Shen:2022lmk,
	author = "Shen, Yizhi and Klymko, Katherine and Sud, James and Williams-Young, David B. and de Jong, Wibe A. and Tubman, Norm M.",
	title = "{Real-Time Krylov Theory for Quantum Computing Algorithms}",
	eprint = "2208.01063",
	archivePrefix = "arXiv",
	primaryClass = "quant-ph",
	doi = "10.22331/q-2023-07-25-1066",
	journal = "Quantum",
	volume = "7",
	pages = "1066",
	year = "2023"
}

@article{Epperly:2021ugt,
	author = "Epperly, Ethan N. and Lin, Lin and Nakatsukasa, Yuji",
	title = "{A Theory of Quantum Subspace Diagonalization}",
	eprint = "2110.07492",
	archivePrefix = "arXiv",
	primaryClass = "quant-ph",
	doi = "10.1137/21M145954X",
	journal = "SIAM J. Matrix Anal. Appl.",
	volume = "43",
	number = "3",
	pages = "1263--1290",
	year = "2022"
}

@article{Cleve:1997dh,
	author = "Cleve, Richard and Ekert, Artur and Macchiavello, Chiara and Mosca, Michele",
	title = "{Quantum algorithms revisited}",
	eprint = "quant-ph/9708016",
	archivePrefix = "arXiv",
	doi = "10.1098/rspa.1998.0164",
	journal = "Proc. Roy. Soc. Lond. A",
	volume = "454",
	pages = "339",
	year = "1998"
}

@article{Kitaev:1995qy,
	author = "Kitaev, A. Yu.",
	title = "{Quantum measurements and the Abelian stabilizer problem}",
	eprint = "quant-ph/9511026",
	archivePrefix = "arXiv",
	month = "11",
	year = "1995"
}

@article{Grimsley:2018wnd,
	author = "Grimsley, Harper R. and Economou, Sophia E. and Barnes, Edwin and Mayhall, Nicholas J.",
	title = "{An adaptive variational algorithm for exact molecular simulations on a quantum computer}",
	eprint = "1812.11173",
	archivePrefix = "arXiv",
	primaryClass = "quant-ph",
	doi = "10.1038/s41467-019-10988-2",
	journal = "Nature Commun.",
	volume = "10",
	pages = "3007",
	year = "2019"
}

@article{Rahman:2022tkr,
	author = "Rahman, Sarmed A. and Lewis, Randy and Mendicelli, Emanuele and Powell, Sarah",
	title = "{Self-mitigating Trotter circuits for SU(2) lattice gauge theory on a quantum computer}",
	eprint = "2205.09247",
	archivePrefix = "arXiv",
	primaryClass = "hep-lat",
	doi = "10.1103/PhysRevD.106.074502",
	journal = "Phys. Rev. D",
	volume = "106",
	number = "7",
	pages = "074502",
	year = "2022"
}

@article{Berry:2005yrf,
	author = "Berry, Dominic W. and Ahokas, Graeme and Cleve, Richard and Sanders, Barry C.",
	title = "{Efficient Quantum Algorithms for Simulating Sparse Hamiltonians}",
	eprint = "quant-ph/0508139",
	archivePrefix = "arXiv",
	doi = "10.1007/s00220-006-0150-x",
	journal = "Commun. Math. Phys.",
	volume = "270",
	number = "2",
	pages = "359--371",
	year = "2007"
}

@article{Low:2016znh,
	author = "Low, Guang Hao and Chuang, Isaac L.",
	title = "{Hamiltonian Simulation by Qubitization}",
	eprint = "1610.06546",
	archivePrefix = "arXiv",
	primaryClass = "quant-ph",
	doi = "10.22331/q-2019-07-12-163",
	journal = "Quantum",
	volume = "3",
	pages = "163",
	year = "2019"
}

@article{Jordan:2012xnu,
	author = "Jordan, Stephen P. and Lee, Keith S. M. and Preskill, John",
	title = "{Quantum Algorithms for Quantum Field Theories}",
	eprint = "1111.3633",
	archivePrefix = "arXiv",
	primaryClass = "quant-ph",
	doi = "10.1126/science.1217069",
	journal = "Science",
	volume = "336",
	pages = "1130--1133",
	year = "2012"
}

@article{Lin:2022vrd,
	author = "Lin, Lin and Nathan Wiebe",
	title = "{Lecture Notes on Quantum Algorithms for Scientific Computation}",
	eprint = "2201.08309",
	archivePrefix = "arXiv",
	primaryClass = "quant-ph",
	month = "1",
	year = "2026",
	urlx = {https://math.berkeley.edu/~linlin/qasc/},
	note = {continuously updated version: \url{https://math.berkeley.edu/~linlin/qasc/}}
}

@article{Babbush:2025eqc,
	author = "Babbush, Ryan and King, Robbie and Boixo, Sergio and Huggins, William and Khattar, Tanuj and Low, Guang Hao and McClean, Jarrod R. and O'Brien, Thomas and Rubin, Nicholas C.",
	title = "{Grand Challenge of Quantum Applications}",
	eprint = "2511.09124",
	archivePrefix = "arXiv",
	primaryClass = "quant-ph",
	doi = "10.1103/6r9l-lynr",
	journal = "PRX Quantum",
	volume = "7",
	number = "2",
	pages = "020101",
	year = "2026"
}

@article{Yin2026, 
	author = {Jian-Wei Yin and Zi-Rong Chen and Shun Peng and Hao-Chen Luo and Chen-Ning Tao and Si-Wei Tan and Li-Qiang Lu},
	title = "{A Review of Quantum Computing Systems and Software}",
	year = {2026},
	journal = {Journal of Computer Science and Technology},
	volume = {41},
	number = {1},
	pages = {147-169},
	url = {https://www.sciopen.com/article/10.1007/s11390-025-5953-3},
	doi = {10.1007/s11390-025-5953-3},
}

@article{Doyle:2026xcd,
	author = "Doyle, Liam and Seifollahi, Fargol and Singh, Chandralekha",
	title = "{Do we have a quantum computer? Expert perspectives on current state and future prospects}",
	eprint = "2602.15217",
	archivePrefix = "arXiv",
	primaryClass = "physics.ed-ph",
	doi = "10.1103/md6x-pfrr",
	journal = "Phys. Rev. Phys. Educ. Res.",
	volume = "22",
	number = "1",
	pages = "010101",
	year = "2026"
}

@article{Endo:2020kro,
	author = "Endo, Suguru and Cai, Zhenyu and Benjamin, Simon C. and Yuan, Xiao",
	title = "{Hybrid Quantum-Classical Algorithms and Quantum Error Mitigation}",
	eprint = "2011.01382",
	archivePrefix = "arXiv",
	primaryClass = "quant-ph",
	doi = "10.7566/JPSJ.90.032001",
	journal = "J. Phys. Soc. Jap.",
	volume = "90",
	number = "3",
	pages = "032001",
	year = "2021"
}

@article{Vovrosh:2021ocf,
	author = "Vovrosh, Joseph and Khosla, Kiran E. and Greenaway, Sean and Self, Christopher and Kim, Myungshik and Knolle, Johannes",
	title = "{Simple mitigation of global depolarizing errors in quantum simulations}",
	eprint = "2101.01690",
	archivePrefix = "arXiv",
	primaryClass = "quant-ph",
	doi = "10.1103/PhysRevE.104.035309",
	journal = "Phys. Rev. E",
	volume = "104",
	number = "3",
	pages = "035309",
	year = "2021"
}

@article{Urbanek:2021oej,
	author = "Urbanek, Miroslav and Nachman, Benjamin and Pascuzzi, Vincent R. and He, Andre and Bauer, Christian W. and de Jong, Wibe A.",
	title = "{Mitigating depolarizing noise on quantum computers with noise-estimation circuits}",
	eprint = "2103.08591",
	archivePrefix = "arXiv",
	primaryClass = "quant-ph",
	doi = "10.1103/PhysRevLett.127.270502",
	journal = "Phys. Rev. Lett.",
	volume = "127",
	number = "27",
	pages = "270502",
	year = "2021"
}

@article{Kane:2025ybw,
	author = "Kane, Christopher F. and Hariprakash, Siddharth and Bauer, Christian W.",
	title = "{Obtaining continuum physics from dynamical simulations of Hamiltonian lattice gauge theories}",
	eprint = "2506.16559",
	archivePrefix = "arXiv",
	primaryClass = "hep-lat",
	month = "6",
	year = "2025"
}

@article{Low:2016sck,
	author = "Low, Guang Hao and Chuang, Isaac L.",
	title = "{Optimal Hamiltonian Simulation by Quantum Signal Processing}",
	eprint = "1606.02685",
	archivePrefix = "arXiv",
	primaryClass = "quant-ph",
	doi = "10.1103/PhysRevLett.118.010501",
	journal = "Phys. Rev. Lett.",
	volume = "118",
	number = "1",
	pages = "010501",
	year = "2017"
}

@article{Motlagh:2023oqc,
	author = "Motlagh, Danial and Wiebe, Nathan",
	title = "{Generalized Quantum Signal Processing}",
	eprint = "2308.01501",
	archivePrefix = "arXiv",
	primaryClass = "quant-ph",
	doi = "10.1103/PRXQuantum.5.020368",
	journal = "PRX Quantum",
	volume = "5",
	number = "2",
	pages = "020368",
	year = "2024"
}

@article{Laneve:2025nvj,
	author = "Laneve, Lorenzo",
	title = "{Generalized Quantum Signal Processing and Non-Linear Fourier Transform are equivalent}",
	eprint = "2503.03026",
	archivePrefix = "arXiv",
	primaryClass = "quant-ph",
	month = "3",
	year = "2025"
}

@article{Hariprakash:2023tla,
	author = "Hariprakash, Siddharth and Modi, Neel S. and Kreshchuk, Michael and Kane, Christopher F. and Bauer, Christian W.",
	title = "{Strategies for simulating the time evolution of Hamiltonian lattice field theories}",
	eprint = "2312.11637",
	archivePrefix = "arXiv",
	primaryClass = "quant-ph",
	doi = "10.1103/PhysRevA.111.022419",
	journal = "Phys. Rev. A",
	volume = "111",
	number = "2",
	pages = "022419",
	year = "2025"
}

@inproceedings{Gilyen:2018khw,
	author = "Gily{\'e}n, Andr{\'a}s and Su, Yuan and Low, Guang Hao and Wiebe, Nathan",
	title = "{Quantum singular value transformation and beyond: exponential improvements for quantum matrix arithmetics}",
	booktitle = "{51st Annual ACM SIGACT Symposium on Theory of Computing}",
	eprint = "1806.01838",
	archivePrefix = "arXiv",
	primaryClass = "quant-ph",
	doi = "10.1145/3313276.3316366",
	month = "6",
	year = "2018"
}

@article{Draper:2026bcj,
	author = "Draper, Patrick",
	title = "{Block Encoding Non-Abelian Lattice Gauge Theory}",
	eprint = "2608.17115",
	archivePrefix = "arXiv",
	primaryClass = "quant-ph",
	month = "8",
	year = "2026"
}

@article{Rhodes:2024zbr,
	author = "Rhodes, Mason L. and Kreshchuk, Michael and Pathak, Shivesh",
	title = "{Exponential Improvements in the Simulation of Lattice Gauge Theories Using Near-Optimal Techniques}",
	eprint = "2405.10416",
	archivePrefix = "arXiv",
	primaryClass = "quant-ph",
	doi = "10.1103/PRXQuantum.5.040347",
	journal = "PRX Quantum",
	volume = "5",
	number = "4",
	pages = "040347",
	year = "2024"
}

@article{Kikuchi:2023qbb,
	author = "Kikuchi, Yuta and Mc Keever, Conor and Coopmans, Luuk and Lubasch, Michael and Benedetti, Marcello",
	title = "{Realization of quantum signal processing on a noisy quantum computer}",
	eprint = "2303.05533",
	archivePrefix = "arXiv",
	primaryClass = "quant-ph",
	doi = "10.1038/s41534-023-00762-0",
	journal = "npj Quantum Inf.",
	volume = "9",
	pages = "93",
	year = "2023"
}

@article{Runge:1895hdo,
	author = "Runge, C.",
	title = {{Ueber die numerische Aufl{\"o}sung von Differentialgleichungen}},
	doi = "10.1007/BF01446807",
	journal = "Math. Ann.",
	volume = "46",
	number = "2",
	pages = "167--178",
	year = "1895"
}

@article{Kutta,
	author = {Kutta, W.},
	journal = {Zeit. Math. Phys.},
	pages = {435-53},
	title = {Beitrag zur n\"aherungsweisen {I}ntegration
	totaler {D}ifferentialgleichungen},
	volume = 46,
	year = 1901,
	url = {https://archive.org/details/zeitschriftfrma12runggoog/page/434/mode/2up}
}

@article{Butcher_1964, title="{On Runge-Kutta processes of high order}", volume={4}, DOI={10.1017/S1446788700023387}, number={2}, journal={Journal of the Australian Mathematical Society}, author={Butcher, J. C.}, year={1964}, pages={179–194}}

@book{Hairer:1993,
	author = {Hairer, Ernst and Norsett, Syvert and Wanner, Gerhard},
	year = {1993},
	month = {01},
	pages = {},
	title = "{Solving Ordinary Differential Equations I: Nonstiff Problems}",
	volume = {8},
	isbn = {\href{https://link.springer.com/book/10.1007/978-3-540-78862-1}{978-3-540-56670-0}},
	doi = {10.1007/978-3-540-78862-1},
	notex = {\url{https://link.springer.com/book/10.1007/978-3-540-78862-1}},
	addendum = {see esp.\ sec.\ II.16}
}

@book{Hairer:1996,
	title="{Solving Ordinary Differential Equations II: Stiff and Differential-Algebraic Problems}",
	author={Ernst Hairer and Gerhard Wanner},
	year={1996},
	publisher={Springer},
	address={Berlin},
	isbn={\href{https://link.springer.com/book/10.1007/978-3-642-05221-7}{978-3-540-60452-5}},
	doi = {10.1007/978-3-642-05221-7},
	notex = {\url{https://link.springer.com/book/10.1007/978-3-642-05221-7}},
	addendumx = {see esp.\ sec.\ II.16}
}

@article{dormand_prince,
	title = "{A family of embedded Runge-Kutta formulae}",
	journal = "Journal of Computational and Applied Mathematics",
	volume = "6",
	number = "1",
	pages = "19 - 26",
	year = "1980",
	issn = "0377-0427",
	doi = "https://doi.org/10.1016/0771-050X(80)90013-3",
	urlx = "http://www.sciencedirect.com/science/article/pii/0771050X80900133",
	author = "J.R. Dormand and P.J. Prince"
}

@article{matlab_ode,
	title = "{Behind and beyond the Matlab ODE suite}",
	journal = "{Computers \& Mathematics with Applications}",
	volume = "40",
	number = "4",
	pages = "491 - 512",
	year = "2000",
	issn = "0898-1221",
	doi = "https://doi.org/10.1016/S0898-1221(00)00175-9",
	urlx = "http://www.sciencedirect.com/science/article/pii/S0898122100001759",
	author = "R. Ashino and M. Nagase and R. Vaillancourt",
	notex = "{Our algorithm ROW23 is called ode23s here.}"
}

@article{Duane:1987de,
	author =        {Duane, S. and Kennedy, A. D. and Pendleton, B. J. and
	Roweth, D.},
	journal =       {Phys. Lett.},
	pages =         {216-222},
	title =         {{Hybrid Monte Carlo}},
	volume =        {B195},
	year =          {1987},
	doi =           {10.1016/0370-2693(87)91197-X},
}

@article{Urbach:2005ji,
	author = "Urbach, C. and Jansen, K. and Shindler, A. and Wenger, U.",
	title = "{HMC algorithm with multiple time scale integration and mass preconditioning}",
	eprint = "hep-lat/0506011",
	archivePrefix = "arXiv",
	reportNumber = "DESY-05-090, SFB-CPP-05-20",
	doi = "10.1016/j.cpc.2005.08.006",
	journal = "Comput. Phys. Commun.",
	volume = "174",
	pages = "87--98",
	year = "2006"
}

@article{PRXQuantum.2.010342,
	title = "{Real- and Imaginary-Time Evolution with Compressed Quantum Circuits}",
	author = {Lin, Sheng-Hsuan and Dilip, Rohit and Green, Andrew G. and Smith, Adam and Pollmann, Frank},
	eprint = "2008.10322",
	archivePrefix = "arXiv",
	primaryClass = "quant-ph",
	journal = {PRX Quantum},
	volume = {2},
	issue = {1},
	pages = {010342},
	numpages = {15},
	year = {2021},
	month = {Mar},
	publisher = {American Physical Society},
	doi = {10.1103/PRXQuantum.2.010342},
	urlx = {https://link.aps.org/doi/10.1103/PRXQuantum.2.010342}
}

@article{tensor_intro,
	author = "Orus, Roman",
	title = "{A Practical Introduction to Tensor Networks: Matrix Product States and Projected Entangled Pair States}",
	eprint = "1306.2164",
	archivePrefix = "arXiv",
	primaryClass = "cond-mat.str-el",
	doi = "10.1016/j.aop.2014.06.013",
	journal = "Annals Phys.",
	volume = "349",
	pages = "117--158",
	year = "2014"
}

@article{Levenberg:1944,
	ISSN = {0033569X, 15524485},
	URL = {http://www.jstor.org/stable/43633451},
	author = {Kenneth Levenberg},
	journal = {Quarterly of Applied Mathematics},
	number = {2},
	pages = {164--168},
	publisher = {Brown University},
	title = {A METHOD FOR THE SOLUTION OF CERTAIN NON-LINEAR PROBLEMS IN LEAST SQUARES},
	urldate = {2025-12-18},
	volume = {2},
	year = {1944}
}

@article{Marquardt:1963,
	author = {Marquardt, Donald W.},
	title = {An Algorithm for Least-Squares Estimation of Nonlinear Parameters},
	journal = {Journal of the Society for Industrial and Applied Mathematics},
	volume = {11},
	number = {2},
	pages = {431-441},
	year = {1963},
	doi = {10.1137/0111030},
	URL = {https://doi.org/10.1137/0111030},
	eprint = {https://doi.org/10.1137/0111030}	
}

@article{Magnus:1954zz,
	author = "Magnus, Wilhelm",
	title = "{On the exponential solution of differential equations for a linear operator}",
	doi = "10.1002/cpa.3160070404",
	journal = "Commun. Pure Appl. Math.",
	volume = "7",
	pages = "649--673",
	year = "1954"
}

@ARTICLE{2009PhR...470..151B,
	author = {{Blanes}, S. and {Casas}, F. and {Oteo}, J.~A. and {Ros}, J.},
	title = "{The Magnus expansion and some of its applications}",
	journal={Physics Reports},
	publisher={Elsevier BV},
	year = 2009,
	month = jan,
	volume = {470},
	number = {5-6},
	pages = {151-238},
	doi = {10.1016/j.physrep.2008.11.001},
	archivePrefix = {arXiv},
	eprint = {0810.5488},
	primaryClass = {math-ph},
}

@ARTICLE{2017CoPhC.220..243B,
	author = {{Blanes}, Sergio and {Casas}, Fernando and {Thalhammer}, Mechthild},
	title = "{High-order commutator-free quasi-Magnus exponential integrators for non-autonomous linear evolution equations}",
	journal = {Computer Physics Communications},
	year = 2017,
	month = nov,
	volume = {220},
	pages = {243-262},
	doi = {10.1016/j.cpc.2017.07.016},
}
